\documentclass[12pt,a4paper]{article}
\usepackage{latexsym}
\usepackage{amsmath}
\usepackage{amsfonts}
\usepackage{amsbsy}
\usepackage{amssymb}
\usepackage{graphicx}
\usepackage{mathrsfs}
\usepackage{epsfig}
\usepackage{bbm}

\def \R{{\mathbb R}}
\def \C{{\mathbb C}}
\def \Z{\mathbb Z}
\def \N{\mathbb{N}}

\def \arccosh {\text{arccosh }}

\def \G{\mathcal{G}}
\def \Ga{\Gamma}
\def \L{\Lambda}

\def \P{\mathcal{P}}

\def \a{\alpha}
\def \b{\beta}

\def \h{\mathfrak{h}}
\def \H{\mathcal{H}}
\def \d{\delta}

\def \l{\lambda}
\def \g{\gamma}

\def \s{\sigma}
\def \Sig{\Sigma}
\def \e{\varepsilon}

\def \Om{\Omega}
\def \t{\tau}

\def \z{\zeta}

\def \m{\mu}

\def \del{\partial}

\def \tr{\text{ tr}}
\def \arccosh{\text{arccosh }}

\newtheorem{Theorem}{Theorem}[section]

\newtheorem{Lemma}{Lemma}[section]

\numberwithin{equation}{section}

\begin{document}

\title{Gauge-invariant coherent states for Loop Quantum Gravity II: Non-abelian gauge groups\\[20pt]}

\author{\it \small Benjamin Bahr, MPI f\"ur Gravitationsphysik, Albert-Einstein
Institut,\\ \it \small Am M\"uhlenberg 1, 14467 Golm, Germany\\[20pt]
\it \small Thomas Thiemann, MPI f\"ur Gravitationsphysik,
Albert-Einstein Institut, \\ \it \small Am M\"uhlenberg 1, 14467
Golm, Germany;\\ \it \small Perimeter Institute for Theoretical
Physics, \\ \it \small 31 Caroline St. N., Waterloo Ontario N2L 2Y5,
Canada}

\maketitle

\abstract{\noindent This is the second paper concerning
gauge-invariant coherent states for Loop Quantum Gravity. Here, we
deal with the gauge group $SU(2)$, this being a significant
complication compared to the abelian $U(1)$ case encountered in the
previous article. We study gauge-invariant coherent states on
certain special graphs by analytical and numerical methods. We find
that their overlap is Gauss-peaked in gauge-invariant quantities, as
long as states are not labeled by degenerate gauge orbits, i.e.
points where the gauge-invariant configuration space has
singularities. In these cases the overlaps are still concentrated
around these points, but the peak profile exhibits a plateau
structure. This shows how the semiclassical properties of the states
are influenced by the geometry of the gauge-invariant phase space.


\section{Introduction}

\noindent In \cite{GICS-I}, which is the first of a pair of papers,
 gauge-invariant coherent states for Loop Quantum Gravity
(LQG) for the abelian gauge group $U(1)$ were considered. It was
found that these states, which are defined by projecting the
complexifier coherent states \cite{GCS1, GCS2} onto the
gauge-invariant sub-Hilbert space, are labeled by points of the
classical gauge-invariant phase space and a semiclassicality
parameter, which encodes how well the state approximates this
classical point. It was furthermore found that the overlap between
two such states decreases exponentially, as the two labeling points
become distinct. This showed that these states are promising tools
for semiclassical approximations in the gauge-invariant sector of
the theory.\\

In the present paper, we will turn to the case $SU(2)$, which is the
gauge group employed in full LQG. There, we aim to establish similar
results as for $G=U(1)$. The plan for this paper is as follows: In
chapter \ref{Ch:KinematicalFramework} we will shortly review the
kinematical framework of LQG. After describing the Hilbert space and
the action of the (Gauss) gauge transformations, we repeat the
definition of the complexifier coherent states (CCS) of LQG
in chapter \ref{Ch:TheCCS}. 
Also, we will derive a formula for the inner product of two CCSs,
which is defined in terms of geodesics on $SL(2,\C)$. A similar
formula was given for the inner product between two $U(1)$-CCS in
\cite{GICS-I}, and this shows that the CCS can be understood
entirely in terms of geometry. We will also comment on this at the
end of the article.

In chapter \ref{Ch:Examples} we will define the gauge-invariant
coherent states as the projection of the complexifier coherent
states onto the gauge-invariant Hilbert space. We show how the
labels of these states can be interpreted as points in
gauge-invariant phase space, and comment on the Ehrenfest properties
of these states. Furthermore, we will investigate these states on
some simple graphs, in particular the $1$-flower-, the $2$-flower-,
the $3$-bridge- and the tetrahedron graph. In particular, we will
compute the overlap between gauge-invariant coherent states on these
graphs, demonstrating their peakedness properties. While on the
$1$-flower graph all calculations can be done analytically, the
shape of the gauge orbits in the two other examples are already too
complicated to allow for an analytical treatment of these cases.
Rather, we will use numerical methods to investigate the overlap of
gauge-invariant coherent states on these graphs, confirming the
qualitative results about their peakedness properties from the
$1$-flower graph. To compute the overlap, in particular the inner
product of the gauge-invariant coherent states accessible on these
graphs, an algorithm is used to separate the gauge-dependent from
the gauge-invariant degrees of freedom, which resembles a
gauge-fixing procedure.

In chapter \ref{Ch:GeneralProperties} we will work with the general
formula of the inner product of gauge-invariant coherent states on
arbitrary graphs, in order to establish some qualitative results
about the peakedness properties of these states. Specifically, we
will be able to relate the inner product of states on arbitrary
graphs to those on flower graphs. Again, we will employ gauge-fixing
methods on this behalf. This will allow for a qualitative
description of the overlaps of gauge-invariant coherent states
labeled by degenerate gauge orbits.

We will close with a summary of the present work, as well as with a
conclusion and an outlook.\\

\section{The kinematical setting of
LQG}\label{Ch:KinematicalFramework}

\noindent We shortly repeat the kinematical framework of LQG.
Detailed expositions can be found in \cite{INTRO, INTRO3, ROVELLISBUCH,  ALLMT} and in the references therein.

The starting point of LQG is the phase space of Ashtekar-connections
$A_a^I(x)$ and electric fluxes $E_J^b(y)$, both fields on a $3$-dim
spatial manifold $\Sig$, which can be thought of as a Cauchy surface
in space-time. The Poisson structure is given by

\begin{eqnarray*}
\big\{A_a^I(x)\,,\,A_b^J(y)\big\}\;&=&\;\big\{E_I^a(x)\,,\,E_J^b(y)\big\}\;=\;0\\[5pt]
\big\{A_a^I(x)\,,\,E_J^b(y)\big\}\;&=&\;8\pi
G\b\;\d_{b}^a\,\d_J^I\;\d(x-y).
\end{eqnarray*}

\noindent The fields are not free, but subject to so-called
constraints, which are phase-space functions, i.e. functions of
$A_a^I(x)$ and $E_J^b(y)$. They encode the diffeomorphism-invariance
of the theory, and the Einstein equations. The reduced phase space
consists of all phase space points $A,\,E$ where the constraints
vanish. On this set, the constraints act as gauge transformations,
and the set of gauge orbits is the physical phase space.
 The set of constraints is divided into the Gauss constraints
$G_I(x)$, the diffeomorphism constraints $D_a(x)$ and the Hamilton
constraints $H(x)$. It is the set of Gauss constraints that is of
particular importance in the rest of this work. \\

The holonomy-flux-algebra generated by holonomies of $A_a^I(x)$
along edges and electric fields $E_J^b(y)$ smeared over $2$-dim
surfaces is the starting point of the quantization programme. There
is a unique representation of this algebra in which the spatial
diffeomorphisms, which are generated by the diffeomorphism
constraints $D_a(x)$, act unitarily \cite{LOST}. This kinematical
Hilbert space $\H_{kin}$, on which the holonomy-flux algebra is
represented, also carries a representation of the constraint
algebra, and is given by
\begin{eqnarray}
\H_{kin}\;=\;\bigoplus_{\g\in\Ga}\;\H_{\g}
\end{eqnarray}

\noindent Here, $\Ga$ is the set of all graphs $\g$ in $\Sig$ which
consist of embedded, regular, analytic edges. Each Hilbert space
$\H_{\g}$ is separable. If $\g$ is a graph with $E$ edges and $V$
vertices, $\H_{\g}$ is isomorphic to
\begin{eqnarray}\label{Gl:IsomorphismBetweenHGammaAndLTwoOverSU2}
\H_{\g}\;\simeq\;L^2\big(G^E,\,d\m_H^{\otimes E}\big),
\end{eqnarray}

\noindent where $G=SU(2)$ is the gauge group acting on the fields
$(A_a^I, E^b_J)$, and $d\m_H$ is the normalized Haar measure on $G$.
Each of these $\H_{\g}$ is left invariant by the gauge
transformations induced by the Gauss-constraints $G_I(x)$. The
restriction $\G_{\g}$ of the set of gauge transformations to
$\H_{\g}$ is isomorphic to
\begin{eqnarray}
\G_{\g}\;\simeq\;G^V,
\end{eqnarray}

\noindent where $V$ is the number of vertices in the graph $\g$. The
action of an element $\vec k\in G^V$ on a square-integrable function
$f:G^E\to \C$ in $\H_{\g}$ is given by the following formula:

\begin{eqnarray}\label{Gl:ActionOfGaugeGroup}
\a_{\vec k} \tilde f\;\big(h_{e_1},\ldots,h_{e_E})\;:=\;\tilde
f\;\big(k_{b(e_1)}h_{e_1}k_{f(e_1)}^{-1},\ldots,k_{b(e_E)}h_{e_E}k_{f(e_E)}^{-1}\big),
\end{eqnarray}

\noindent where $b(e_m)$ and $f(e_m)$ are the beginning- and
end-point of the edge $e_m$. So, the gauge transformations act only
at the vertices of a graph.

In particular, one can write down the projector onto the
gauge-invariant Hilbert space for functions in $\H_{\g}$:

\begin{eqnarray}\label{Gl:Projector}
\P
f(h_{e_1},\ldots,h_{e_E})\;&:=&\;\int_{G^V}d\m_H(k_1,\ldots,k_V)\a_{k_1,\ldots
k_V}\,f(h_{e_1}\ldots,h_{e_E})\\[5pt]\nonumber
&=&\;\int_{G^V}d\m_H(k_1,\ldots,k_V)f\Big(k_{b(e_1)}h_{e_1}k_{f(e_1)}^{-1},\ldots,k_{b(e_E)}h_{e_E}k_{f(e_E)}^{-1}\Big)
\end{eqnarray}

\noindent Since $G^V$ is compact, the integral exists and defines a
projector
\begin{eqnarray*}
\P:\;\H_{\g}\;\longrightarrow\;\H_{\g}
\end{eqnarray*}

\noindent onto a sub-Hilbert space of $\H_{\g}$. In particular, the
gauge-invariant functions on a graph form a subset of all
square-integrable functions on a graph. The gauge-invariant Hilbert
spaces can be described using intertwiners between irreducible
representations of $SU(2)$, and a basis for the gauge-invariant
Hilbert spaces $\P\H_{\g}$ can be written down in terms of
gauge-invariant spin network functions \cite{SNF}.\\


\section{Complexifier coherent states}\label{Ch:TheCCS}

\noindent The complexifier coherent states (CCS) are states first
constructed for quantum mechanics on arbitrary compact Lie groups
\cite{HALL1, HALL3}. They are natural generalizations of the
harmonic oscillator coherent states (HOCS) for quantum mechanics on
a real line, which are given by
\begin{eqnarray}
|z\rangle\;=\;\sum_{n=0}^{\infty}\,\frac{z^n}{\sqrt{n!}}\;|n\rangle.
\end{eqnarray}

\noindent The HOCS can be seen as minimal uncertainty states, or
states that correspond to the system of being in a quantum state
close to a
classical phase space point $z=q+ip$.\\

Complexifier coherent states for quantum mechanics on a compact Lie
group $G$ are given by a choice of a complexifier $\hat C$, a
positive number $t>0$ and a point $g\in G^{\C}$. If one chooses the
complexifier to be the negative Laplacian on $G$, then the states
are given by

\begin{eqnarray}\nonumber
\psi^t_g(h)\;&=&\;\sum_{\pi}e^{-\l_{\pi}\frac{t}{2}}d_{\pi}\,\tr\;\pi(g
h^{-1})\\[5pt]\label{Gl:DefinitionOfComplexifierCoherntStates-2}
\end{eqnarray}

\noindent where the sum runs over all irreducible finite-dimensional
representations $\pi$ of $G$. In the case of $G=SU(2)$, this reads

\begin{eqnarray}
\psi^t_g(h)&=&\;\sum_{j\in\frac{1}{2}\N}\;e^{-j(j+1)\frac{t}{2}}\,(2j+1)\;\tr_j(g^{-1}h)
\end{eqnarray}

\noindent where $\tr_j$ is the trace in the spin-$j$ representation,
and $g\in SU(2)^{\C}=SL(2,\C)$. These states and their properties
have been investigated in \cite{GCS1, GCS2}. It could be shown that
these states are sharply peaked around their labels $g\in SL(2,\C)$,
i.e. the overlap

\begin{eqnarray}\label{Gl:CCS-Overlap}
i^t(g_1,g_2)\;=\;\frac{|\langle\,\psi^t_{g_1}|\psi^t_{g_2}\rangle|^2}{\|\psi^t_{g_1}\|^2\,\|\psi_{g_2}^t\|^2}
\end{eqnarray}

\noindent equals $1$ for $g_1=g_2$, but goes to $0$ faster than any
power of $t$ as $t\to 0$, i.e. is $O(t^{\infty})$. Furthermore, the
$SU(2)$-CCS reproduce classical values of quantized phase space
functions. For example, let $f:SL(2,\C)\to\R$ be a function on phase
space, and $\hat f$ the corresponding quantized operator. Then

\begin{eqnarray}\label{Gl:CCS-Ehrenfest}
\frac{\langle\,\psi_{g_1}^t|\,\hat
f\,|\psi_{g_2}^t\,\rangle}{\langle\,\psi_{g_1}^t|\psi_{g_2}^t\,\rangle}\;=\;f(g_2)\;(1+F(g_1,g_2,
t))
\end{eqnarray}

\noindent where $F$ is a function of $g_1,g_2\in SL(2,\C)$ growing
only polynomially in the complex directions, and is of order $O(t)$.
This gives an immediate interpretation of the labeling parameter
$g\in SL(2,\C)$: it corresponds to a point in classical phase space,
and (\ref{Gl:CCS-Overlap}), (\ref{Gl:CCS-Ehrenfest}) show that
$\psi^t_g$ defines a quantum state being close to the classical
state $g$, with quantum fluctuations determined by $t$. In
particular, one can see that the limit $t\to 0$ corresponds to the
semiclassical limit of the theory, being classical mechanics on
$SU(2)$.

The CCS are useful for the LQG framework, since, restricting the
gravitational degrees of freedom to a finite graph $\g\subset\Sig$
with $E$ edges and $V$ vertices, general relativity in the Ashtekar
formulation can be formulated with a classical configuration space
$SU(2)^E$, i.e. phase space $SL(2,\C)^E$. The corresponding CCS for
this system is given by a tensor product

\begin{eqnarray}\label{Gl:CCSWithTensorProduct}
\psi^t_{g_1,\ldots, g_E}(h_1,\ldots,
h_E)\;=\;\prod_{m=1}^E\,\psi^t_{g_m}(h_m).
\end{eqnarray}

\noindent These states are special cases for the semiclassical
states employed in LQG. Generalized forms correspond to different
complexifiers $\hat C$ on $\H_{kin}$, which make the dependence of
the graph topology more explicit, or superpositions of states
(\ref{Gl:CCSWithTensorProduct}) over different graphs. Details can
be found in \cite{CE, TINA1}

\subsection{Geometric version of the inner product}

\noindent In the following we will give a geometric interpretation
of the inner product of the $SU(2)$-CCS. This has already been done
for their $U(1)$-counterparts in \cite{GICS-I}. We start from the
form of the CCS (\ref{Gl:DefinitionOfComplexifierCoherntStates-2}),
from which the inner product of two CCS can be computed:

\begin{eqnarray}\label{Gl:Geometric-I}
\langle\,\psi^t_g|\psi^t_{g'}\,\rangle\;=\;\sum_{j\in\frac{1}{2}\N}e^{-j(j+1)t}(2j+1)\,\tr_j(g^{\dag}g')
\end{eqnarray}

\noindent In \cite{GCS1}, an application of the Poisson summation
formula was used to bring (\ref{Gl:Geometric-I}) into the form
\begin{eqnarray}\label{Gl:Geometric-II}
\langle\,\psi^t_g|\psi^t_{g'}\,\rangle\;=\;\frac{2e^{\frac{t}{4}}}{\pi}\sqrt\frac{\pi}{t}^3\,\sum_{n\in\Z}\,\frac{z-2\pi
 n}{\sin (z\,-\,2\pi  n)}\;\exp\left[-\frac{(z\,-\,2\pi
n)^2}{t}\right]
\end{eqnarray}

\noindent where
\begin{eqnarray}\label{Gl:Geometric-III}
\cos z\;=\;\frac{1}{2}\tr\;(g^{\dag}g').
\end{eqnarray}

\noindent Note that by (\ref{Gl:Geometric-III}) $z$ is only defined
up to a sign and a shift $z+2\pi n$ for some $n\in \Z$. But
(\ref{Gl:Geometric-II}) has exactly the corresponding symmetries,
such that the formula is well-defined. In \cite{GCS2} the form
(\ref{Gl:Geometric-II}) was chosen for convenience: In the limit
$t\to 0$ one can, if one chooses $\text{Re } z\in(-\pi,\,\pi)$,
neglect all the terms with $n\neq 0$ in (\ref{Gl:Geometric-II}),
since they are exponentially damped compared to the $n=0$ term. This
simplified calculations immensely. We will use the same form, partly
for the same reasons, but also in order to show how the inner
product between the CCS on $SU(2)$ can be interpreted via the
geometry on its complexification $SU(2)^{\C}=SL(2,\C)$. In order to
do this, we have to talk about geodesics on $SL(2,\C)$.

\subsection{Geodesics on $SL(2,\C)$}

\noindent The exponential map
\begin{eqnarray}\label{Gl:CoolsteKarteAufSL(2,C)}
\C^3\,\ni\,\vec z\;\longmapsto\;e^{i \vec z\cdot\vec\s}\;=\;\cos
z\;+\;i\frac{\sin z}{z}\vec z\cdot\vec\s\;\;\in\;SL(2,\C)
\end{eqnarray}

\noindent where $\s_I$, $I=1,\,2,\,3$ are the Pauli spin matrices
and $\vec z\in \C^3$, such that $i\vec z\cdot\vec
\s\,\in\,\mathfrak{sl}_2\C$. Furthermore
$z:=\sqrt{z_1^2+z_2^2+z_3^2}$ is determined up to a sign, but the
functions in (\ref{Gl:CoolsteKarteAufSL(2,C)}) are symmetric in $z$,
so everything is well-defined.

This does not only define a chart that covers all of $SL(2,\C)$, but
allows one also to write down the geodesics immediately. To do so,
we first note that the group structure on $SL(2,\C)$ determines a
(pseudo-)metric. On its Lie algebra $\mathfrak{sl}_2\C$, the Killing
form

\begin{eqnarray}\label{Gl:KillingForm}
(A,\,B)\;\longmapsto\;-\frac{1}{2}\tr\;(AB)
\end{eqnarray}

\noindent can be defined and is a bilinear non-degenerate form on
$\mathfrak{sl}_2\C\;=\;T_1SL(2,\C)$. Since $SL(2,\C)$ is a group,
one can pull back (\ref{Gl:KillingForm}) to every other tangent
space on $SL(2,\C)$ by right (or, which gives the same result, left)
multiplication. This defines a non-degenerate, bi-invariant
pseudo-Riemannian metric called the Killing metric, which is given
in coordinates by

\begin{eqnarray}\label{Gl:HolomorpheMetrikAufSL(2,C)InKoordinaten}
h_{IJ}\;:=\;\frac{1}{2}\tr\big((g^{-1}\del_Ig)(g^{-1}\del_Jg)\big).
\end{eqnarray}

\noindent From this a connection can be formed, and geodesics can be
defined. In particular, the geodesics going through $1\in SL(2,\C)$
are given by

\begin{eqnarray}\label{Gl:TheGeodesicsThrough1OnSL2C}
t\;\longmapsto\;e^{it z_I\s^I}
\end{eqnarray}

\noindent where $\vec z\in \C^3$, such that $i\vec z\cdot\vec
\s\,\in\,\mathfrak{sl}_2\C$ is the velocity vector at $t=0$. An
immediate consequence of the bi-invariance of the Killing metric is
that, given a geodesic $t\to\g(t)$ on $SL(2,\C)$, for any
$g_1,\,g_2\in SL(2,\C)$, also $t\to g_1\g(t)g_2$ is a geodesic,
which allows to compute all geodesics from $g_1$ to $g_2$ by
computing all from $1$ to $g=g_1^{-1}g_2$. In general, there will be
more than one, and from (\ref{Gl:TheGeodesicsThrough1OnSL2C}) one
can see that any geodesic from $1$ to $g=\exp\,i\vec z\cdot\vec\s$
is given by
\begin{eqnarray}
[0,\,1]\;\ni\;t\;\longmapsto\;e^{it\vec z\cdot\vec\s}.
\end{eqnarray}

\noindent Thus, different geodesics arise from the fact that
$\exp\,i\vec z\cdot\vec\s=\exp\,i\vec w\cdot\vec\s$ for $\vec
z\neq\vec w$, i.e. $\vec z\in\C^3$ is not uniquely determined by
$g\in SL(2,\C)$ We will classify the different possible cases as
follows:

\begin{Lemma}\label{Lem:FourCases}
Let $\vec z\in \C^3$. Then $z=\sqrt{\vec z\cdot\vec z}$ is
determined only up to a sign. Define $g=\exp\,i\vec z\cdot\vec\s$,
then exactly one of the following is true:
\begin{itemize}
\item \text{Case 1: }$z=0\;$\text{and}$\;\vec z\neq 0$
\item \text{Case 2: }$z=2\pi n\;$\text{for}$\;n\in Z\backslash\{0\}$ or $\vec z=0$
\item \text{Case 3: }$z=\pi\,+\,2\pi n\;$\text{for}$\;n\in Z$
\item \text{Case 4: }$z\notin 2\pi\Z$ and $z\notin\pi+2\pi\Z$
\end{itemize}

\noindent In case $1$, there is exactly one geodesic from $1$ to
$g$, in cases $2$ and $3$ there are uncountably many, and in case
$4$ there are countably many geodesics from $1$ to $g$.
\end{Lemma}

\noindent\textbf{Proof:} In the first case, where
$z_1^2+z_2^2+z_3^2=0$, we have

\begin{eqnarray}
g\;=\exp\,i\vec z\cdot\vec\s\;=\;1\;+\;i\vec z\cdot\vec \s.
\end{eqnarray}

\noindent Let now $\vec w\in \C^3$ with $\exp\,i\vec w\cdot\s=g$.
Then it follows from (\ref{Gl:CoolsteKarteAufSL(2,C)}) that $\cos
w=1$. In particular, $w=2\pi n$ for some $n\in \Z$. But if $n\neq
0$, then $\sin w\,/w=0$ and $g=1$, hence $\vec z=0$, which is
excluded in this case. So $n=0$ and hence $i\vec w\cdot\vec\s =
i\vec z\cdot\vec\s$, so $\vec z=\vec w$. In particular, the vector
$\vec z$ is unique. Hence, there is only one
geodesic from $1$ to $g$.\\

The second and third case can readily be seen to correspond to $g=1$
and $g=-1$. Also, it can easily be seen that all geodesics from $1$
to itself are given by

\begin{eqnarray}
[0,1]\;\ni\;t\;\longmapsto\;e^{2\pi n t \vec\phi\cdot\vec\s}
\end{eqnarray}

\noindent for arbitrary $n\in \Z$ and $\vec\phi\in S^2$, i.e.
$\|\vec\phi\|=1$. Similarly, all geodesics from $1$ to $-1$ are
given by

\begin{eqnarray}\label{Gl:AllGeodesicsFrom1Tog-}
[0,1]\;\ni\;t\;\longmapsto\;e^{(2\pi n+\pi) t \vec\phi\cdot\vec\s}.
\end{eqnarray}

\noindent So, there are uncountably many geodesics in both cases.\\

Case 4 is the generic case: Let $\vec w,\vec z\in\C^3$ such that
$e^{ i\vec z\cdot\vec\s}=e^{ i\vec w\cdot\vec\s}$. Thus, by
(\ref{Gl:CoolsteKarteAufSL(2,C)}) we have $\cos w=\cos z$, so
\begin{eqnarray}\label{Gl:VerschiedeneHolomorpheLaengen}
w^2\,=\,(z+2\pi n)^2
\end{eqnarray}

\noindent for some $n\in \Z$. Furthermore, from the linear
independence of the $\s$-matrices, it follows that
\begin{eqnarray}\label{Gl:HowToGetAllGeodesics}
\vec w\;=\;\frac{z+2\pi n}{z}\,\vec z.
\end{eqnarray}

\noindent Thus we see that, if one chooses $\vec z\in\C^3$ such that
$g=\exp\,i\vec z\cdot\vec\s$, then all other $\vec w$ with
$g=\exp\,i\vec w\cdot\vec\s$ can be obtained by
(\ref{Gl:HowToGetAllGeodesics}), via letting $n$ go through $\Z$.
Thus, there are countably many geodesics from $1$ to $g$.

This finishes the proof.\\

\noindent Knowing this, we can turn to defining the complex length of a geodesic, or more generally of an $h$-regular curve.
Let $t\to\g(t)$ be an $h$-regular curve. That is, $\g$ does not
necessarily have to be a geodesic, but the Killing form $h$ shall
never annihilate the velocity vector, i.e. $h(\dot\g,\,\dot\g)\neq
0$ along the curve. Then, with the help of the Killing metric, the
complex length of $\g$ can be defined via

\begin{eqnarray}\label{Gl:DefinitionComplexLength}
l(\g)\;:=\;\int_{\g}\;\sqrt{h(\dot\g,\,\dot\g)}.
\end{eqnarray}

\noindent Note that this is in complete analogy to the definition of
the length of a curve with the help of a Riemannian metric, the only
difference being a sign issue: Since generically
$h(\dot\g,\,\dot\g)$ will be complex on the path, its square-root is
determined up to a sign, but since the integrand never vanishes,
this sign can be chosen such that $\sqrt{h(\dot\g,\,\dot\g}$ is
smooth along the curve, and there are exactly two such choices.
Thus, $l(\g)$ is only defined up to a sign. \\

Now, to compute the complex length of a geodesic (\ref{Gl:AllGeodesicsFrom1Tog-}) from $1$ to $g=\exp\,i\vec z\cdot\s$ for
$\vec z\in\C^3$ is easy: Since the velocity vector is parallely transported along geodesics, the integrand
in (\ref{Gl:DefinitionComplexLength}) is constant along $[0,\,1]$. In particular, it equals its
value at $t=0$, which is nothing but the Killing form on $\mathfrak{sl}_2\C$, i.e.
\begin{eqnarray}
l(\g)\;=\;\sqrt{\vec z\cdot\vec z}\;=\;z.
\end{eqnarray}

\noindent Note that there are also geodesics $t\to \g(t)$ where
$\dot\g(t)$ vanishes. But since the velocity vector of a geodesic is
parallely transported, $\dot \g$ vanishes identically for these
curves, and thus we can consistently define the complex length of
such a geodesic to be $0$.\\

Knowing this, we arrive at the main part of this section.

\begin{Lemma}
Let $g_1,\,g_2\in SL(2,\C)$ and
\begin{eqnarray}
j^t(g_1,\,g_2)\;=\;\langle\psi_{g_1}^t|\psi^t_{g_2}\rangle
\end{eqnarray}

\noindent the inner product between complexifier coherent states.
Apart from the cases $g_1^c=g_2$ or $g_1^c=-g_2$, when there are
uncountably many geodesics between  $g_1^c$ and $g_2$, the inner
product is given by

\begin{eqnarray}\label{Gl:InnerProductAsSumOverGeodesics}
j^t(g_1,\,g_2)\;=\;
\frac{2e^{\frac{t}{4}}}{\pi}\sqrt{\frac{\pi}{t}}^3\;
\sum_{\begin{array}{c}\text{\scriptsize \rm $\g$
 geodesic }\\\text{\scriptsize \rm from $g_1^c$ to
$g_2$}\end{array}}\frac{l(\g)}{\sin l(\g)}e^{-\frac{l(\g)^2}{t}}
\end{eqnarray}
\end{Lemma}

\noindent\textbf{Proof:} Write $g=g^{\dag}_1g_2$. Since neither
$g_1^c=g_2$ nor $g_1^c=-g_2$, it follows that neither $g=1$ nor
$g=-1$. Then we know by Lemma \ref{Lem:FourCases} that either there
is only one geodesic from $1$ to $g$, or there are countably many.
Assume there only to be one, i.e.
\begin{eqnarray}
g\;=\;\exp\,i\vec z\cdot\vec\s
\end{eqnarray}

\noindent with a unique $\vec z\in\C^3$ with $z=0$. It follows from
(\ref{Gl:CoolsteKarteAufSL(2,C)}) that $\frac{1}{2}\tr\;g\;=\;0$.
So, by carefully performing the limit $z\to 0$ in
(\ref{Gl:Geometric-II}), one can see that all terms cancel apart
from the $n=0$ term, which equals $1$, so
\begin{eqnarray}
j^t(g_1,\,g_2)\;=\;\frac{2e^{\frac{t}{4}}}{\pi}\sqrt{\frac{\pi}{t}}^3\;
\end{eqnarray}

\noindent Since there is only one geodesic from $1$ to $g$ of
complex length $0$, the same holds true for $g_1^c$ and $g_2$ by the
bi-invariance of the Killing metric. Thus, we have

\begin{eqnarray}
j^t(g_1,\,g_2)\;=\;
\frac{2e^{\frac{t}{4}}}{\pi}\sqrt{\frac{\pi}{t}}^3\;
\sum_{\begin{array}{c}\text{\scriptsize \rm $\g$
 geodesic }\\\text{\scriptsize \rm from $g_1^c$ to
$g_2$}\end{array}}\frac{l(\g)}{\sin l(\g)}e^{-\frac{l(\g)^2}{t}},
\end{eqnarray}

\noindent where the sum consists only of one term, with $l(\g)=0$.

Now assume the generic case, i.e. there are infinitely many
geodesics from $g_1^c$ to $g_2$, or equivalently from $1$ to
$g=g_1^{\dag}g_2$. Choose a $\vec z\in\C^3$ such that $g=\exp\,i
\vec z\cdot\vec\s$. We have already seen that this amounts to
choosing a geodesic from $1$ to $g$. Then $z=\sqrt{\vec z\cdot\vec
z}$ is the complex length of this geodesic, which is determined up
to a sign. By (\ref{Gl:VerschiedeneHolomorpheLaengen}) we see that
all other complex lengths are determined by letting $n$ run through
$\Z$, which tells us that

\begin{eqnarray}\nonumber
j^t(g_1,\,g_2)\;&=&\;\frac{2e^{\frac{t}{4}}}{\pi}\sqrt{\frac{\pi}{t}}^3\;
\sum_{n\in\Z}\;=\;\frac{z-2\pi n}{\sin(z-2\pi
n)}\exp\left[-\frac{(z-2\pi
n)^2}{t}\right]\\[5pt]
\;&=&\;\frac{2e^{\frac{t}{4}}}{\pi}\sqrt{\frac{\pi}{t}}^3\;
\sum_{\begin{array}{c}\text{\scriptsize \rm $\g$
 geodesic }\\\text{\scriptsize \rm from $g_1^c$ to
$g_2$}\end{array}}\frac{l(\g)}{\sin l(\g)}e^{-\frac{l(\g)^2}{t}}.
\end{eqnarray}

\noindent which proves the Lemma.\\

\noindent The inner product between two coherent states is thus a
function intimately related to the geometry on $SL(2,\C)$.
Unfortunately, the formula (\ref{Gl:InnerProductAsSumOverGeodesics})
is valid only if $g_1^{\dag}g_2\neq\pm 1$, since in these two cases
the sum over geodesics makes no sense, because there are uncountably
many. On the other hand, these two cases are of measure zero in all
of $(g_1,g_2)\in SL(2,\C)\times SL(2,\C)$, and the formula is
holomorphic in both variables, and can be regarded to have removable
singularities at $g_1^{\dag}g_2=\pm 1$.\\

The conjugation $g\to g^c$ can, in the polar decomposition of $g=Hu$
into hermitean $H$ and unitary $u$, be written as
\begin{eqnarray*}
H u\;\longmapsto\;H^{-1}u.
\end{eqnarray*}

\noindent For the norm of a complexifier coherent state we thus get
\begin{eqnarray*}
\big
\|\psi^t_g\big\|^2\;&=&\;\frac{2e^{\frac{t}{4}}}{\pi}\sqrt{\frac{\pi}{t}}^3\;\sum_{\begin{array}{c}\text{\scriptsize
$\g$
 geodesic }\\\text{\scriptsize from $H^{-1}u$ to
$H u$}\end{array}}\frac{l(\g)}{\sin l(\g)}\;\;e^{-\frac{l(\g)^2}{t}}
=\;\frac{2e^{\frac{t}{4}}}{\pi}\sqrt{\frac{\pi}{t}}^3\;\sum_{\begin{array}{c}\text{\scriptsize
$\g$
 geodesic }\\\text{\scriptsize from $1$ to
$H^2$}\end{array}}\frac{l(\g)}{\sin
l(\g)}\;\;e^{-\frac{l(\g)^2}{t}}\\[5pt]
&=&\;\frac{2e^{\frac{t}{4}}}{\pi}\sqrt{\frac{\pi}{t}}^3\;\sum_{n\in\Z}\frac{l+2\pi
in}{\sinh (l+2\pi in)}\;\;e^{\frac{(l+2\pi
in)^2}{t}}
=\;\frac{2e^{\frac{t}{4}}}{\pi}\sqrt{\frac{\pi}{t}}^3\;\frac{l}{\sinh
l}\;\;e^{\frac{l^2}{t}}\big(1+O(t^{\infty})\big)
\end{eqnarray*}

\noindent with $l=|\vec l|$, $\vec l\in\R^3$ and  $H=e^{l_I\s^I}$.
For two complexifier coherent states peaked on elements $g\in
SU(2)\subset SL(2,\C)$, we thus have

\begin{eqnarray}
\|\psi^t_g\|=\sqrt{\frac{2e^{\frac{t}{4}}}{\pi}\sqrt{\frac{\pi}{t}}^3}(1+O(t^{\infty}))
\end{eqnarray}

\noindent Note that  the complex length
(\ref{Gl:DefinitionComplexLength}) of a curve $\g$ lying entirely in
$SU(2)\subset SL(2,\C)$ is real, and its square $l(\g)^2$ coincides
with the geodesic distance $d(\g)^2$ on $SU(2)$ determined by the
Killing metric. Thus, the overlap of two complexifier coherent
states peaked on elements $g_1,g_2\in SU(2)\subset SL(2,\C)$ is
given by

\begin{eqnarray*}
\frac{\langle\psi_{g_1}^t|\psi^t_{g_2}\rangle}{\|\psi^t_{g_1}\|\;\|\psi^t_{g_2}\|}\;=\;\sum_{\begin{array}{c}\text{\scriptsize
$\g$
 geodesic }\\\text{\scriptsize from $g_1$ to
$g_2$}\end{array}}\;\frac{d(\g)}{\sin
d(\g)}\;e^{-\frac{d(\g)^2}{t}}\;\big(1+O(t^{\infty})\big),
\end{eqnarray*}

\noindent since $g^c=g$ for $g\in SU(2)$.

For states labeled by elements $g_1,g_2\in SU(2)$, this immediately
shows the nice peakedness properties these states have: For two
states being labeled by different elements, the overlap is,
basically, a sum over terms being proportional to Gaussians in the
lengths of geodesics from one to the other elements. For small
$t>0$, the term with the shortest distance dominates all other
terms, and the overlap is nearly proportional to a Gaussian in the
geodesic distance on $SU(2)$.

\section{Gauge-invariant coherent states for
$G=SU(2)$}\label{Ch:Examples}

\noindent In the following, we will construct the gauge-invariant
coherent states on a graph by projecting the complexifier coherent
states on the gauge-invariant subspace. This will result in states
labeled by gauge-equivalence classes of phase space points which
will be identified with points gauge-invariant phase space.
Afterwards, we will examine these states for some particular graphs.
There we will show that gauge-invariant coherent states labeled by
points in gauge-invariant phase space have an overlap that vanishes
exponentially as the two points become distinct. This will
demonstrate the peakedness properties for these states, which make
these states useful for semiclassical analysis in the
gauge-invariant sector.

\subsection{Gauge-invariant
functions}\label{Ch:GaugeInvariantFunctions}

\noindent  The Hilbert space $\H_{\g}$ consists of functions on
$SU(2)^E$, square-integrable with respect to the $E$-fold Haar
measure $d\m^{\otimes E}_H$.
\begin{eqnarray*}
H_{\g}\;\simeq\;L^2\left(G^E,\,d\m_{H}^{\otimes E}\right),
\end{eqnarray*}

 In chapter
\ref{Ch:KinematicalFramework}, the gauge action of the Gauss
constraints on the kinematical Hilbert space has been discussed, in
particular the gauge transformation of a function cylindrical on a
graph (\ref{Gl:ActionOfGaugeGroup}). Denote by $\pi:
SU(2)^E\;\to\;SU(2)^E/SU(2)^V$ the projection map of $SU(2)^E$ to
the space of orbits under the action

\begin{eqnarray}\label{Gl:Chapter5Gauge}
\a\;:\;SU(2)^V\,&\times&\,\H_{\g}\;\longrightarrow\;\H_{\g}\\[5pt]
\left(\a_{k_{v_1},\ldots,k_{v_V}}f\right)(h_{e_1},\ldots,h_{e_E})\;&=&\;f\big(k_{b(e_1)}^{-1}h_{e_1}k_{f(e_1)},\;\ldots\;,k_{b(e_E)}^{-1}h_{e_E}k_{f(e_E)}\big).
\end{eqnarray}

\noindent of the gauge group $SU(2)^V$. Then
\begin{eqnarray}
\P\H_{\g}\;\simeq\;L^2\Big(SU(2)^E/SU(2)^V,\,\pi_*d\m^{\otimes
E}_H\Big).
\end{eqnarray}

\noindent So, the orbifold $SU(2)^E/SU(2)^V$ is treated as
gauge-invariant configuration space. This space can be formulated
nicely in terms of cohomology, in particular

\begin{eqnarray*}
SU(2)^E/SU(2)^V\;\simeq\;H^1(\g, SU(2))
\end{eqnarray*}

\noindent where $H^1(\g, SU(2))$ is the first cohomology group on
the graph $\g$ with values in $SU(2)$. See appendix
\ref{Ch:CohomologyNonabeilan} for
details.\\

\subsection{Gauge-invariant coherent states and gauge orbits}

\noindent In the following sections we will describe the projections
of the complexifier coherent states (CCS)
\begin{eqnarray}\label{Gl:Chapter5CCS}
\psi^t_{g_1,\ldots,g_E}(h_1,\ldots,h_E)\;=\;\prod_{e\in
E(\g)}\;\sum_{j_e\in\frac{1}{2}\N}e^{-j_e(j_e+1)\frac{t}{2}}(2j_e+1)\,\tr_{j_e}\,\left(g_eh_e^{-1}\right)
\end{eqnarray}

\noindent onto the gauge-invariant subspace
$\P\H_{\g}\subset\H_{\g}$. This will define the
\emph{gauge-invariant coherent states}, and we will investigate some
of their properties. Since the gauge integral will be too
complicated to perform exactly, we will have to rely on numerical
investigations in some cases, where analytical methods are not
enough.

By the explicit form (\ref{Gl:Chapter5CCS}), the action
(\ref{Gl:Chapter5Gauge}) also induces an action on the set of
coherent states, hence an action on $SL(2,\C)^E$ via
\begin{eqnarray}
\a_{\vec k}\psi_{\vec
g}^t\;=\;\a_{k_{v_1},\ldots,k_{v_V}}\psi^t_{g_{e_1},\ldots,g_{e_E}}\;=\;\psi^t_{k_{f(e_1)}g_{e_1}k^{-1}_{b(e_1)},\;\ldots\;,k_{f(e_E)}g_{e_E}k^{-1}_{b(e_E)}}\;=:\;\psi^t_{\a_{\vec
k}\vec g}.
\end{eqnarray}

The gauge-invariant coherent states are the image of the
complexifier coherent states (\ref{Gl:Chapter5CCS}) under the action
of $\P$ (\ref{Gl:Projector}). 
\begin{eqnarray}\label{Gl:DefinitionGaugeInvariantCoherentState}
\Psi^t_{[\vec g]}\;:=\;\P
\psi^t_{\vec{g}}\;=\;\int_{SU(2)^V}d\m_h(\vec k)\,\psi^t_{\a_{\vec
k}\vec g}.
\end{eqnarray}

\noindent Naively, one could think that the labeling of the
gauge-invariant coherent state is now $[\vec g]\in\big\{\a_{\vec
k}\vec g,\;\;\vec k=(k_{v_1},\ldots,k_{v_V})\in SU(2)^{V}\big\}$.
However, this is not the case! $\P$ can map states labeled by $\vec
g_1,\vec g_2\in SL(2,\C)^E$ to the same state in $\P\H_{\g}$
although there is no $\vec k\in SU(2)^V$ such that $\a_{\vec k}\vec
g_1=\vec g_2$. The reason for this is the holomorphic dependence of
the CCS on their labeling parameter. To shed some light on this
issue, we need the following lemma.

\begin{Lemma}\label{Lem:ComplexShiftOfIntegrationOnLieGroups}
Let $G$ be a compact Lie group and $G^{\C}$ its complexification.
Let $f:G^{\C}\to\C$ be an analytic function (i.e. holomorphic w.r.t
to the complex structure on $G^{\C}$. Then, for any $g,g'\in
G^{\C}$, he have
\begin{eqnarray*}
\int_{G}d\m_H(h)\;f(h)\;=\;\int_{G}d\m_H(h)\;f(ghg')
\end{eqnarray*}
\end{Lemma}

\noindent\textbf{Proof:} Since $f$ is analytic on $G^{\C}$, it is in
particular continuous on $G$, which is compact. The restriction of
$f$ on $G$, and hence $|\tilde f|^2$ are uniformly bounded functions
on $G$. So the restriction is square-integrable, and by the
Peter-Weyl theorem it can be decomposed into its Fourier
coefficients
\begin{eqnarray*}
f(h)\;=\;\sum_{\pi}\sum_{m,n}\;\sqrt{d_\pi}\;c_{\pi m
n}\;\pi(h)_{mn}
\end{eqnarray*}

\noindent for all $h\in G$. The functions $h\to\pi(h)_{mn}$ are all
anayltic, and is the matrix element function of an irreducible
representation of $G^{\C}$. Then, for any $g,g'$
\begin{eqnarray*}
\int_Gd\m_H(h)\,f(ghg')\;&=&\;\sum_{\pi}\,\sum_{m,n}\sqrt{d_{\pi}}\,c_{\pi
m n}\int_{G}d\m_H(h)\,\pi(g h g')_{mn}\\[5pt]
&=&\;\sum_{\pi}\sum_{m,m',n',n}\sqrt{d_{\pi}}\pi(g)_{mm'}c_{\pi m
n}\pi(g')_{n'n}\int_Gd\m_H(h)\pi(h)_{mn}
\end{eqnarray*}

\noindent But the integral only gives a contribution for the trivial
representation $\pi_0$, which is $1$-dimensional, the corresponding
Fourier coefficient corresponding to the integral of $f$ over $G$.
So the only term remaining is
\begin{eqnarray*}
\int_Gd\m_H(h)\,f(ghg')\;&=&\;\pi_0(g)\,c_{0}\,\pi_0(g')\;=\;c_{0}\;=\;\int_Gd\m(h)\,f(h)
\end{eqnarray*}

\noindent This completes the proof.\\

\noindent Of course, the above lemma carries over to integrals over
the $V$-fold product $SU(2)^V$. The gauge action $\a_{\vec k}$ of
$SU(2)^V$
 on the label set $SL(2,\C)^E$
\begin{eqnarray}\label{Gl:GaugeActionOnLabelSet}
\a_{k_{v_1},\ldots,k_{v_V}}\left(g_{e_1},\ldots,g_{e_E}\right)\;=\;\left(k_{f(e_1)}g_{e_1}k^{-1}_{b(e_1)},\;\ldots\;,k_{f(e_E)}g_{e_E}k^{-1}_{b(e_E)}\right)
\end{eqnarray}

\noindent can obviously continued analytically to an action of
$SL(2,\C)^V$ on $SL(2,\C)^E$ simply by taking the same formula
(\ref{Gl:GaugeActionOnLabelSet}), but allowing for $\vec k\in
SL(2,\C)^V$. Let $\vec g_1,\vec g_2\in SL(2,\C)^E$ such that they
can be related by such a gauge transformation with elements $\vec
l\in SL(2,\C)^V$:
\begin{eqnarray}
\vec g_1\in SL(2,\C)^E,\qquad \vec g_2\,=\,\a_{\vec l}\vec
g_1,\quad\vec l\in SL(2,\C)^V.
\end{eqnarray}

\noindent Then, since the coherent states (\ref{Gl:Chapter5CCS})
depend analytically on their labels, i.e. for each $h\in SU(2)$ the
function
\begin{eqnarray}
SL(2,\C)\,\ni\,g\;\longmapsto\;\psi_g^t(h)\;\in\;\C
\end{eqnarray}

\noindent is analytic, also the gauge-integrand is analytic. In
particular, by (\ref{Gl:Chapter5Gauge}), for each $\vec h\in
SU(2)^E$ and $\vec g\in SL(2,\C)^E$, the function
\begin{eqnarray}
SU(2)^V\;\ni\vec k\;\longmapsto\;\psi^t_{\a_{\vec k}\vec g}(\vec
h)\;\in\;\C
\end{eqnarray}

\noindent is analytic, and can in an obvious way be extended to an
analytic function on $SL(2,\C)^V$. Hence, by lemma
\ref{Lem:ComplexShiftOfIntegrationOnLieGroups}, we get
\begin{eqnarray}
\P\psi^t_{\vec g_1}\;&=&\;\int_{SU(2)^V}d\m_H(\vec
k)\;\psi_{\a_{\vec k}\vec g_1}^t\\[5pt]\nonumber
\;&=&\;\int_{SU(2)^V}d\m_H(\vec k)\;\psi_{\a_{\vec k}\a_{\vec l}\vec
g_1}^t\;=\;\int_{SU(2)^V}d\m_H(\vec k)\;\psi_{\a_{\vec k}\vec
g_2}^t\;=\;\P\psi_{\vec g_2}^t,
\end{eqnarray}

\noindent where the shift of integration variables $k_{v}\to
k_v\,l_v$ has been used.\\

The gauge invariant coherent states are thus labeled by a
semiclassicality parameter $t>0$ and an equivalence class $[\vec g]$
which is given by
\begin{eqnarray*}
[\vec g]\;:=\;\big\{\a_{\vec k}\vec g,\;\;\vec
k=(k_{v_1},\ldots,k_{v_V})\in SL(2,\C)^{V}\big\}.
\end{eqnarray*}

\noindent Note that, at a second glance, it is quite natural that
the gauge-invariant states are labeled by orbits of the complexified
gauge action on $SL(2,\C)^E$, i.e. on the orbifold
$SL(2,\C)^E/SL(2,\C)^V$ rather than $SL(2,\C)^E/SU(2)^V$, for
dimensional reasons. The complexifier coherent states $\psi^t_{\vec
g}$ are functions on $SU(2)^E$, which can be  seen as configuration
space, while the states are labeled by elements in $SL(2,\C)^E$, a
space which has twice the number of real dimensions than $SU(2)^E$.
In particular, since $SL(2,\C)^E$ is diffeomorphic to the tangent
bundle of $SU(2)^E$, it can be identified with the phase space of a
system whose configuration space is $SU(2)^E$.

The gauge-invariant coherent states are, as we have seen in chapter
\ref{Ch:GaugeInvariantFunctions}, functions on the set
$SU(2)^E/SU(2)^V$ of gauge orbits of $SU(2)^E$ under the gauge
action $SU(2)^V$. This is no manifold any more, since it contains
singular points. So, it is not clear what the tangent bundle of it
might be. However, the set $SL(2,\C)^E/SL(2,\C)^V$ of orbits of the
tangent bundle $SL(2,\C)^E$ under the complexified gauge action of
$SL(2,\C)^V$ serves as a natural candidate: It is an orbifold of
twice the (real) dimension than $SU(2)^E/SU(2)^V$. This is not the
case for $SL(2,\C)^E/SU(2)^V$, as one can readily see. So, if we
view $SU(2)^E/SU(2)^V$ as the gauge-invariant configuration space,
then $SL(2,\C)^E/SL(2,\C)^V$ is the natural candidate for its
gauge-invariant phase space.

\subsection{On semiclassical properties}

\noindent We are interested in the peakedness properties of the
gauge-invariant coherent states, in particular the inner product
\begin{eqnarray}\label{Gl:GICSInnerProduct}
J^t\left([\vec g_1],\,[\vec
g_2]\right)\;:=\;\big\langle\;\Psi^t_{[\vec g_1]}\big|\Psi^t_{[\vec
g_2]}\;\big\rangle,
\end{eqnarray}

\noindent as well as their overlap

\begin{eqnarray}\label{Gl:GICSOverlap}
I^t\left([\vec g_1],\,[\vec
g_2]\right)\;:=\;\frac{\big|\big\langle\;\Psi^t_{[\vec
g_1]}\big|\Psi^t_{[\vec
g_2]}\;\big\rangle\big|^2}{\big\|\Psi^t_{[\vec
g_1]}\big\|^2\,\big\|\Psi^t_{[\vec g_2]}\big\|^2}.
\end{eqnarray}

\noindent  If the states $\Psi^t_{[\vec g]}$ are to be good
semiclassical states, their overlap (\ref{Gl:GICSOverlap}) should be
sharply peaked at $g_1\approx g_2$ in the semiclassical limit $t\to
0$, as well as they should approximate operators corresponding to
gauge-invariant obervables well, i.e. they should satisfy the
gauge-invariant version of the Ehrenfest property
(\ref{Gl:CCS-Ehrenfest}). In these two conditions, the semiclassical
properties of the states are encoded. They simply amount to the fact
that, in the limit $t\to 0$, a state labeled by a classical,
gauge-invariant  phase space point $[\vec g]$ approach the classical
state given by this point. In particular, taking expectation values
of a quantum observable in this state amounts to evaluation of the
corresponding classical observable. Furthermore, the overlap between
states labeled with different phase space points $[\vec g_1],
\,[\vec g_2]$ is nearly zero, i.e. quantum fluctuations between
different states become small.

As we have already reported in chapter \ref{Ch:TheCCS}, the
complexifier coherent states have these properties, and thus are
viable semiclassical states for the gauge-variant, i.e. kinematical
Hilbert space, approximating gauge-variant classical observables.
Establishing analogous properties for the gauge-invariant coherent
states, resembling the gauge-invariant physical systems, is the main
purpose of this article. The rest of this work will be devoted to
investigate the gauge-invariant inner product, in order to show the
peakedness properties of the gauge-invariant overlap
(\ref{Gl:GICSOverlap}). Note that, as soon as peakedness properties
are established, the Ehrenfest property follow immediately from the
fact that the corresponding property (\ref{Gl:CCS-Ehrenfest}) holds
for the CCS. Let $f$ denote a gauge-invariant observable, i.e. a
function on phase space being invariant under the action of the
gauge group. In particular, $f$ is a function of $\vec g\in
SL(2,\C)^E$ that depends only on the gauge orbits $[\vec g]$. The
corresponding quantization should yield an operator $\hat f$ on
$\h_{\g}$ that leaves the gauge-invariant Hilbert space $\P\H_{\g}$
invariant, i.e. can be restricted to an operator to $\P\H_{\g}$.
This implies
\begin{eqnarray}\label{Gl:GaugeInvariantObservable}
\big[\P,\,\hat f\big]\;=\;0.
\end{eqnarray}

\noindent From (\ref{Gl:GaugeInvariantObservable}) and
(\ref{Gl:DefinitionGaugeInvariantCoherentState}), we get

\begin{eqnarray}
\big\langle\,\Psi^t_{[\vec g_1]}\,\big|\,\hat f\,|\,\Psi^t_{[\vec
g_2]}\,\big\rangle\;&=&\;\langle \,\psi^t_{\vec g_1}|\,\hat
f\,|\P\,\psi^t_{\vec g_2} \rangle\\[5pt]\nonumber
&=&\; \int_{SU(2)^V}d\m^{\otimes
V}_H(k_1,\ldots,k_V)\;\big\langle\psi^t_{\vec g_1}\,|\,\hat
f\,|\,\psi^t_{\a_{\vec k}\vec g_2}\,\rangle.
\end{eqnarray}

\noindent But the expectation value can, since the complexifier
coherent states have the Ehrenfest property
(\ref{Gl:CCS-Ehrenfest}), be written as
\begin{eqnarray}\nonumber
\big\langle\,\Psi^t_{[\vec g_1]}\,\big|\,\hat f\,|\,\Psi^t_{[\vec
g_2]}\,\big\rangle\;&=&\;\int_{SU(2)^V}d\m^{\otimes
V}_H(k_1,\ldots,k_V)\,f(\a_{\vec k}\vec g_2)\;\langle\psi^t_{\vec
g_1}\,|\,\psi^t_{\a_{\vec k}\vec
g_2}\,\rangle\;(1\;+\;O(t))\\[5pt]\nonumber
&=&\;f\big([\vec g_2]\big)\int_{SU(2)^V}d\m^{\otimes
V}_H(k_1,\ldots,k_V)\;\langle\psi^t_{\vec g_1}\,|\,\psi^t_{\a_{\vec
k}\vec
g_2}\,\rangle\\[5pt]\label{Gl:GICS-Ehrenfest}
&=&\;f\big([\vec g_2])\;\big\langle\;\psi^t_{\vec
g_1}\big|\P\psi^t_{\vec g_2}\;\big\rangle\;(1+O(t))\\[5pt]\nonumber
&=&\;f\big([\vec g_2])\;\big\langle\;\Psi^t_{[\vec
g_1]}\big|\Psi^t_{[\vec g_2]}\;\big\rangle\;(1+O(t)).
\end{eqnarray}

\noindent where
\begin{eqnarray}\label{Gl:EichinvariantesSkalarprodukt-2}
\langle\,\Psi^t_{[\vec g_1]}\,|\,\Psi^t_{[\vec g_2]}\rangle\;&=&\;\langle\,\P\psi^t_{\vec g_1}\,|\,\P\psi^t_{\vec g_2}\rangle\\[5pt]\nonumber
&=&\;\langle\,\psi^t_{\vec g_1}\,|\P\psi^t_{\vec
g_2}\rangle\;=\;\int_{G^V}d\m_H^{\otimes V}(\vec
k)\;\langle\,\psi^t_{\vec g_1}\,|\,\psi^t_{\a_{\vec k}\vec
g_2}\rangle
\end{eqnarray}

\noindent has been used.

Note that in this considerations, we have pulled the error terms
$O(t)$ out of the integral, which is allowed, since the error term
is of order $O(t)$ on all of the integration range $SU(2)^V$, and
$SU(2)^V$ is compact. Hence we can replace the errors by their
maximal absolute value, pull that out of the integral and still have
made only an $O(t)$-error.

These considerations show that as soon as we have established the
peakedness properties for the gauge-invariant coherent states, the
corresponding Ehrenfest properties automatically follow. We will try
to establish these
peakedness properties in the rest of this article.\\

\noindent The gauge-invariant coherent states can be put into an
explicit form \cite{GCS2}. Starting from (\ref{Gl:Chapter5CCS}), one
can perform the gauge-integrals (\ref{Gl:Projector}) and arrive at
\begin{eqnarray}\label{Gl:EichinvarianteKohaerenteZustaendeNachIntertwinernZerlegt}
\Psi^t_{[\vec g]}(\vec h)\;=\;\sum_{\vec j,\vec
I}e^{-j_{e_1}(j_{e_1}+1)\frac{t}{2}-\ldots-j_{e_E}(j_{e_E}+1)\frac{t}{2}}T_{\g,\vec
j,\vec I}(\vec g)\overline{T_{\g,\vec j,\vec I}(\vec h)}.
\end{eqnarray}

\noindent Here, the labels denote a distribution $\vec j$ of
irreducible representations of $SU(2)$ among the edges of the graph
$\g$, and $\vec I$ denote a distribution of intertwiners among the
vertices of the graph. These $T_{\g,\vec j,\vec I}(\vec h)$ form an
orthonormal basis for the gauge-invariant Hilbert space $\P\H_{\g}$. The intertwiners $\vec I$, in particular the basis
functions $T_{\g,\vec\pi,\vec I}$ can be found by employing the
coupling scheme for angular momenta, in particular they contain the
$3Nj$-symbols \cite{INTRO, SNF}. Although these symbols are known
in principle, they become arbitrarily complicated for large large
graphs. This makes the expressions
(\ref{Gl:EichinvarianteKohaerenteZustaendeNachIntertwinernZerlegt})
rather difficult to handle, in particular it is not clear how to
extract information about peakedness properties, apart from the simplest
 example. This is the reason why we will pursue another way, already suggested in \cite{GCS2}:\\

Remember that (\ref{Gl:EichinvariantesSkalarprodukt-2}) holds:

\begin{eqnarray}\label{Gl:EichinvariantesSkalarprodukt}
\langle\,\Psi^t_{[\vec g_1]}\,|\,\Psi^t_{[\vec
g_2]}\rangle\;&=&\;\int_{G^V}d\m_H^{\otimes V}(\vec
k)\;\langle\,\psi^t_{\vec g_1}\,|\,\psi^t_{\a_{\vec k}\vec
g_2}\rangle.
\end{eqnarray}

 \noindent So the inner product between gauge-invariant
coherent states can be obtained by an integral over the inner
product between complexifier coherent states. This inner product is
given by (\ref{Gl:Geometric-II}) and (\ref{Gl:Geometric-III}).

\begin{eqnarray}\label{Gl:Geometric-IV}
\langle\,\psi^t_{g_1}|\psi^t_{g_2}\,\rangle\;=\;\frac{2e^{\frac{t}{4}}}{\pi}\sqrt\frac{\pi}{t}^3\,\sum_{n\in\Z}\,\frac{z-2\pi
 n}{\sin (z\,-\,2\pi  n)}\;\exp\left[-\frac{(z\,-\,2\pi
n)^2}{t}\right]
\end{eqnarray}

\noindent with
\begin{eqnarray}
\cos z\;=\;\frac{1}{2}\tr\;(g_1^{\dag}g_2).
\end{eqnarray}

\noindent From the gauge-invariant inner product, we can immediately
obtain the overlap between  gauge-invariant coherent states by

\begin{eqnarray}\nonumber
\frac{\big|\langle\,\Psi^t_{[\vec g_1]}\,|\,\Psi^t_{[\vec
g_2]}\rangle\big|^2}{\|\Psi^t_{[\vec g_2]}\|^2\|\Psi^t_{[\vec
g_2]}\|^2}\;=\;\frac{\left[\int_{G^V}d\m_H^{\otimes V}(\vec
k)\;\langle\,\psi^t_{\vec g_1}\,|\,\psi^t_{\a_{\vec k}\vec
g_2}\rangle\right]\left[\int_{G^V}d\m_H^{\otimes V}(\vec
k)\;\langle\,\psi^t_{\vec g_2}\,|\,\psi^t_{\a_{\vec k}\vec
g_1}\rangle\right]}{\left[\int_{G^V}d\m_H^{\otimes V}(\vec
k)\;\langle\,\psi^t_{\vec g_1}\,|\,\psi^t_{\a_{\vec k}\vec
g_1}\rangle\right]\left[\int_{G^V}d\m_H^{\otimes V}(\vec
k)\;\langle\,\psi^t_{\vec g_2}\,|\,\psi^t_{\a_{\vec k}\vec
g_2}\rangle\right]}.\\[5pt]\label{Gl:DieVierHeiligenIntegrale}
\end{eqnarray}

\noindent This form of the inner product between gauge-invariant
coherent states is  not explicit, since it still contains the gauge
integral. But this form has some advantages over the corresponding formula
using the explicit form of the gauge-invariant coherent states
(\ref{Gl:EichinvarianteKohaerenteZustaendeNachIntertwinernZerlegt}):
First, we will see that we can, by some gauge-fixing procedure,
extract the gauge-invariant

quantities that label the gauge orbits. These not only have a
geometrical interpretation, the procedure also allows for a
numerical investigation that enables us to show that the
gauge-invariant coherent states are peaked on these gauge-invariant
quantities. Second, starting from
(\ref{Gl:DieVierHeiligenIntegrale}), we will be able to prove a
general theorem about the peakedness of these states on the singular
points of the space of gauge-invariant
quantities.\\

\noindent So, four integrals of the type of
(\ref{Gl:EichinvariantesSkalarprodukt}) have to computed. It is
exactly these integrals that turn out to be not solvable in a closed
form for most graphs, apart from the simplest ones. In particular,
we will be able to compute the overlap analytically for the
$1$-flower graph, but on the $2$-flower-, the 3-bridge- and the
tetrahedron graph, we will employ numerical integration of
(\ref{Gl:DieVierHeiligenIntegrale}), in order to investigate the
overlap. After that, we will demonstrate some qualitative properties
of the overlap (\ref{Gl:DieVierHeiligenIntegrale}) in chapter
\ref{Ch:GeneralProperties}.

\subsection{The 1-flower graph}

\noindent In the following we will consider the gauge-invariant
coherent states on simple graphs for the case of $G=SU(2)$. In
particular, we will have to evaluate integrals of the kind
(\ref{Gl:EichinvariantesSkalarprodukt}). This will not be possible
analytically for most cases, but we will do this where we can, and
use numerics in all other cases.\\

We start with the one-flower graph. Here we will be able to perform
the results analytically, since the intertwiner for this graph are
just the traces.

With
(\ref{Gl:EichinvarianteKohaerenteZustaendeNachIntertwinernZerlegt}),
we get
\begin{eqnarray}
\Psi^t_{[g]}(h)\;=\;\sum_{j\in\frac{1}{2}\N}e^{-j(j+1)\frac{t}{2}}\;\tr_j(g)\;\tr_j(h).
\end{eqnarray}

\noindent The inner product between two such coherent states labeled
by $[g_1]$ and $[g_2]$ is then given by
\begin{eqnarray}
\big\langle\Psi^t_{[g_1]}\,\big|\,\Psi^t_{[g_2]}\big\rangle\;=\;\sum_{j\in\frac{1}{2}\N}e^{-j(j+1)t}\;\overline{\tr_j(g_1)}\,\tr_j(g_2).
\end{eqnarray}

\noindent Consider the $j=\frac{1}{2}$-representation of $SL(2,\C)$.
There are invertible $2\times 2$-matrices $\Phi_1$,$\Phi_2$, such
that
\begin{eqnarray}\label{Gl:DiagonalFormVonSL(2,C)Matrizen}
\Phi_2\pi_{\frac{1}{2}}(g_2)\Phi_2^{-1}\;=\;\left(\begin{array}{ll}\l&\;0\\0&\;\l^{-1}\end{array}\right)\;,
\qquad\Phi_1\pi_{\frac{1}{2}}(g_1)\Phi_1^{-1}\;=\;\left(\begin{array}{ll}\m&\;0\\0&\;\m^{-1}\end{array}\right)
\end{eqnarray}

\noindent for nonzero $\l,\m\in\C$. By an explicit formulation of
the irreducible representations of $SL(2,\C)$,  one gets \cite{GCS1}
\begin{eqnarray*}
\tr\,\pi_j(g_2)\;=\;\frac{\l^{2j+1}-\l^{-2j-1}}{\l-\l^{-1}}\;,\qquad\tr\,\pi_j(g_1)\;=\;\frac{\m^{2j+1}-\m^{-2j-1}}{\m-\m^{-1}}.
\end{eqnarray*}

\noindent Note that these expressions are invariant under the change
of $\l\to\l^{-1}$ or $\m\to\mu^{-1}$, as are the choices of
$\Phi_1,\,\Phi_2$ in  (\ref{Gl:DiagonalFormVonSL(2,C)Matrizen}).

Write $\l=e^{i z}$ and $\m=e^{i w}$, then we get
\begin{eqnarray}\nonumber
\big\langle\Psi^t_{[g_1]}\,\big|\,\Psi^t_{[g_2]}\big\rangle\;&=&\;\sum_{j\in\frac{1}{2}\N}e^{-j(j+1)t}\;\frac{e^{-i(2j+1)\bar
w}-e^{i(2j+1)\bar w}}{e^{-i\bar w}-e^{i\bar
w}}\;\frac{e^{i(2j+1)z}-e^{-i(2j+1)z}}{e^{iz}-e^{-iz}}\\[5pt]\label{Gl:Gleichungsschritt}
&=&\;e^{-\frac{1}{4}}\sum_{n=1}^{\infty}e^{-n^2t}\;\frac{e^{-in\bar
w}-e^{in\bar w}}{e^{-i\bar w}-e^{i\bar
w}}\;\frac{e^{inz}-e^{-inz}}{e^{iz}-e^{-iz}}\\[5pt]\nonumber
&=&\;\frac{1}{2}e^{-\frac{1}{4}}\sum_{n\in\Z}e^{-n^2t}\;\frac{e^{-in\bar
w}-e^{in\bar w}}{e^{-i\bar w}-e^{i\bar
w}}\;\frac{e^{inz}-e^{-inz}}{e^{iz}-e^{-iz}}
\end{eqnarray}

\noindent By the Poisson summation formula,
\begin{eqnarray}\label{Gl:PoissonSummationExponential}
\sum_{n\in\Z}\;e^{-n^2t}e^{inA}\;=\;\sqrt{\frac{\pi}{t}}\sum_{n\in\Z}\;e^{-\frac{(A+2\pi
n)^2}{4t}}.
\end{eqnarray}

\noindent With (\ref{Gl:PoissonSummationExponential}),
(\ref{Gl:Gleichungsschritt}) can be rewritten as follows:
\begin{eqnarray}
\big\langle\Psi^t_{[g_1]}\,\big|\,\Psi^t_{[g_2]}\big\rangle\;&=&\;2\,e^{-\frac{1}{4}}\,\sqrt\frac{\pi}{t}\sum_{n\in\Z}\frac{e^{-\frac{(\bar
w-z-2\pi n)^2}{4t}}-e^{-\frac{(\bar w+z-2\pi n)^2}{4t}}}{(e^{-i\bar
w}-e^{i\bar w})(e^{iz}-e^{-iz})}.
\end{eqnarray}

\noindent if we choose $z$ and $w$ to lie inside the strip
$[-\pi,\,\pi]\times i\R$, the inner product can be approximated by

\begin{eqnarray}\label{Gl:FinalerUeberlapp}
\big\langle\Psi^t_{[g_1]}\,\big|\,\Psi^t_{[g_2]}\big\rangle\;&=&\;2\,e^{-\frac{1}{4}}\,\sqrt\frac{\pi}{t}\frac{e^{-\frac{(\bar
w-z)^2}{4t}}-e^{-\frac{(\bar w+z)^2}{4t}}}{(e^{-i\bar w}-e^{i\bar
w})(e^{iz}-e^{-iz})}\;\big(1+O(t^{\infty})\big).
\end{eqnarray}

\noindent With
\begin{eqnarray*}
e^A\,-\,e^B\;=\;2\,e^{\frac{A+B}{2}}\,\sinh{\frac{A-B}{2}},
\end{eqnarray*}

\noindent this can be put into the form

\begin{eqnarray*}
\big\langle\Psi^t_{[g_1]}\,\big|\,\Psi^t_{[g_2]}\big\rangle\;&=&\;\,e^{-\frac{1}{4}}\,\sqrt\frac{\pi}{t}\frac{e^{-\frac{\bar
w^2+z^2}{4t}} \,\sinh\frac{\bar w z}{2t}}{\sinh\bar w\sinh
z}\;\big(1+O(t^{\infty})\big),
\end{eqnarray*}

\noindent which, for the overlap, gives us
\begin{eqnarray}\label{Gl:FinalerUeberlapp}
I^t\big([g_1],\,[g_2]\big)\;=\;
\frac{\big|\langle\Psi^t_{[g_1]}\,\big|\,\Psi^t_{[g_2]}\big\rangle\big|^2}{\big\|\Psi^t_{[g_1]}\big\|\;\big\|\Psi^t_{[g_2]}\big\|}\;=
\;\frac{\sinh\frac{\bar w z}{2t}\sinh\frac{ w \bar
z}{2t}}{\sinh\frac{|w|^2}{2t}\sinh\frac{|z|^2}{2t}}\;\big(1+O(t^{\infty})\big)
\end{eqnarray}

\noindent Note that the complex numbers $z,\,w$ appearing here are
related to the labels of the coherent states via $2\cos
z=\tr\;{g_2}$ and $2\cos w=\tr\; g_1$. Note also how the
gauge-invariant coherent states only depend on the trace $\tr_j(g)$.
Since the only invariant information under conjugation in
$SL(2,\C)$: $g\to kgk^{-1}$ is the trace, we see that the
gauge-invariant coherent states on the one-flower graph only depend
on the conjugation classes, i.e. elements in $SL(2,\C)/SL(2,\C)$, as
has been stated in the last section.
\newpage

We plot the overlap between $\Psi^t_{\cos w}$
and $\Psi^t_{\cos z}$, for $w=1+1i$, depending on $z$, with $t=0.25$.\\

\begin{figure}[hbt!]
\begin{center}
    \includegraphics[scale=0.75]{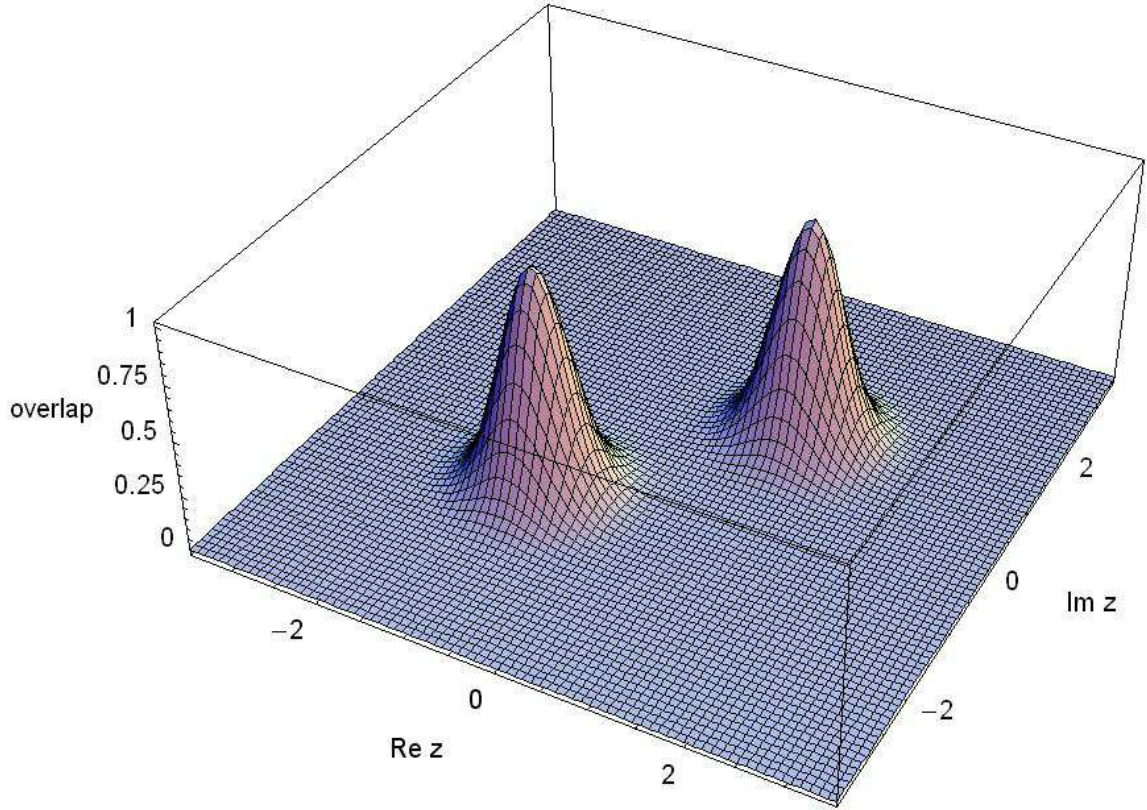}
    \end{center}
    \label{fig:Fig1}
    \caption{\small Overlap between gauge-invariant coherent states $\Psi^t_{\cos z}$ and $\Psi^t_{\cos w}$ with $w=1+i$, $t=0.25$, depending on $z$.}
\end{figure}

\noindent As one can see, the overlap is peaked at $z=\pm w$, as
should be the case, since $z$ and $w$ are, if both in the region
$[-\pi,\,\pi]\times i\R$, only determined up to a sign. The overlap
profile is a gaussian, as can readily be seen.\\

\begin{figure}[hbt!]
\begin{center}
    \includegraphics[scale=0.75]{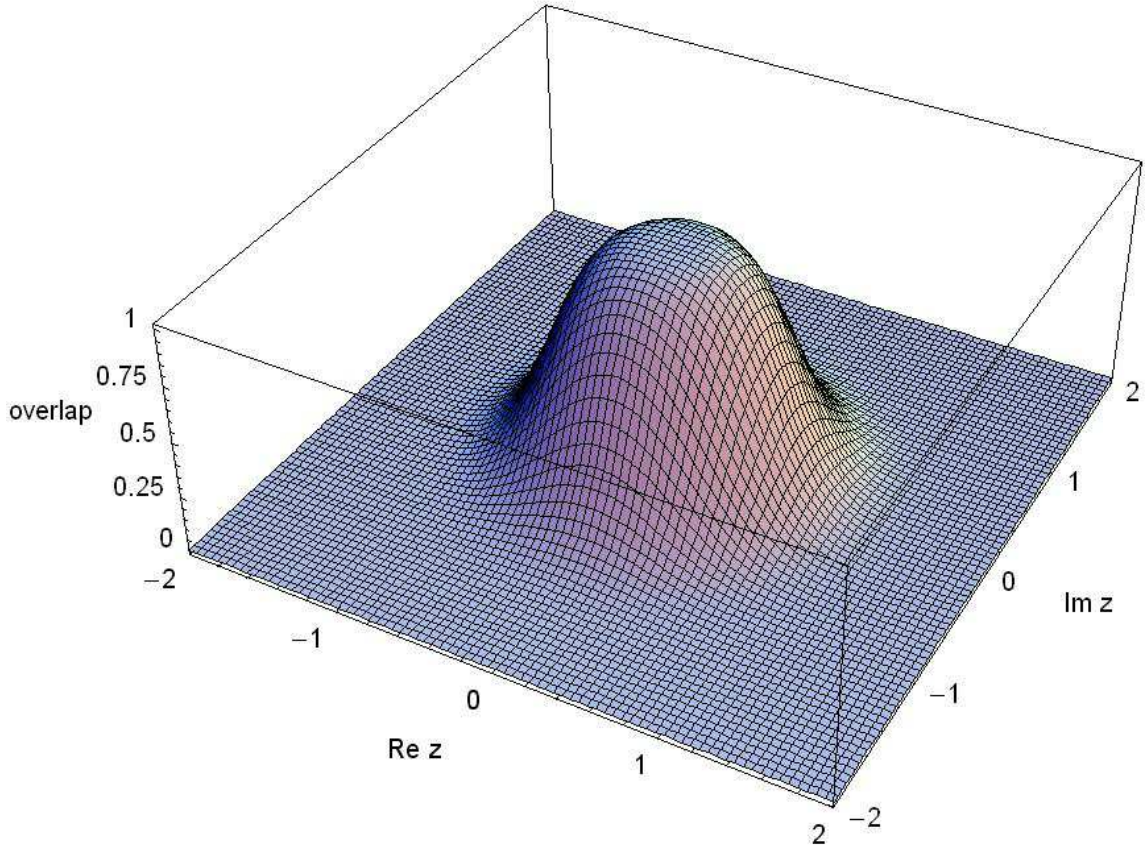}
    \end{center}
    \label{fig:Figureexample}
    \caption{\small Overlap between gauge-invariant coherent states $\Psi^t_{\cos z}$ and $\Psi^t_{\cos w}$ with $w=0$, $t=0.25$, depending on $z$.}
\end{figure}

\noindent In the second plot, we note the significant broader peak
around $z=0$. This is no Gaussian anymore. Rather, by performing
limits carefully in the expression (\ref{Gl:FinalerUeberlapp}), one
can see that it is actually a $|z|^2/(2t\,\sinh(|z|^2/2t))$-profile,
which is much broader than a Gaussian. For $|z|\to 0$, this goes as
$1-\frac{1}{24t^2}|z|^4$, rather than as $1-k|z|^2$, as would have
been expected from a Gaussian.
\newpage
This significant change in the peak profile is simply due to the
fact that the space of equivalence classes is no manifold any
longer, but contains singularities: Namely the ones at $g=\pm 1\in
SL(2,\C)$, that are the only points where the gauge group does not
act effectively on the orbits. It is exactly this feature that we
will also encounter in the other examples. We will also be able to
provide a general result concerning this property of gauge-invariant
coherent states labeled by degenerate gauge orbits.

\subsection{The 2-flower graph}

\noindent We now turn to the more complicated case of a 2-flower
graph, which consists of one vertex $v$ and 2 edges, both starting
and ending at $v$. Although this graph is still much simpler than
any graph of relevance for Loop Quantum Gravity, the gauge-invariant
coherent states are sufficiently complicated, such that the overlap
cannot be calculated analytically any more. Starting with the form
(\ref{Gl:EichinvarianteKohaerenteZustaendeNachIntertwinernZerlegt})
for the gauge-invariant coherent states, the intertwiner for the
2-flower graph can be computed, as can the basis $T_{\vec j,\vec I}$
for the gauge-invariant coherent states. In particular, the
gauge-invariant coherent states are given by
\begin{eqnarray}\label{Gl:GaugeInvariantStatesOn2FlowerWithIntertwiners}
\Psi_{[g_1,g_2]}^t(h_1,h_2)\;&=&\;\sum_{j_1,j_2\in\frac{1}{2}\Z}\;e^{-j_1(j_1+1)\frac{t}{2}-j_2(j_2+1)\frac{t}{2}}(2j_1+1)(2j_2+1)\\[5pt]\nonumber
&&\times\;\sum_{J=|j_1-j_2|}^{j_1+j_2}\sum_{M,N=-J}^J
C^{JM}_{j_1j_2m_1m_2}\,C^{JM}_{j_1j_2n_1n_2}\;\pi_{j_1}(h_1)_{m_1n_1}\,\pi_{j_2}(h_2)_{m_2n_2}\\[5pt]\nonumber
&&\times\;C^{JN}_{j_1j_2\tilde m_1\tilde m_2}\,C^{JN}_{j_1j_2\tilde
n_1\tilde n_2}\;\pi_{j_1}(g^c_1)_{\tilde m_1\tilde
n_1}\,\pi_{j_2}(g^c_2)_{\tilde m_2\tilde n_2}
\end{eqnarray}

\noindent Unfortunately, the occurring Clebsh-Gordan coefficients
make the use of the Poisson summation formula too complicated, such
that one cannot hope to rewrite
(\ref{Gl:GaugeInvariantStatesOn2FlowerWithIntertwiners}) as a sum
over $n$, such that only the $n=0$ term dominates in the $t\to
0$-limit, which simplified the analysis of the overlap tremendously
in the case of $SU(2)$-complexifier- and $U(1)$-gauge-invariant
coherent states.

So, we will perform a different route: We will start with formula
 (\ref{Gl:DieVierHeiligenIntegrale}) for the overlap of the gauge-invariant coherent states. Thus,
we need to perform the gauge integrals

\begin{eqnarray}\nonumber
\left\langle\Psi_{[g_1,g_2]}^t\big|\Psi_{[h_1,h_2]}^t\right\rangle\;&\sim&\;\int_{SU(2)}d\m_H(k)\sum_{n_1,n_2\in\Z}\frac{f_1(k)-2\pi
i n_1}{\sinh (f_1(k)-2\pi i n_1)}\frac{f_2(k)-2\pi i n_2}{\sinh
(f_2(k)-2\pi
i n_2)}\\[5pt]\label{Gl:GaugeInvariantInnerProductOn2Flower}
&&\qquad\times \exp\left({\frac{(f_1(k)-2\pi i n_1)^2+(f_2(k)-2\pi i
n_2)^2)}{t}}\right)
\end{eqnarray}

\noindent with
\begin{eqnarray}\nonumber
\cosh
f_1(k)=\frac{1}{2}\tr\,(g_1^{\dag}kh_1k^{-1})\\[5pt]\label{Gl:WhatAreF1AndF2OnThe2Flower}
\cosh f_2(k)=\frac{1}{2}\tr\,(g_2^{\dag}kh_2k^{-1}).
\end{eqnarray}

\noindent Unfortunately, this gauge integral is, although only an
integral over $SU(2)$, still too complicated to compute
analytically. Even the attempt to obtain an asymptotic expression
for the limit $t\to 0$, i.e. by employing the method of stationary
phase, fails, because of the complicated structure of
(\ref{Gl:WhatAreF1AndF2OnThe2Flower}). In trying to find the points
where the exponent $(f_1(k)-2\pi i n_1)^2+(f_2(k)-2\pi i n_2)^2$
becomes stationary, one can proceed quite far, in fact one can
locate these points to be in certain one-dimensional submanifolds of
$SU(2)$, depending on the $g_{1,2},h_{1,2}$. But calculating the
exact position of the stationary points eventually lead to
transcendent equations, which could not be solved.

What remains to do for us is to compute the integrand
(\ref{Gl:GaugeInvariantInnerProductOn2Flower}) numerically, and thus
perform a numerical analysis for the overlap
(\ref{Gl:DieVierHeiligenIntegrale}). In order to show that the
gauge-invariant coherent states are really peaked on gauge-invariant
quantities, we first have to identify these quantities, i.e.
gauge-fix the integrand in
(\ref{Gl:GaugeInvariantInnerProductOn2Flower}). \\

Lemma (\ref{Lem:ComplexShiftOfIntegrationOnLieGroups}) allows us to
shift the integrand in
(\ref{Gl:GaugeInvariantInnerProductOn2Flower}) not only by elements
of $SU(2)$ from the right and from the left, which is clear from the
bi-invariance of the Haar measure, but also allows us to shift the
integrand by elements of $SL(2,\C)$. Note the analogy to shifting
the integral of an analytic function over $\R$ into the complex
plane.

Now let us investigate the expressions
(\ref{Gl:WhatAreF1AndF2OnThe2Flower}). By writing
\begin{eqnarray*}
g_j\;=\;\exp(i\vec\s\cdot\vec w_j),\qquad h_j=\exp(i\vec\s\cdot\vec
z_j),\qquad j=1,\,2
\end{eqnarray*}

\noindent with $\vec z_i,\vec w_i\in\C^3$, we see that shifting the
integration variable by $k\to\tilde g^{\dag}k\tilde h$ changes
(\ref{Gl:WhatAreF1AndF2OnThe2Flower}) to
\begin{eqnarray*}
\frac{1}{2}\tr\,(g_j^{\dag}kh_jk^{-1})\;\longrightarrow\;\frac{1}{2}\tr\,((g'_j)^{\dag}k(h'_j)k^{-1})
\end{eqnarray*}

\noindent with
\begin{eqnarray*}
g'_j\;&=&\;\tilde gg_j\tilde g^{-1}\;=\;\exp(i\vec\s\cdot G\,\vec w_j)\\[5pt]
h'_j\;&=&\;\tilde hh_j\tilde h^{-1}\;=\;\exp(i\vec\s\cdot H\vec
z_j),\qquad j=1,\,2
\end{eqnarray*}

\noindent with $G,H\in O(3,\C)$. Note that here we encounter the
vector representation of $SL(2,\C)$ as orthogonal rotations in
$\C^3$. In particular, we have
\begin{eqnarray*}
G\;=\;\pi_1(\tilde g),\qquad H\;=\;\pi_1(\tilde h).
\end{eqnarray*}

\noindent Here we get a first glimpse at the gauge-invariant
information contained in a pair $(g_1,\,g_2)$. A gauge-invariant
state is not labeled by this pair, but rather by this pair modulo
conjugation with elements in $\tilde g\in SL(2,\C)$. If
$(g_1,\,g_2)$ are well away from the negative hermitian elements in
$SL(2,\C)$, we can talk about vectors in $\C^3$ rather than
$SL(2,\C)$-elements, which illustrates the facts better: Instead of
$g_j=\exp(i\vec\s\cdot\vec z_j)\in SL(2,\C)$, consider the $\vec
z_j\in\C^3$ themselves. Rather than by two vectors $\vec z_1,\,\vec
z_2\in\C^3$, the gauge-invariant states are labeled by these vectors
modulo rotation in $\C^3$. Here we mean rotations that leave $\vec
z_j\cdot\vec z_j$, rather than $\bar{\vec z}_j\cdot\vec z_j$
invariant, i.e. we talk about orthogonal rotations, not unitary
rotations.

Geometrically, this can be seen as the description of a
parallelogram in $\C^3$, where one point is fixed at $\vec
0\in\C^3$, modulo rotations around that point. Such a parallelogram
is just given by three (complex) numbers: The complex length of two
of its sides and the complex angle between them. In the following we
will gauge-fix the integrand
(\ref{Gl:GaugeInvariantInnerProductOn2Flower}) such that these
gauge-invariant quantities will explicitly be visible. We will then
present a number of plots which confirm that the overlap between
gauge-invariant coherent states is indeed peaked on these three
numbers, i.e. the overlap between two gauge-invariant coherent
states where only one of these three parameters differs, is close to
zero.\\

To simplify our analysis, we only consider elements $g_j,\,h_j$,
such that the complex length of neither of the associated vectors is
zero:
\begin{eqnarray*}
\vec z_j\cdot\vec z_j\;\neq\;0\neq\;\vec w_j\cdot\vec w_j,\qquad
j=1,\,2.
\end{eqnarray*}

\noindent Note that $O(3,\C)$ does not only map the set of vectors
$\{\vec z\cdot\vec z=a\}$ into itself for each $a\in \C$, but also
acts simply transitively on them, except for the case of $a=0$,
where the zero vector has to be excluded. Thus, we use the freedom
of $G,H\in O(3,\C)$ to rotate $\vec w_1$ and $\vec z_2$ into the
$3-$direction. In particular, there are $G,H\in O(3,\C)$ such that
\begin{eqnarray}\label{Gl:FirstVectorsGaugeFixed}
G\,\vec
w_1\;=\;\left(\begin{array}{c}0\\0\\w_1\end{array}\right),\qquad
H\,\vec z_2\;=\;\left(\begin{array}{c}0\\0\\z_2\end{array}\right)
\end{eqnarray}

\noindent with
\begin{eqnarray*}
w_1^2\;=\;\vec w_1\cdot \vec w_1,\qquad z_2^2\;=\;\vec z_2\cdot \vec
z_2.
\end{eqnarray*}

\noindent Of course, $\vec z_1,\,\vec w_2$ have also been changed by
this transformation, to $\vec z'_1,\,\vec w'_2$. The remaining
freedom to rotate these vectors without destroying the gauge fixed
(\ref{Gl:FirstVectorsGaugeFixed}), is effectively an
$O(2,\C)$-rotation in the complex $1-2$-plane. We can use this
remaining freedom to  rotate both $\vec z'_1,\,\vec w'_2$ out of the
$1$-direction. In particular, there are rotations $G,H \in O(3,\C)$
such that
\begin{eqnarray*}
G\left(\begin{array}{c}0\\0\\w_1\end{array}\right)\;=\;\left(\begin{array}{c}0\\0\\w_1\end{array}\right)\;=:\;\vec
w_1,&\qquad&
H\left(\begin{array}{c}0\\0\\z_2\end{array}\right)\;=\;\left(\begin{array}{c}0\\0\\z_2\end{array}\right)\;=:\;\vec z_2\\[5pt]
G\vec
w_2'\;=\;\left(\begin{array}{c}0\\w_2\sin\chi\\w_2\cos\chi\end{array}\right)\;=:\;\vec
w_2,&\qquad& G\vec
z_1'\;=\;\left(\begin{array}{c}0\\z_1\sin\theta\\z_1\cos\theta\end{array}\right)\;=:\;\vec
z_1
\end{eqnarray*}

\noindent with $\theta,\chi\in\C$. With this, the integrand is
completely gauge-fixed, apart from discrete symmetries having to do
with the fact that the $z_j,\,w_j$ and $\theta,\,\chi$ are defined
only up to a sign. With this, the integral
(\ref{Gl:GaugeInvariantInnerProductOn2Flower}) reads
\begin{eqnarray}\nonumber
\left\langle\Psi_{[g_1,g_2]}^t\big|\Psi_{[h_1,h_2]}^t\right\rangle\;&\sim&\;\int_{SU(2)}d\m_H(k)\sum_{n_1,n_2\in\Z}\frac{f_1(k)-2\pi
i n_1}{\sinh (f_1(k)-2\pi i n_1)}\frac{f_2(k)-2\pi i n_2}{\sinh
(f_2(k)-2\pi i
n_2)}\\[5pt]\label{Gl:GaugeInvariantInnerProductOn2FlowerGaugeFixed}
&&\qquad\times \exp\left({\frac{(f_1(k)-2\pi i n_1)^2+(f_2(k)-2\pi i
n_2)^2)}{t}}\right)
\end{eqnarray}

\noindent with
\begin{eqnarray}\nonumber
\cosh f_1(k)=\cos z_1\cos \bar w_1\,+\,\sin z_1\,\sin \bar
w_1\cos(\tilde\theta(k))
\\[5pt]\label{Gl:WhatAreF1AndF2OnThe2FlowerGaugeFixed}
\cosh f_2(k)=\cos z_2\cos \bar w_2\,+\,\sin z_2\,\sin \bar
w_2\cos(\bar{\tilde\chi}(k))
\end{eqnarray}

\noindent where $\tilde\theta(k)$ is the (complex) angle between
$\bar{\vec w}_1$ and $\pi_1(k)\vec z_1$, and $\tilde\chi(k)$ the one
between $\bar{\vec w}_2$ and $\pi_1(k)\vec z_2$.\\

Now the inner product between two coherent states is only dependent
of the triples $(z_1,z_2,\theta)$ and $(w_1,w_2,\chi)$, which
constitutes the gauge-invariant information of the two
gauge-invariant coherent states. With
(\ref{Gl:DieVierHeiligenIntegrale}) one can compute the overlap from
the inner products given by
(\ref{Gl:GaugeInvariantInnerProductOn2FlowerGaugeFixed}) with
(\ref{Gl:WhatAreF1AndF2OnThe2FlowerGaugeFixed}). We have done this
numerically for a couple of examples. In each case, we have fixed
the triple $(w_1,w_2,\chi)$ and four of the six real parameters in
$(z_1,z_2,\theta)$, and plotted the overlap
\begin{eqnarray}\label{Gl:Overlap}
\mbox{overlap
}\;=\;\frac{\left|\left\langle\Psi_{(w_1,w_2,\chi)}^t|\Psi_{(z_1,z_2,\theta)}^t\right\rangle\right|^2}
{\left\|\Psi_{(w_1,w_2,\chi)}^t\right\|^2\;\left\|\Psi_{(z_1,z_2,\theta)}^t\right\|^2}.
\end{eqnarray}

\begin{figure}[hbt!]
\begin{center}
    \includegraphics[scale=0.75]{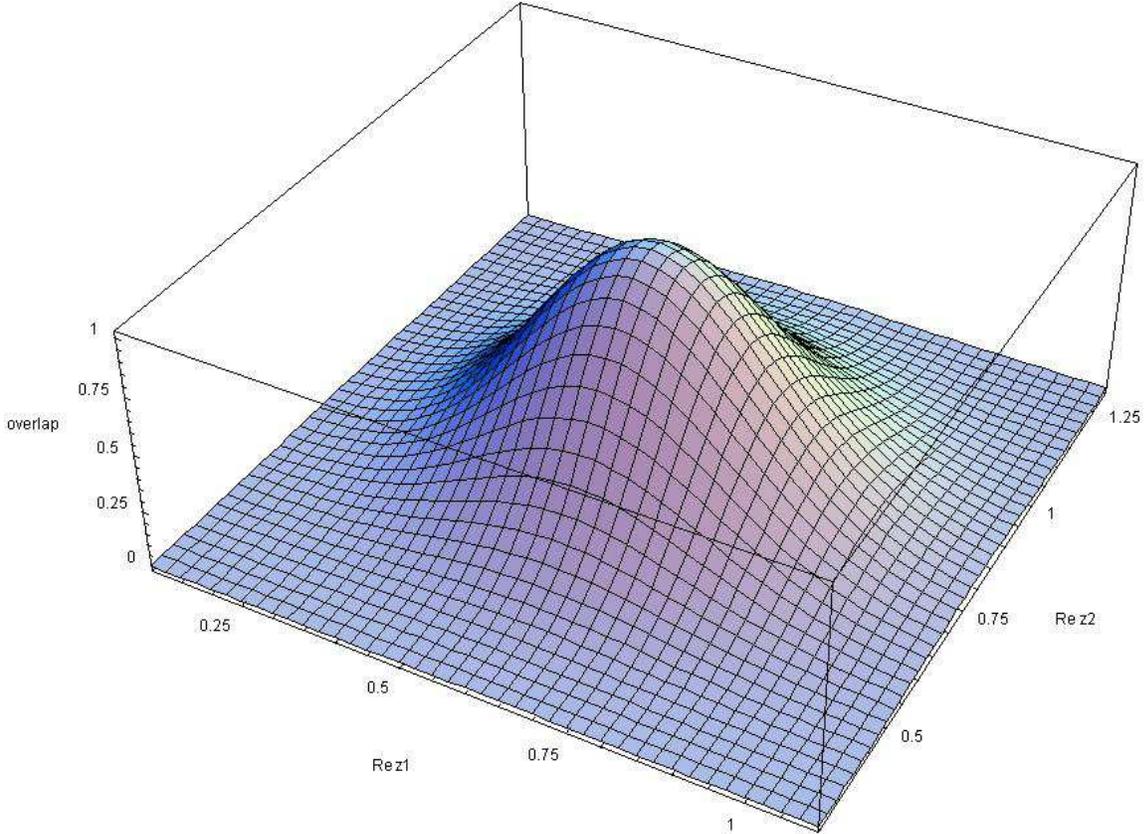}
    \end{center}
    \label{fig:w106w208chi0theta0t02}
    \caption{\small Overlap between two coherent states $\Psi^t_{[\vec g]}$, one with gauge-invariant
    data $w_1=0.6$, $w_2=0.8$, $\chi=0$, the other one with
    $\theta=0$, depending on $z_1,$ and $z_2$}
\end{figure}

\noindent The first plot shows the overlap between a coherent state
labeled by $w_1=0.6,\,w_2=0.8,\chi=0$ and one labeled by $\theta=0$
and variable $z_1,\,z_2$. The width of both states have been chosen
to be $t=0.2$. As one can see, the overlap is peaked at the point
where both label sets coincide. Also, the shape resembles that of a
Gaussian quite well.\\

\noindent The second plot shows that the gauge-invariant coherent
states are also peaked in the complex angles $\theta,\,\chi$. The
plot shows the overlap between a coherent state labeled with
$w_1=1,\,w_2=1,\,\chi=0.2$, and one labeled by $z_1=1,\,z_2=1$ and
$\theta$, where the overlap depending on the complex parameter
$\theta$ is shown. The semiclassicality parameter was again chosen
to be $t=0.2$.

\begin{figure}[hbt!]
\begin{center}
    \includegraphics[scale=0.75]{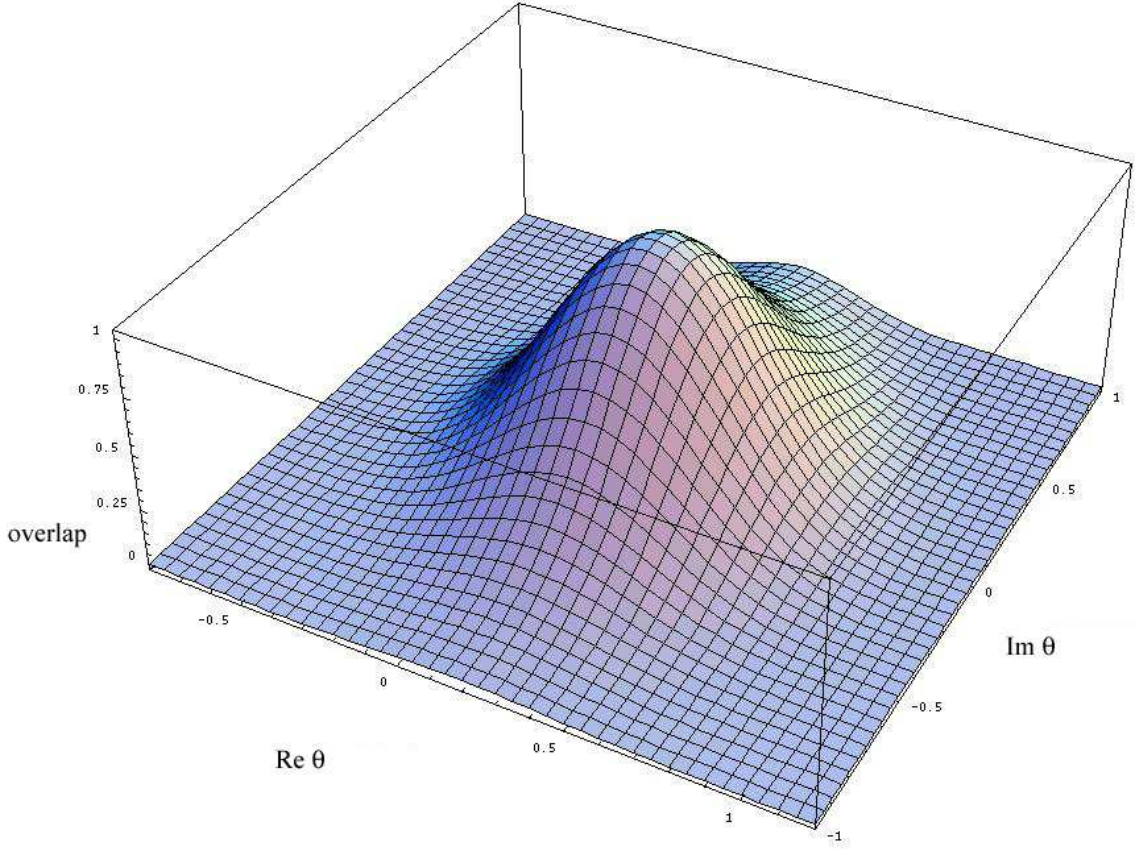}
    \end{center}
    \label{fig:Figureexample}
    \caption{\small Overlap between two coherent states $\Psi^t_{[\vec g]}$, one with gauge-invariant
    data $w_1=1$, $w_2=1$, $\chi=0.2$, the other one with $z_1=1$ and
    $z_2=1$,
    depending on $\theta$}
\end{figure}

\noindent Again, one can see that in the vicinity of  the point
where the label sets coincide, the overlap behaves nearly as a
Gaussian. Further away, though, the overlap differs slightly from a
Gaussian (note the 'bump' at $\theta\approx I$).

Apparently, this is due to the fact that $(z_1,z_2,\theta)$ are just
one set of gauge-invariant labels of a state. There are, of course,
infinitely many other equivalent ones, depending on how one
gauge-fixes the integrand
(\ref{Gl:GaugeInvariantInnerProductOn2Flower}). In most of them, the
overlap will not exactly be a Gaussian, but will have a rather
generic form like
\begin{eqnarray*}
\text{overlap}\;=\;\exp\left(-\frac{F(w_1,w_2,\chi,z_1,z_2,\theta)}{t}\right)
\end{eqnarray*}

\noindent with $F=O((z_1-w_1)^2,\;(z_2-w_2)^2,\;(\theta-\chi)^2)$.
Of course, as $t\to 0$, $F$ can be more and more approximated by its
series up to the quadratic order, hence as $t\to 0$, the overlap
becomes more and more a Gaussian.\\

\noindent The third plot does not show a Gaussian peak: In this
plot, we have chosen one state to be peaked at $[1,1]$, i.e.
$w_1=w_2=\chi=0$, as well as $\theta=0$ and $t=0.2$.

\begin{figure}[hbt!]
\begin{center}
    \includegraphics[scale=0.75]{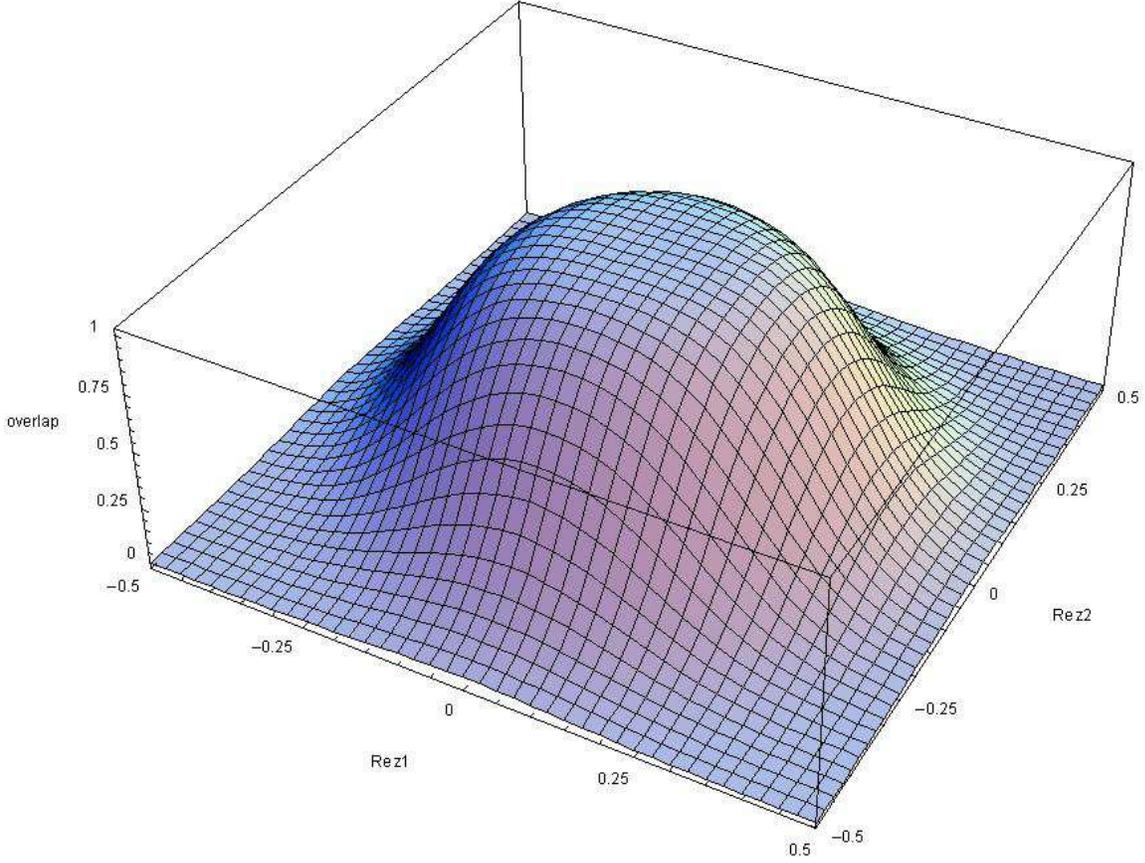}
    \end{center}
    \label{fig:Figureexample}
    \caption{\small Overlap between two coherent states $\Psi^t_{[\vec g]}$, one with gauge-invariant
    data $w_1=0$, $w_2=0$, $\chi=0$ (corresponding to $[\vec g]=[1,1]$), the other one with
    $\theta=0$, depending on $z_1,$ and $z_2$}
\end{figure}

\noindent One can immediately see the 'plateau' around the point
where both label sets coincide. Here, although the overlap decreases
as the two label sets start to disagree, it decreases qualitatively
slower than in the other cases. In particular, the overlap behaves
rather like $\exp(-x^4)$, than an $\exp(-x^2)$. This behavior does
not change as $t\to 0$. Rather, the function $F$ in
(\ref{Gl:Overlap}) goes like $O(|z_1|^4,\,|z_2|^4,\,|\theta|^4)$.

Note that this phenomenon has already been encountered in the case
of the 1-flower graph. The overlap (\ref{Gl:FinalerUeberlapp}) is,
in the limit of $w=0$, equal to
\begin{eqnarray*}
\frac{\Big|\langle\Psi_{0}^t|\Psi_{z}\rangle\Big|^2}{\left\|\Psi^t_0\right\|^2\,\left\|\Psi^t_z\right\|^2}\;=\;\frac{|z|^2/2t}{\sinh(|z|^2/2t)},
\end{eqnarray*}

\noindent which goes as $\sim 1-|z|^4/24t^2$ as $t\to 0$.

So, although peaked at the point indicated by the label set, the
state $\Psi^t_{[1,1]}$, with $(1,1)\in SL(2,\C)^2$, is much less
peaked than generic gauge-invariant states. This will be proven
explicitly in chapter \ref{Ch:GaussianPimple}.


\subsection{The 3-Bridge graph}\label{Ch:TheThreeBridgeGraph}

\noindent After the 2-flower graph, we will proceed with the
3-bridge (or sunset-) graph, which consists of two vertices $v_1$
and $v_2$, and three edges $e_1,\, e_2,\, e_3$, all beginning at
$v_1$ and ending at $v_2$.

As for the 2-flower graph, the gauge-invariant coherent states on
the 3-bridge graph are too complicated to compute the overlap such
that statements about peakedness properties can be made. Therefore,
we will again rely on numerics to show some qualitative features,
which will strengthen the results from the last sections, and again
hint towards section \ref{Ch:GeneralProperties}.\\

\noindent Up to a $t$-dependent factor, the scalar product
(\ref{Gl:EichinvariantesSkalarprodukt}) between two gauge-invariant
coherent states $\Psi^t_{[h_1, h_2, h_3]}$ and $\Psi^t_{[g_1, g_2,
g_3]}$ is given by

\begin{eqnarray}\label{Gl:GaugeInvariantInnerProductOnTreeBridge}
&&\Big\langle\Psi^t_{[h_1, h_2, h_3]}\Big|\Psi^t_{[g_1, g_2,
g_3]}\Big\rangle\\[5pt]\nonumber
\;&\sim
&\;\int_{SU(2)^2}d\m_H(k_1,\,k_2)\;\prod_{l=1}^3\sum_{n_l\in\Z}\frac{f_l(k_1,k_2)-2\pi
n_l}{\sinh (f_l(k_1,k_2)-2\pi
n_l)}\;\exp\left(\frac{(f_l(k_1,k_2)-2\pi n_l)^2}{t}\right)
\end{eqnarray}

\noindent with
\begin{eqnarray}
\cosh f_l(k_1,k_2)\;=\;\frac{1}{2}\tr\;\Big(g_l^{\dag}\,k_1\,h_l\,
k_2^{-1}\Big).
\end{eqnarray}

\noindent As in the case of the 2-flower graph, we start to
gauge-fix the integrand by applying Lemma
\ref{Lem:ComplexShiftOfIntegrationOnLieGroups}. If we write
\begin{eqnarray}
h_l\;=\;\exp\big(i\vec\s\cdot\vec
w'_l\big),\qquad\,g_l\;=\;\exp\big(i\vec\s\cdot\vec z'_l\big)
\end{eqnarray}

\noindent for $\vec z'_l,\vec w'_l\in\C^3$, and choose the
$h_l,\,g_l$ to be well away from the negative hermitean elements, we
can choose the real part of the complex vectors to be in a ball
around zero with radius $\pi$: ${\rm Re}\,\vec w'_l,\,{\rm Re}\,\vec
z'_l\,\in B_{\pi}(0)$, and thus we can neglect all terms in the
infinite sum (\ref{Gl:GaugeInvariantInnerProductOnTreeBridge}) apart
from $n_1=n_2=n_3=0$. Changing the integration variables

\begin{eqnarray}
k_1\;\longmapsto\;(g_1^{\dag})^{-1}k_1,\qquad
k_2\;\longmapsto\;k_2h_1
\end{eqnarray}

\noindent amounts to a change of the terms occurring in
(\ref{Gl:GaugeInvariantInnerProductOnTreeBridge}) as
\begin{eqnarray}\nonumber
\frac{1}{2}\,\tr\Big(g_1^{\dag}k_1\,h_1\,k_2^{-1}\Big)\;&\longmapsto&\;\frac{1}{2}\,\tr\Big(k_1\,k_2^{-1}\Big)\\[5pt]\label{Gl:ThreeBridgeFirstGaugeChange}
\frac{1}{2}\,\tr\Big(g_2^{\dag}k_1\,h_2\,k_2^{-1}\Big)\;&\longmapsto&\;\frac{1}{2}\,\tr\Big(\tilde
g_2^{\dag}k_1\,\tilde h_2\,k_2^{-1}\Big)\\[5pt]\nonumber
\frac{1}{2}\,\tr\Big(g_3^{\dag}k_1\,h_3\,k_2^{-1}\Big)\;&\longmapsto&\;\frac{1}{2}\,\tr\Big(\tilde
g_3^{\dag}k_1\,\tilde h_3\,k_2^{-1}\Big).
\end{eqnarray}

\noindent with

\begin{eqnarray}
\tilde g_l\;=\;g_1^{-1}g_l,\qquad \tilde h_l\;=\;h_l h_1^{-1}.
\end{eqnarray}

\noindent Write

\begin{eqnarray}
\tilde h_l\;=\;\exp\big(i\vec\s\cdot\vec w_l\big),\qquad\,\tilde
g_l\;=\;\exp\big(i\vec\s\cdot\vec z_l\big)
\end{eqnarray}

\noindent with vectors $\vec w_l,\,\vec z_l\in\C^3$, $l=2,3$. The
only transformations of the integration variables $k_1,\,k_2$ that
leave the form of the terms (\ref{Gl:ThreeBridgeFirstGaugeChange})
invariant are simultaneously gauging
\begin{eqnarray}
k_{1,2}\;\longmapsto\;g^{\dag}\,k_{1,2}\,h^{-1}
\end{eqnarray}

\noindent for arbitrary $g,\,h\in SL(2,\C)$. As with the coherent
states on the 2-flower graph, this induces an $O(3,\C)-$ action on
the vectors $\vec w_l,\,\vec z_l$, which can be used to gauge $\vec
w_2$ and $\vec z_3$ to point into the $3$-direction, and $\vec
w_3,\,\vec z_2$ out of the $1$-direction.

\begin{eqnarray*}
\vec
w_2\;=\;\left(\begin{array}{c}0\\0\\w_2\end{array}\right),&\qquad&
\vec z_3\;=\;\left(\begin{array}{c}0\\0\\z_3\end{array}\right)\\[5pt]
\vec
w_3\;=\;\left(\begin{array}{c}0\\w_3\sin\chi\\w_3\cos\chi\end{array}\right),&\qquad&
\vec
z_2\;=\;\left(\begin{array}{c}0\\z_2\sin\theta\\z_2\cos\theta\end{array}\right).
\end{eqnarray*}

\noindent Note that the gauge-invariant information for
gauge-invariant states on the 3-bridge graph are the same than the
ones on the 2-flower graph. This was to be expected, since both have
the same first fundamental group (see appendix
\ref{Ch:GaugeInvariantFunctions}). The gauge-invariant inner
product, on the other hand, looks slightly different:

\begin{eqnarray}
\Big\langle\Psi^t_{[h_1, h_2, h_3]}\Big|\Psi^t_{[g_1, g_2,
g_3]}\Big\rangle\;&\approx
&\;\int_{SU(2)^2}d\m_H(k_1,\,k_2)\;\prod_{l=1}^3\frac{f_l(k_1,k_2)}{\sinh
f_l(k_1,k_2)}\;\exp\left(\frac{f_l(k_1,k_2)^2}{t}\right)
\end{eqnarray}

\noindent with
\begin{eqnarray}\nonumber
\cosh f_2(k)=\cos z_2\cos \bar w_2\,+\,\sin z_2\,\sin \bar
w_2\cos(\tilde\theta(k_1,\,k_2))
\\[5pt]\label{Gl:WhatAreF1AndF2OnThe3BridgeGaugeFixed}
\cosh f_3(k)=\cos z_3\cos \bar w_3\,+\,\sin z_3\,\sin \bar
w_3\cos(\bar{\tilde\chi}(k_1,\,k_2))
\end{eqnarray}

\noindent where $\tilde\theta(k_1,\,k_2)$ is the (complex) angle
between $\pi_1(k_1)\bar{\vec w}_2$ and $\pi_1(k_2)\vec z_2$, and
$\tilde\chi(k_1,\,k_2)$ the one
between $\pi_1(k_1)\bar{\vec w}_3$ and $\pi_1(k_2)\vec z_3$.\\

\noindent The gauge-fixed integrand can be evaluated numerically. We
show the result for two different gauge-invariant label sets.

\newpage

 \begin{figure}[hbt!]
\begin{center}
    \includegraphics[scale=0.75]{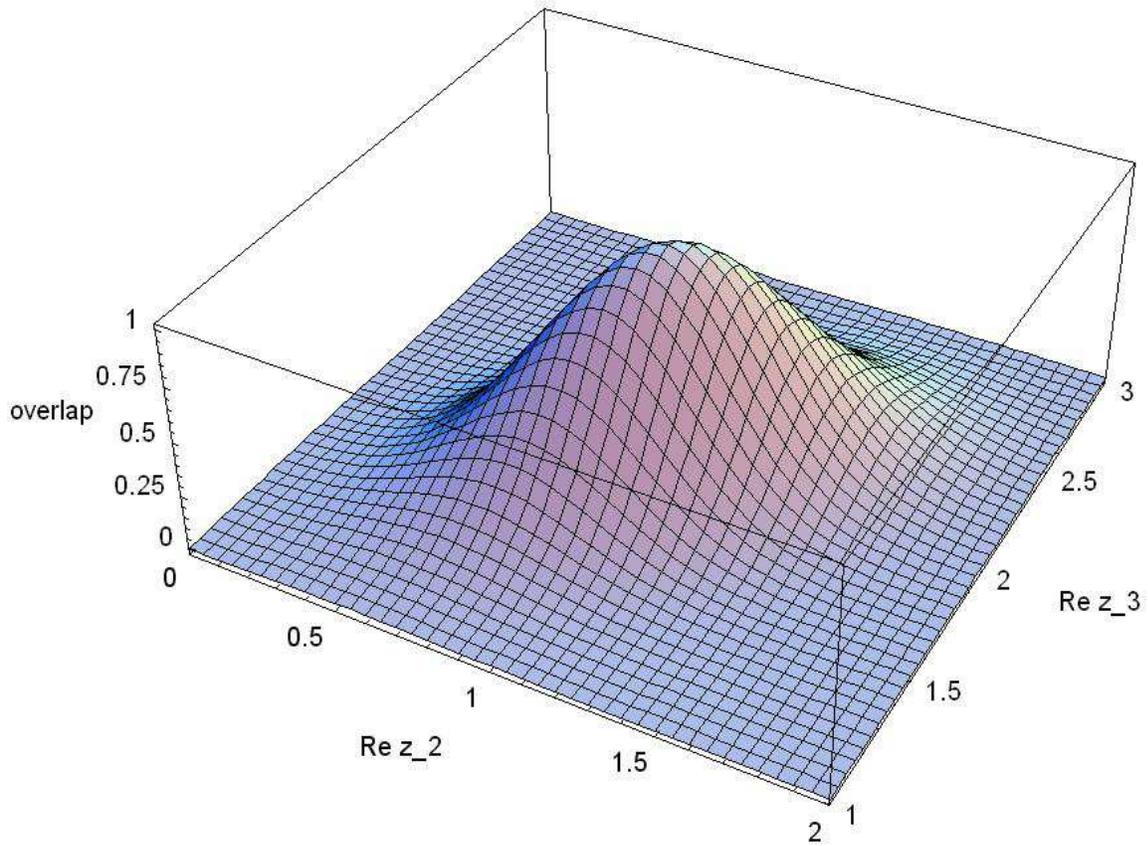}
    \end{center}
    \label{fig:Tetrahedron1}
    \caption{\small The overlap between two gauge-invariant coherent states on the 3-bridge graph.
    One is labeled by the gauge-invariant data $w_2=1$, $w_3=2$, $\chi=0.3$. The other one was labeled at $\theta=0.3$, depending on $z_2,\, z_3$}
\end{figure}

\noindent The first plot shows, again, that the overlap is peaked at
the point where both gauge-invariant data sets coincide.

\newpage

The second plot shows the overlap of a state with one labeled by
$[1, 1, 1]$.
 \begin{figure}[hbt!]
\begin{center}
    \includegraphics[scale=0.75]{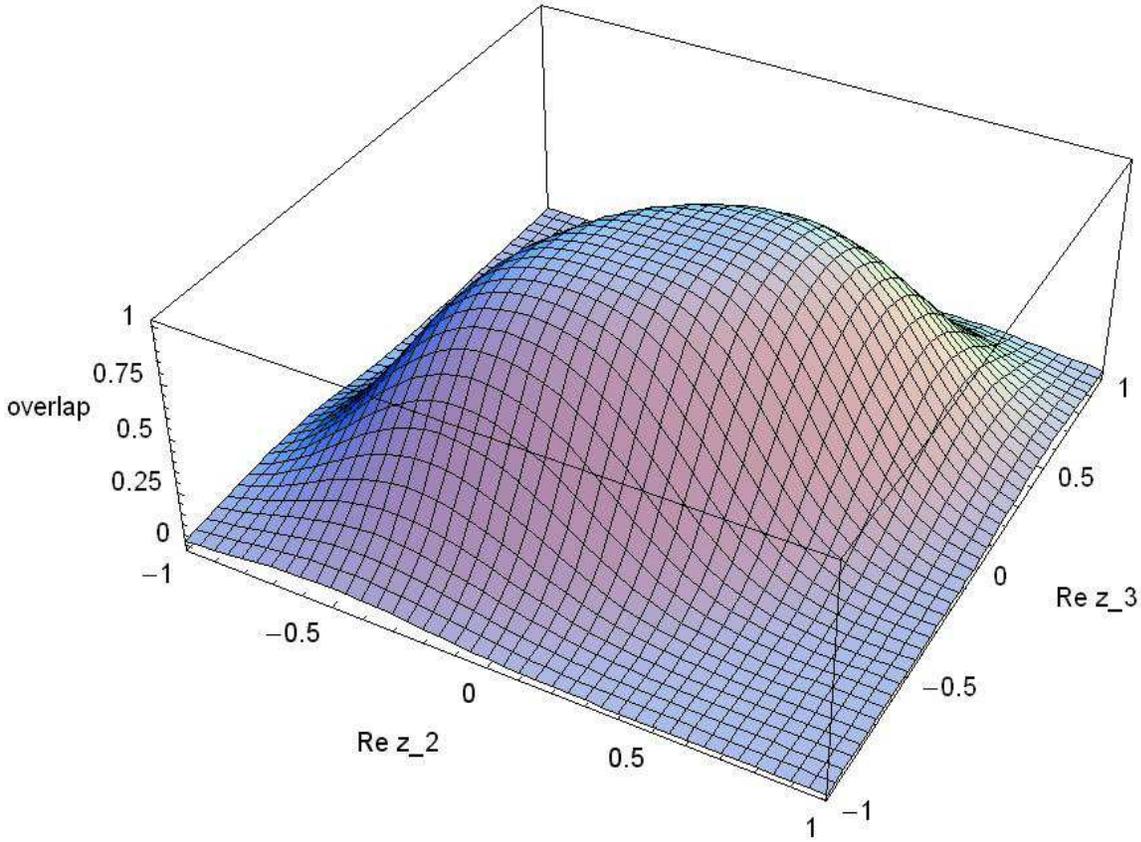}
    \end{center}
    \label{fig:Tetrahedron1}
    \caption{\small The overlap between two gauge-invariant coherent states on the 3-bridge graph.
    One is labeled by the gauge-invariant data $w_2=0$, $w_3=0$, $\chi=0$, corresponding to
    $\g_l=1\in SL(2,\C)$ for $l=1,\,2,\,3$. The other one was labeled at $\theta=0$, depending on $z_2,\, z_3$}
\end{figure}

\noindent Again, one can see the plateau structure of the overlap.
This feature was also visible with the 1-flower and the 2-flower
graph and shows that this phenomenon is not just a feature of the
flower graphs. We will discuss this in chapter
\ref{Ch:GeneralProperties}.

\subsection{The tetrahedron graph}\label{Ch:Tetrahedron}

\noindent As well as the gauge-invariant overlap on the other
graphs, the gauge-invariant overlap on the tetrahedron graph is
completely out of reach of any analytical computations. 
Still, one can, in complete analogy to the case of the 2-flower or
the 3-bridge graph, describe the gauge-invariant degrees of freedom
in $SL(2,\C)^6$ by a gauge-fixing procedure, and show numerically
that the overlap of the gauge-invariant coherent states is peaked at
the points where the two label sets coincide.

The tetrahedron is given by four vertices and six edges. The
vertices are labeled by $v_I$ with $I=1,\ldots, 4$. There is an edge
between any two vertices, $e_{IJ}$, oriented from $v_I$ to $v_J$ for
 $I<J$.\\


The gauge-invariant overlap will again be calculated from the inner
product between two gauge-invariant coherent states, by
(\ref{Gl:DieVierHeiligenIntegrale}). The gauge-invariant inner
product is given by
\begin{eqnarray}\label{Gl:InnerProductTetrahedronGraph}
&&\left\langle\Psi_{[g_{IJ}]}^t\big|\Psi_{[h_{IJ}]}^t\right\rangle\;\\[5pt]\nonumber
&\sim&\;\int_{SU(2)^4}d\m^{\otimes
4}_H(k)\left[\prod_{I<J}\sum_{n_{IJ}\in\Z}\frac{f_{IJ}(k)-2\pi i
n_{IJ}}{\sinh (f_{IJ}(k)-2\pi i n_{IJ})}
\exp\left(\frac{(f_{IJ}(k)-2\pi i n_{IJ})^2}{t}\right)\right]
\end{eqnarray}

\noindent with

\begin{eqnarray}\label{Gl:SixTermsInTetrahedronGraph}
\cosh
f_{IJ}(k)\;=\;\frac{1}{2}\tr\,\left(g_{IJ}^{\dag}k_Ih_{IJ}k_J^{-1}\right).
\end{eqnarray}

\noindent A crucial point in the gauge fixing procedure used to
extract the gauge-invariant degrees of freedom is the use of a
maximal tree $\t$ in the tetrahedron graph $\g$. From usual lattice
gauge theory, on knows that by successive gauging one can gauge a
distribution of group elements along the edges of $\g$ such that
there is a $1$ along each edge of $\t$. The resulting distributions
of elements among the (three) leaves of $\g$, i.e. the edges not
belonging to $\t$, contain (modulo global conjugation) the
gauge-invariant information of the distribution of elements.

The remaining global conjugation freedom in  $g_k=e^{i\vec
z_k\cdot\vec\s}\,\in\,SL(2,\C)$, with $\vec z_k\in\C^3$, corresponds
to a similar $O(3,\C)$-rotation of the $\vec z_k$. This rotation can
- exactly as in the previous examples - be used to rotate one of the
vectors into the $3$-direction, another into the $2-3$-plane, while
the third vector is fixed then. The remaining  degrees of freedom
are then two complex lengths, a complex angle, and a complex
$3$-component vector. These six parameters, i.e. $z_1,\, z_2, \,
\theta,\,\vec z_3$, determine the gauge-invariant data set, which
corresponds to the geometry of a tetrahedron in $SL(2,\C)$.\\

Unfortunately, since the integral
(\ref{Gl:InnerProductTetrahedronGraph}) ranges over $SU(2)^4$ which
is 12-dimensional, the numerical integration becomes quite involved.
In particular, to compute overlaps with a sufficient precision is
very time-consuming. Still, we were able to produce some integrals,
which show the peakedness of the overlap:

\begin{figure}[hbt!]
\begin{center}
    \includegraphics[scale=0.7]{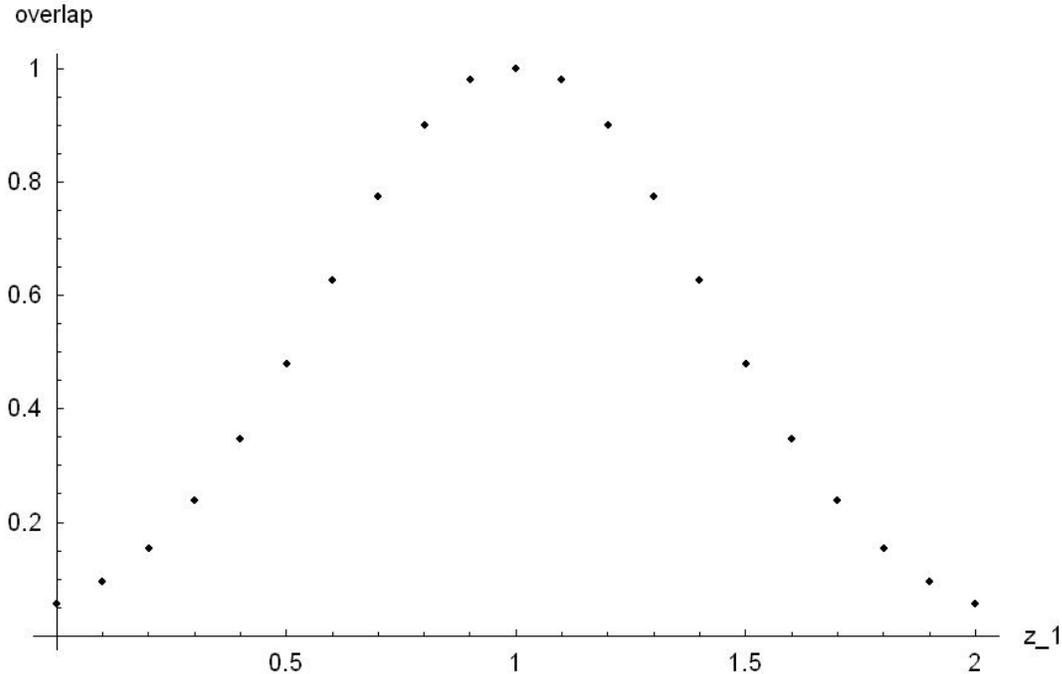}
    \end{center}
    \label{fig:Figureexample}
    \caption{\small The overlap between gauge-invariant coherent states labeled by gauge-invariant data $w_1=1,\,z_2=w_2=2,\,\chi=\theta=0.3,\,\vec z_3=\vec w_3= (1,1,1)$, depending on $z_1$.}
\end{figure}

\noindent As one can see, the overlap is Gaussian-peaked at the
point $z_1=1$, i.e. where the two label sets coincide.

\newpage

\section{Gauge-invariant coherent states for $G=SU(2)$: General
properties}\label{Ch:GeneralProperties}

\noindent In the last sections we have investigated the
gauge-invariant coherent states for the gauge group $SU(2)$
analytically and numerically for some simple examples.
Unfortunately, the formula for the inner product between these
states is too complicated in order to prove peakedness properties of
the overlap for arbitrary graphs.

    Still, in the following sections we will investigate some
properties of the gauge-invariant states on arbitrary graphs. In
particular, we will be able, by an appropriate gauge-fixing
procedure, to connect the inner product between two gauge-invariant
coherent states on an arbitrary graph with $E$ edges and $V$
vertices to the inner product between gauge-invariant coherent
states on an $E-V+1$-flower graph.

The same procedure is in principle also possible for the case of
$G=U(1)$, which leads to a gauge-invariant coherent state on a
flower graph, which can be, due to the abelianess of the gauge
group, explicitly written down. Remember that for abelian gauge
groups, all functions on flower graphs are automatically
gauge-invariant. This is, of course, no longer true for non-abelian
gauge groups, so relating the gauge-invariant coherent states on
arbitrary graphs to gauge-invariant coherent states on flower graphs
is the best one can do.

Still, this formula will allow us to prove a theorem about the
peakedness properties of states labeled by $[1,\ldots ,1]$. In the
last sections, we have already seen that the overlap between two
states, one of them being labeled by $[1,\ldots ,1]$, does not
behave like a Gaussian. Rather, the profile of the overlap looks
like a $e^{-x^4}$-curve. In the following, we will prove this for
arbitrary graphs.

We will start with a theorem about the peakedness of
$\Psi_{[1,\ldots,1]}^t$ on flower graphs. Then, we will derive a
formula relating the inner products of gauge-invariant coherent
states on arbitrary graphs to those on flower graphs. This will
ultimately enable us to formulate a corresponding theorem about the
peakedness properties of $\Psi_{[1,\ldots,1]}^t$ on arbitrary
graphs.

\subsection{Peakedness of $\Psi_{[1,\ldots,1]}^t$ on $E$-flower graphs}\label{Ch:GaussianPimple}

\noindent In the previous sections, we have seen that the peakedness
of the state $\Psi_{[1,\ldots,1]}^t$, i.e. the gauge-invariant state
labeled by the equivalence class of $(1,\ldots,1)\in SL(2,\C)^E$, is
qualitatively different than the peakedness of states that are
labeled by generic elements. In particular, the overlap
(\ref{Gl:FinalerUeberlapp}) on the 1-flower graph
\begin{eqnarray*}
z\;\longmapsto\;\frac{\Big|\left\langle\Psi_{[1]}^t\Big|\Psi_{\cos\,z}^t\right\rangle\Big|^2}{\Big\|\Psi_{\cos\,z}^t\Big\|^2\;\Big\|\Psi_{[1]}^t\Big\|^2}
\:=\;\frac{|z^2|/2t}{\sinh(|z|^2/2t)}(1+O(t^{\infty}))
\end{eqnarray*}

\noindent behaves like
$\sim\,1-\frac{1}{12t}|z|^4\sim\exp(-|z|^4/12t)$, as $z\to 0$,
rather than a Gaussian. The plots of overlaps on the 2-flower graph
support the conjecture that this is a general feature of states on
$E$-flowers peaked at the equivalence class of $(1,\ldots,1)\in
SL(2,\C)^E$.

This conjecture is in fact true, as we will show now. Despite the
notoriously complicated structure of the overlap, in this particular
case we are able to perform the gauge integrals in the limit of the
labelings being close to $(1,\ldots,1)$.

\begin{Theorem}
Let $\g$ be the $E$-flower graph, i.e. the graph with one vertex and
 $E$ edges all starting and ending at that vertex.
 Define $h(\vec z_j)\in SL(2,\C)^E$ by $h_j=\exp(i \vec \s
\cdot\vec z_j)\in SL(2,\C)$, $j=1,\ldots, E$. Then one has
\begin{eqnarray}\label{GL:GaussianPimple}
\frac{\Big|\left\langle\Psi_{[1,\ldots,1]}^t\Big|\Psi_{[h(\vec
z_j)]}^t\right\rangle\Big|^2}
{\Big\|\Psi_{[1,\ldots,1]}^t\Big\|^2\;\Big\|\Psi_{[h(\vec
z_j)]}^t\Big\|^2}\;=\;1-O(\|\vec z_j\|^4).
\end{eqnarray}
\end{Theorem}

\noindent\textbf{Proof:} Of course, $h(\vec z_j\equiv 0)=1$, it is
clear that the overlap between $\Psi_{[1,\ldots,1]}^t$ and
$\Psi_{[h(\vec z_j\equiv 0)]}$ is equal to $1$. What we will do now
to prove equation (\ref{GL:GaussianPimple}) is to expand the two
inner products
\begin{eqnarray}\label{Gl:TheseTwoContributionsWeNeed}
\left\langle\Psi_{[1,\ldots,1]}^t\Big|\Psi_{[h(\vec
z_j)]}^t\right\rangle \qquad\qquad\mbox{and}\qquad\qquad
\Big\|\Psi_{[h(\vec z_j)]}^t\Big\|^2\;=\;\left\langle\Psi_{[h(\vec
z_j)]}^t\Big|\Psi_{[h(\vec z_j)]}^t\right\rangle
\end{eqnarray}

\noindent into second order around $\vec z_j=0$. The odd orders all
vanish, and we will be able to show that the second order of the
$\vec z_j$ will cancel in numerator and denominator of
(\ref{GL:GaussianPimple}), such that the first nontrivial order will
be that of $\|\vec z_j\|^4$, obviously with a negative sign, since
the
overlap has to be at most 1.\\

\noindent Remember that the overlap between two gauge-invariant
coherent states on the $E$-flower graph, labeled by $[\tilde h]$ and
$[h]$, $h,\tilde h\in SL(2,\C)^E$, is given by
\begin{eqnarray}\nonumber
\Big\langle\Psi_{[\tilde
h]}^t\Big|\Psi_{[h]}^t\Big\rangle\;&=&\;\int_{SU(2)}d\m_{H}(k)\;\sum_{n_1,\ldots
n_E\in \Z}\prod_{j=1}^E\left[\frac{f_j(k)-2\pi i
n_j}{\sinh(f_j(k)-2\pi
i n_j)}\right]\\[5pt]\label{Gl:RememberThisIsTheOverlap}
&&\qquad\qquad\times\;\exp\left(\sum_{j=1}^E\frac{(f_j(k)-2\pi i
n_j)^2}{t}\right)
\end{eqnarray}

\noindent where
\begin{eqnarray}
\cosh f_j(k)\;=\;\frac{1}{2}\tr \;\big(\tilde
h_j^{\dag}k\,h_j\,k^{-1}\big).
\end{eqnarray}

\noindent We calculate the inner product between two gauge-invariant
coherent states on the $E$-flower graph, labeled by $h(\vec w_j)$
and $h(\vec z_j)$, expanding the expression in quadratic powers of
the $\vec z_j$ and $\vec w_j$. The two contributions
(\ref{Gl:TheseTwoContributionsWeNeed}) are then just obtained by
setting $\vec w_j\equiv 0$ and $\vec w_j=-i \bar{\vec z}_j$
respectively.\\

We write

\begin{eqnarray*}
k\;&=&\;\exp\big(i\vec \s\cdot\vec
\phi\big)\;=\;\cos\phi\,+\,i\frac{\sin\phi}{\phi}\,\vec \s\cdot\vec
\phi
\end{eqnarray*}

\noindent where, as usual,
\begin{eqnarray*}
\phi\;:=\;\|\vec\phi\|\;=\;\sqrt{(\phi^1)^2+(\phi^2)^2+(\phi^3)^2}.
\end{eqnarray*}

\noindent  Additionally, we get
\begin{eqnarray*}
h(\vec z_j)\;&=&\;\exp\big(i\vec\s\cdot\vec z_j\big)\;\;=\;\cos
z_j\,+\,i\frac{\sin z_j}{z_j}\,\vec\s\cdot \vec
z_j\;\approx\;\left(1-\frac{1}{2}z_j^2\right)\,+\,i\vec\s\cdot z_j.
\end{eqnarray*}

\noindent again with
\begin{eqnarray*}
z_j^2\,:=\,\vec z_j\cdot\vec z_j\;=\;(z_j^1)^2+(z_j^2)^2+(z_j^3)^2,
\end{eqnarray*}

\noindent and according expressions for $h(\vec w_j)$. With these
expressions, we get

\begin{eqnarray}\nonumber
\cosh\,f_j(k)\;&\approx&\;\frac{1}{2}\tr\left[\left(1-\frac{1}{2}w_j^2\,+\,i\vec\s\cdot
\vec w_j\right)\left(\cos\phi\,+\,i\frac{\sin\phi}{\phi}\,\vec\s\cdot\vec\phi\right)\right.\\[5pt]\nonumber
&&\qquad\times\left.\left(1-\frac{1}{2}z_j^2\,+\,i\vec\s\cdot
\vec z_j\right)\left(\cos\phi\,-\,i\frac{\sin\phi}{\phi}\,\vec\s\cdot\vec\phi\right)\right]\\[5pt]\nonumber
&\approx&\;1-\frac{z_j^2+
w^2_j}{2}\,+\,2\cos\phi\frac{\sin\phi}{\phi}\vec \phi\cdot(\vec
z_j\times\vec
w_j)\\[5pt]\label{Gl:LinearizedIntegrandForEFlowers}
&&\qquad+(\sin^2\phi-\cos^2\phi)\vec w_j\cdot\vec
z_j\;-\;2\frac{\sin^2\phi}{\phi^2}\,(\vec\phi\cdot\vec
w_j)(\vec\phi\cdot\vec z_j)\\[5pt]\nonumber
&=:&\;1\;+\;I_j\left(\vec\phi,\vec z_j,\vec w_j\right).
\end{eqnarray}

\noindent Here, the algebraic relations
$\s_I\s_J=\d_{IJ}+i\e_{IJK}\s_K$ and $\tr\,\s_I=0$ have been used,
furthermore, only terms of the form $z^2,\,zw,\,w^2$ have been kept.
Before we continue to calculate, is pays to think about which terms
will play a role at all. First note that, since $h_j(\vec w_j)$ and
$h_j(\vec z_j)$ are both close to $1\in SL(2,\C)$, they are also
close to each other, hence also $h_j(\vec w_j)^c$ and $h_j(\vec
z_j)$ are close to each other. Thus, the main contribution of the
infinite sums in (\ref{Gl:RememberThisIsTheOverlap}) will come from
the geodesics going directly from $h_j(\vec w_j)^c$ to $h_j(\vec
z_j)$ rather than the longer ones. Technically, this means that only
the term with $n_1=\ldots=n_E=0$ will contribute significantly to
the integral. All other terms will be of order $O(t^{\infty})$
compared to it.

Secondly, we note that $z\to(\arccosh(1+z))^2\approx 2z+O(z^4)$ and
$z/\sinh z\approx 1-z^2/6+O(z^4)$. As $I_j(\vec\phi, \vec z_j,\vec
w_j)\to 0$ as $\vec z_j,\vec w_j\to 0$, we can expand the terms in
the exponential and the $z/\sinh z$-function, as well as the
exponential itself into quadratic orders of $\vec z_j,\vec w_j$, and
get
\begin{eqnarray}\nonumber \Big\langle\Psi_{[h(\vec
w_j)]}^t\Big|\Psi_{[h(\vec
z_j)]}^t\Big\rangle\;&\approx&\;\int_{B_{\pi}(0)}\left[\frac{\sin^2\phi}{\phi^2}d^3\phi\right]\;\left(1-\frac{\sum_{j=1}^E2I_j\left(\vec\phi,\vec
z_j,\vec
w_j\right)}{6}\right)\\[5pt]\label{Gl:CooleLineareFormal}
&&\qquad\times\left(1+\frac{\sum_{j=1}^E2I_j\left(\vec\phi,\vec
z_j,\vec w_j\right)}{t}\right)\\[5pt]\nonumber
&\approx&\;\int_{B_{\pi}(0)}\left[\frac{\sin^2\phi}{\phi^2}d^3\phi\right]\;\left(1+2\left(1-\frac{t}{3}\right)\sum_{j=1}^E\frac{I_j\left(\vec\phi,\vec
z_j,\vec w_j\right)}{t}\right)
\end{eqnarray}

\noindent where

\begin{eqnarray}\label{Gl:IntegrationInPolarCoordinates}
\int_{SU(2)}d\m_H(k)\,f(k)\;=\;\int_{B_{\pi}(0)}\frac{\sin^2\|\vec\phi\|}{\|\vec\phi\|^2}d^3\phi\;f\left(\exp(i\vec\s\cdot\vec\phi)\right)
\end{eqnarray}

\noindent has been used.\\

We immediately note the following interesting feature of the
integral (\ref{Gl:CooleLineareFormal}): It is linear in all the
$I_j$-terms. Thus, we can perform the integration over each $I_j$,
moreover, over each term in the $I_j$, separately! This huge
simplification is due to the fact that $(1,\ldots,1)\in SL(2,\C)^E$
is in fact a fixed point under the action of the gauge group, and
for small $\vec w_j,\vec z_j$, the orbit always stays close to
$(1,\ldots, 1)$.

So we only have to compute the integrals over $I_j$. In particular,
writing
\begin{eqnarray*}
\vec\phi\;=\;\phi\left(\begin{array}{c}\cos\varphi\sin\vartheta\\\sin\varphi\sin\vartheta\\\cos\vartheta\end{array}\right),
\end{eqnarray*}

\noindent it follows that
\begin{eqnarray*}
\int_{B_{\pi}(0)}\frac{\sin^2\phi}{\phi^2}d^3\phi\;&=&\;\frac{1}{2\pi^2}\int_0^{\pi}\sin^2\phi\,d\phi\int_{S^2}d\Om(\vartheta,\,\varphi)\;\\[5pt]
\;&=&\;
\;\frac{1}{2\pi^2}\int_0^{\pi}\sin^2\phi\,d\phi\int_{0}^{\pi}\sin\vartheta
\,d\vartheta\;\int_0^{2\pi}d\varphi,
\end{eqnarray*}

\noindent and with this one can calculate
\begin{eqnarray*}
\int_{B_{\pi}(0)}\frac{\sin^2\phi}{\phi^2}d^3\phi&&\;\;I_j(\vec\phi,\vec
w_j,\vec z_j)\;=\;-\frac{ z^2_j+ w_j^2}{2}\\[5pt]
&&+\:\frac{1}{\pi^2}\int_0^{\pi}d\phi\,\sin^3\phi\,\cos\phi\int_{S^2}d\Om(\vartheta,\varphi)\left(\begin{array}{c}
\cos\varphi\sin\vartheta\\\sin\varphi\sin\vartheta\\\cos\vartheta\end{array}\right)\cdot(\vec
z_j\times\vec w_j)\\[5pt]
&&+\;\frac{\vec w_j\cdot\vec
z_j}{2\pi^2}\int_0^{\pi}d\phi\,\sin^2\phi\,(\sin^2\phi-\cos^2\phi)\int_{S^2}d\Om(\vartheta,\varphi)\\[5pt]
&&-\;\frac{1}{\pi^2}\int_0^{\pi}d\phi\,\sin^4\phi\,\int_{S^2}d\Om(\vartheta,\varphi)\left[\left(\begin{array}{c}
\cos\varphi\sin\vartheta\\\sin\varphi\sin\vartheta\\\cos\vartheta\end{array}\right)\cdot\vec
z_j\right]\,\left[\left(\begin{array}{c}
\cos\varphi\sin\vartheta\\\sin\varphi\sin\vartheta\\\cos\vartheta\end{array}\right)\cdot\vec
w_j\right].
\end{eqnarray*}

\noindent The first integral in this expression vanishes, simply
because
\begin{eqnarray*}
\int_{S^2}d\Om(\vec u)\;\vec u\cdot\vec x \;=\;0
\end{eqnarray*}

\noindent for all $\vec x\in\C^3$. The second integral over $S^2$ is
trivial, and the integral over $\phi$ is elementary. To evaluate the
third integral we write
\begin{eqnarray*}
\left[\left(\begin{array}{c}
\cos\varphi\sin\vartheta\\\sin\varphi\sin\vartheta\\\cos\vartheta\end{array}\right)\cdot\vec
z_j\right]&&\left[\left(\begin{array}{c}
\cos\varphi\sin\vartheta\\\sin\varphi\sin\vartheta\\\cos\vartheta\end{array}\right)\cdot\vec
w_j\right]\\[5pt]
\;&=&\;\Big(z_1w_1\cos^2\varphi\sin^2\vartheta\;+\;(z_1w_2+z_2w_1)\cos\varphi\sin\varphi\sin\vartheta\\[5pt]
&&+\;w_2z_2\sin^2\varphi\sin^2\vartheta\;+\;(z_1w_3+z_3w_1)\cos\varphi\sin\vartheta\sin\varphi\\[5pt]
&&+\;(w_2z_3+w_3z_2)\sin\varphi\sin\vartheta\cos\vartheta\;+\;z_3w_3\,\cos^2\vartheta\Big).
\end{eqnarray*}

\noindent With elementary integrals, we get
\begin{eqnarray*}
\int_{B_{\pi}(0)}\frac{\sin^2\phi}{\phi^2}d^3\phi\;I_j(\vec\phi,\vec
w_j,\vec z_j)\;&=&\;-\frac{ z^2_j+w_j^2}{2}\;+\;\frac{1}{2}\vec
w_j\cdot\vec z_j-\;\frac{1}{2}\vec
w_j\cdot\vec z_j\\[5pt]
&=&\;-\frac{ z^2_j+w_j^2}{2}.
\end{eqnarray*}

\noindent Thus, we arrive at the result

\begin{eqnarray*}
\Big\langle\Psi_{[h(\vec w_j)]}^t\Big|\Psi_{[h(\vec
z_j)]}^t\Big\rangle\;&=&\;1\;-\;\left(1-\frac{t}{3}\right)\sum_{j=1}^E\frac{z_j^2+w_j^2}{t}\;+\;O(|z_j|^4,\;|w_j|^4).
\end{eqnarray*}

\noindent With this, the claim can be proven directly: Expanding the
overlap between $\Psi^t_{[1,\ldots 1]}$ and $\Psi^t_{[h(\vec z_j)]}$
into quadratic order reveals

\begin{eqnarray*}
\frac{\Big|\left\langle\Psi_{[1,\ldots,1]}^t\Big|\Psi_{[h(\vec
z_j)]}^t\right\rangle\Big|^2}
{\Big\|\Psi_{[1,\ldots,1]}^t\Big\|^2\;\Big\|\Psi_{[h(\vec
z_j)]}^t\Big\|^2}\;&=&\;1\;-\;\left(1-\frac{t}{3}\right)\sum_{j=1}^E\left(\frac{z_j^2}{t}+\frac{\bar
z_j^2}{t}\;-\;\frac{z_j^2+\bar z_j^2}{t}\right)\;-\;O(|z_j|^4)\\[5pt]
&=&\;1\;-\;O(|z_j|^4).
\end{eqnarray*}

\noindent This proves the claim of the theorem.\\

\noindent The previous theorem shows the effect degenerate gauge
orbits have on the peakedness properties of gauge-invariant coherent
states. Although the states are still concentrated around the gauge
orbits, the overlap function is no Gaussian anymore. Rather, the
function decreases much slower, revealing a plateau around the phase
space point they are labeled with.

\subsection{Inner product of gauge-invariant coherent states}

\noindent In this section we will relate the inner product of
gauge-invariant coherent states on arbitrary graphs to the inner
product of a corresponding state on a flower graph. A gauge-fixing
procedure closely related to gauge-fixing in lattice gauge theory
will be employed (see, e.g. \cite{FREI}).\\


\noindent The inner product between two gauge-invariant coherent
states on a graph $\g$ with $E$ edges and $V$ vertices is given by

\begin{eqnarray}\label{Gl:EichinvariantesInneresProdukt2}
\Big\langle\Psi^t_{[g_1,\ldots,g_E]}\Big|\Psi^t_{[h_1,\ldots,h_E]}\Big\rangle\;
&=&\;\frac{e^{E/t}}{\pi^E}\;\sqrt\frac{\pi}{t}^{3E}\int_{SU(2)^V}d\m_H(k_1,\ldots
k_V)\\[5pt]\nonumber
&&\qquad \sum_{n_1,\ldots
n_E\in\Z}\;\prod_{m=1}^E\,\frac{z_m\,-\,2\pi i
n_m}{\sinh(z_m\,-\,2\pi i n_m)}\;e^{\frac{(z_m\,-\,2\pi i
n_m)^2}{t}}
\end{eqnarray}

\noindent with
\begin{eqnarray}\label{Gl:EichinvariantesInneresProdukt3}
\cosh z_m\;=\;\frac{1}{2}\tr\;\big( g_m^{\dag}\,k_{b(m)}\,
h_m\,k_{f(m)}^{-1}\big).
\end{eqnarray}

\noindent The key to the procedure is lemma
\ref{Lem:ComplexShiftOfIntegrationOnLieGroups}. It enables us to
shift an integration variable $k_l$ in
(\ref{Gl:EichinvariantesInneresProdukt3}) by elements in $SL(2,\C)$:
\begin{eqnarray*}
k_l\;\longrightarrow\; G\,k_l\,H\;\qquad \text{with arbitrary
}H,G\in SL(2,\C).
\end{eqnarray*}

\noindent Now choose a maximal tree $\t$ in the graph $\g$. Remember
that if $\g$ has $V$ vertices and $E$ edges, then $\t$ has $V$
vertices and $V-1$ edges. Choose a vertex $\tilde v$. For each other
vertex $v_l\in \g$ define
\begin{eqnarray*}
G_l\;&:=&\;g_{e_1}^{\pm1}\cdot\ldots\cdot g_{e_n}^{\pm1}\\[5pt]
H_l\;&:=&\;h_{e_1}^{\pm1}\cdot\ldots\cdot h_{e_n}^{\pm1}.
\end{eqnarray*}

\noindent Here, the edges $e_1,e_2,\ldots,e_n$ are the edges one
needs to go in the maximal tree $\t$ from $v_l$ to $\tilde v$. This
path is unique, as $\t$ contains no loops. For each edge $e_k$
encountered, if the path from $v_l$ to $\tilde v$ goes against the
orientation of the edge $e_k$, then take $g_{e_k}^{-1}$ and
$h_{e_k}^{-1}$, if the path goes with the orientation of $e_k$, take
$g_{e_k}$ and $h_{e_k}$. By this procedure, for each vertex $v_l$
two elements $G_l,H_l$ are defined (Note that by definition the
elements for the vertex $\tilde v$ are both $1\in SL(2,\C)$). Then,
shift the integration in (\ref{Gl:EichinvariantesInneresProdukt3})
by
\begin{eqnarray*}
k_l\;\longrightarrow\;G_l^{\dag}\,k_l\,H_l.
\end{eqnarray*}

\noindent Let $e_m$ be an edge in $\t$. The corresponding function
(\ref{Gl:EichinvariantesInneresProdukt3}) changes to
\begin{eqnarray*}
\frac{1}{2}\tr\;\Big(g_m^{\dag}\,k_{b(m)}\,h_m\,k_{f(m)}^{-1}\Big)\;\longrightarrow\;\frac{1}{2}\tr\;\Big(k_{b(m)}\,k_{f(m)}^{-1}\Big),
\end{eqnarray*}

\noindent whereas for some $e_m$ not being in $\t$, the term
(\ref{Gl:EichinvariantesInneresProdukt3}) changes to

\begin{eqnarray*}
\frac{1}{2}\tr\;\Big(g_m^{\dag}\,k_{b(m)}\,h_m\,k_{f(m)}^{-1}\Big)\;\longrightarrow\;\frac{1}{2}\tr\;\Big(\tilde
g_m^{\dag}\,k_{b(m)}\,\tilde h_m\,k_{f(m)}^{-1}\Big)
\end{eqnarray*}

\noindent with $\tilde g_m$ and $\tilde h_m$ being nontrivial
products of various $g$'s and $h$'s respectively. In particular,
these products are given by
\begin{eqnarray*}
\tilde h_m\;&=&\;h_{e_1}^{\pm 1}\ldots h^{\pm
1}_{e_n}\,h_m\,h_{e_{n+1}}^{\pm 1}\ldots h_{e_m}^{\pm 1} \\[5pt]
\tilde g_m\;&=&\;g_{e_1}^{\pm 1}\ldots g^{\pm
1}_{e_n}\,g_m\,g_{e_{n+1}}^{\pm 1}\ldots g_{e_m}^{\pm 1}.
\end{eqnarray*}

\noindent Here, the sequence of edges
$e_1,\ldots,e_n,e_m,e_{n+1},\ldots e_N$ is a loop in $\g$, starting
at $\tilde v$, going to the beginning of the edge $e_m$ in $\t$,
going along $e_m$, and then going back to $\tilde v$, again along
edges in $\t$. Note that this path is unique. As usual, $h_{e_l}$ is
taken if the path is going along the orientation of the edge, and
$h_{e_l}^{-1}$ is taken if the path goes against the orientation of
$e_l$. Similarly for the $\tilde g$.

This gives two sets of $E-V+1$ elements in $SL(2,\C)$, which are not
gauge-invariant, but behave quite simple under a global
gauge-transformation. It is quite easy to se that under some gauge
transformation $k\;\equiv\;k_{v_1}\,=\,\ldots\,=\,k_{v_V}$ the new
elements change as
\begin{eqnarray*}
\tilde h_{e_l}\;\longrightarrow\;k\,\tilde h_{e_l}\,k^{-1}.
\end{eqnarray*}

\noindent A similar formula holds for the $\tilde g$. 
With
\begin{eqnarray*}
\cos\frac{1}{2}\tr\;\Big(k_{v}^{-1}\,k_{v'}\Big)\;=\;d(k_v,\,k_{v'}),
\end{eqnarray*}

\noindent where $d$ denotes the geodesic distance on $SU(2)$ (for
the shortest geodesic connecting $k_v,\,k_{v'}\in SU(2)$, and
$\cosh(i x)=\cos(x)$, the above considerations give the following
formula for the inner product between two gauge-invariant coherent
states:
\begin{eqnarray}\nonumber
\Big\langle\Psi^t_{[g_1,\ldots,g_E]}\Big|\Psi^t_{[h_1,\ldots,h_E]}\Big\rangle\;
&=&\;\frac{e^{E/t}}{\pi^E}\;\sqrt\frac{\pi}{t}^{3E}\int_{SU(2)^V}d\m_H(k_1,\ldots
k_V)\\[5pt]\nonumber
&& \hspace{-50pt}\times\sum_{n_1,\ldots
n_{V-1}\in\Z}\prod_{m=1}^{V-1}\,\frac{d(k_{b(m)},\,k_{f(m)})\,-\,2\pi
n_m}{\sin
d(k_{b(e_m)},\,k_{f(e_m)})}\;\exp\left[{-\frac{(d(k_{b(e_m)},\,k_{f(e_m)})\,-\,2\pi
n_m)^2}{t}}\right]\\[5pt]\label{Gl:ElementsOnTreeCanBeGaugedTo1}
&&\hspace{-50pt}\times\sum_{n_{V},\ldots,n_E\in\Z}\prod_{m=V}^E\,\frac{z_m\,-\,2\pi
i n_m}{\sinh(z_m\,-\,2\pi i n_m)}\;\exp\left[{\frac{(z_m\,-\,2\pi i
n_m)^2}{t}}\right]
\end{eqnarray}

\noindent with
\begin{eqnarray*}
\cosh z_m\;=\;\frac{1}{2}\tr\;\big(\tilde
g_m^{\dag}\,k_{b(m)}\,\tilde h_m\,k_{f(m)}^{-1}\big)
\end{eqnarray*}

\noindent for $m=V,\ldots,E$. Note that this can be rewritten as

\begin{eqnarray}\label{Gl:ElementsOnTreeCanBeGaugedTo1-2}
\Big\langle\Psi^t_{[g_1,\ldots,g_E]}\Big|\Psi^t_{[h_1,\ldots,h_E]}\Big\rangle\;=\;\Big\langle\Psi^t_{[1,\ldots,1,\tilde
g_V,\ldots \tilde g_E]}\Big|\Psi^t_{[1,\ldots, 1,\tilde
h_V,\ldots,\tilde h_E]}\Big\rangle.
\end{eqnarray}

\noindent Thus, we have seen that the inner product between
gauge-invariant coherent states labeled with arbitrary elements in
$SL(2,\C)$ is equal to the inner product of states where all labels
corresponding to edges in a maximal tree $\t$ are equal to $1\in
SL(2,\C)$. This procedure can, in an obvious way, carried over to
the states themselves, showing that one can always gauge the labels
corresponding to edges in $\t$ to $1$.

To make contact with the inner product of gauge-invariant coherent
states on flower graphs, we make further modifications in formula
(\ref{Gl:ElementsOnTreeCanBeGaugedTo1}). We now shift the
integrations in (\ref{Gl:ElementsOnTreeCanBeGaugedTo1}) one after
another.\\

First, order the set of vertices by the following method: Choose a
vertex and call it $v_1$. For all $l=2,\ldots V$, define $v_{l+1}$
such that the path from $v_{l+1}$ to $v_1$ through $\t$ only passes
the vertices $v_1,\ldots,v_l$, i.e. already ordered vertices. This
gives, in the end, a numeration $v_1,\ldots, v_V$ of the vertices,
which we will keep for the rest of this section.\\

The integration variables $k_1,\ldots,k_V$ appear in the integral
(\ref{Gl:ElementsOnTreeCanBeGaugedTo1}) in the following
combinations:

\begin{eqnarray}\nonumber
k_{b(1)}\,&&k_{f(1)}^{-1}\\[5pt]\nonumber
k_{b(2)}\,&&k_{f(2)}^{-1}\\[5pt]\nonumber
\vdots\quad&&\\[5pt]\label{Gl:IntegrationVariablesInGaugeShiftedIntegrand}
k_{b(V-1)}\,&&k_{f(V-1)}^{-1}\\[5pt]\nonumber
\tilde g^{\dag}_{V}\,k_{b(V)}\,&&\tilde h_{V}\,k_{f(V)}^{-1}\\[5pt]\nonumber
\vdots\quad&&\\[5pt]\nonumber
\tilde g^{\dag}_{E}\,k_{b(E)}\,&&\tilde h_{E}\,k_{f(E)}^{-1}
\end{eqnarray}

\noindent We shift the elements $k_1,\ldots,k_V$ in ascending order,
where the order of $k_l$ is determined by the numeration of vertices
$v_1,\ldots v_V$ defined above. First, we do not shift $k_1$.
Second, we shift $k_2$ by
\begin{eqnarray*}
k_2\;\longrightarrow\;k_2k_1.
\end{eqnarray*}

\noindent We proceed by shifting the $k_l$ by
\begin{eqnarray}\label{Gl:DefinitionOfTildeK}
k_l\;\longrightarrow\;k_lk_ak_b\ldots k_1\;=:\;\tilde k_l
\end{eqnarray}

\noindent such that $v_l\to v_a\to v_b\to\ldots\to v_1$ is a path
from the vertex $v_l$ to $v_1$ in the maximal tree $\t$. So, as soon
as a $k_l$ in (\ref{Gl:IntegrationVariablesInGaugeShiftedIntegrand})
is replaced by the appropriate $\tilde k_l$, the single elements in
it will not be altered by the following shifts, since the product of
the $\tilde k_l$ only consists of elements $k_1,\ldots k_{l-1}$ that
already have been shifted by the above procedure. It follows that,
after this procedure, the expressions
(\ref{Gl:IntegrationVariablesInGaugeShiftedIntegrand}), that
ultimately appear in (\ref{Gl:ElementsOnTreeCanBeGaugedTo1}), will
have changed to

\begin{eqnarray}\nonumber
\tilde k_{b(1)}\,&&\tilde k_{f(1)}^{-1}\\[5pt]\nonumber
\tilde k_{b(2)}\,&&\tilde k_{f(2)}^{-1}\\[5pt]\nonumber
\vdots\quad&&\\[5pt]\label{Gl:IntegrationVariablesInGaugeShiftedIntegrand-Shifted}
\tilde k_{b(V-1)}\,&&\tilde k_{f(V-1)}^{-1}\\[5pt]\nonumber
\tilde g^{\dag}_{V}\,\tilde k_{b(V)}\,&&\tilde h_{V}\,\tilde k_{f(V)}^{-1}\\[5pt]\nonumber
\vdots\quad&&\\[5pt]\nonumber
\tilde g^{\dag}_{E}\,\tilde k_{b(E)}\,&&\tilde h_{E}\,\tilde
k_{f(E)}^{-1},
\end{eqnarray}

\noindent where the $\tilde k_l$, defined by
(\ref{Gl:DefinitionOfTildeK}), contain products of various $k_l$.
But these products have the following properties:

First, for $1\leq m\leq V-1$, we have that
\begin{eqnarray}\label{Gl:DecisionWhichElementIsTaken}
\tilde k_{b(m)}\,&&\tilde
k_{f(m)}^{-1}\;=\;\left\{\begin{array}{cl}k_{f(m)}^{-1}& \text{ if }e_m\text{ is part of the path }v_{f(m)}\,\to\,v_1\\[5pt]
k_{b(m)}&\text{ if }e_m\text{ is part of the path
}v_{b(m)}\,\to\,v_1\end{array}\right.
\end{eqnarray}

\noindent One can easily see that the two cases are mutually
excluding. From this we can also immediately deduce that in the
first $V-1$ terms in
(\ref{Gl:IntegrationVariablesInGaugeShiftedIntegrand-Shifted}),
every $k_l$ from $k_2$ to $k_V$ appears exactly once, either as
$k_l$ or as $k_l^{-1}$. Assume that a $k_l$ is appearing twice,
either as $k_l$, or as $k_l^{-1}$ in the first $V-1$ terms in
(\ref{Gl:IntegrationVariablesInGaugeShiftedIntegrand-Shifted}), say
in terms $m$ and $m'$. But from
(\ref{Gl:DecisionWhichElementIsTaken}), the $k_l$ appears only if
the corresponding edge is part of the path from $v_l$ to $v_1$ in
$\t$. Having $k_l$ occurring a place $m$ and $m'$ means that the
path from $v_l$ to $v_1$ in $\t$ contains both $e_m$ and $e_{m'}$.
But, since $m\neq m'$, this means that there are two different paths
from $v_l$ to $v_1$, one containing $e_m$, the other one containing
$e_{m'}$. But this is a contradiction, since the paths all lie
entirely in $\t$, which contains no loops, hence from any two
vertices there is a unique path between them. This shows that in the
first $V-1$ places in
(\ref{Gl:IntegrationVariablesInGaugeShiftedIntegrand-Shifted}),
every element $k_l$ appears at most once. And, since every $\tilde
k_l$ ends with $k_1$, as one sees from
(\ref{Gl:DefinitionOfTildeK}), $k_1$ is the only one that is not
appearing. So,
(\ref{Gl:IntegrationVariablesInGaugeShiftedIntegrand-Shifted}) in
fact looks like

\begin{eqnarray}\nonumber
k_2\\[5pt]\nonumber
k_3\\[5pt]\nonumber
\vdots\quad&&\\[5pt]\label{Gl:IntegrationVariablesInGaugeShiftedIntegrand-Shifted-AndRecomputed}
k_V\\[5pt]\nonumber
\tilde g^{\dag}_{V}\,\tilde k_{b(V)}\,&&\tilde h_{V}\,\tilde k_{f(V)}^{-1}\\[5pt]\nonumber
\vdots\quad&&\\[5pt]\nonumber
\tilde g^{\dag}_{E}\,\tilde k_{b(E)}\,&&\tilde h_{E}\,\tilde
k_{f(E)}^{-1},
\end{eqnarray}

\noindent where we have changed the order of the first $V-1$ terms,
and have replaced all elements appearing in these terms at its
inverse by the elements itself, which is allowed, since all terms in
(\ref{Gl:IntegrationVariablesInGaugeShiftedIntegrand-Shifted-AndRecomputed})
appear inside a trace in (\ref{Gl:ElementsOnTreeCanBeGaugedTo1}),
and $\tr\; k=\tr\;k^{-1}$ for all $k\in SU(2)$.\\

The fact that in all $\tilde k_l$, $k_1$ appears only at the last
position of the product (\ref{Gl:DefinitionOfTildeK}) lets us
rewrite the lase $E-V+1$ terms in
(\ref{Gl:IntegrationVariablesInGaugeShiftedIntegrand-Shifted-AndRecomputed})
as follows:

\begin{eqnarray}\nonumber
k_2\\[5pt]\nonumber
k_3\\[5pt]\nonumber
\vdots\quad&&\\[5pt]\label{Gl:IntegrationVariablesInGaugeShiftedIntegrand-Shifted-AndRecomputed-AndRewritten}
k_V\\[5pt]\nonumber
(\bar k_V^{-1}\,\tilde
g_V\,\bar{\bar{k}}_V)^{\dag}\, k_1\,&&\tilde h_{V}\,k_1^{-1}\\[5pt]\nonumber
\vdots\quad&&\\[5pt]\nonumber
 (\bar k_E^{-1}\,\tilde
g_E\,\bar{\bar{k}}_E)^{\dag}\,k_1\,&&\tilde h_{E}\, k_1^{-1},
\end{eqnarray}

\noindent where
\begin{eqnarray*}
\bar k_m\;&:=&\;k_{b(m)}\,k_a\,k_b\,\cdots\,k_z\\[5pt]
\bar{\bar k}_m\;&:=&\;k_{f(m)}\,k_{a'}\,k_{'b}\,\cdots\,k_{z'},
\end{eqnarray*}

\noindent where the path $v_{b(m)}\to v_a\to v_b\to\cdots\to v_z$ is
the path from $v_{b(f)}$ to $v_1$ (in $\t$), excluding $v_1$ at the
last position. Consequently, $v_{f(m)}\to v_{a'}\to
v_{b'}\to\cdots\to v_{z'}$ is the path from $v_{f(m)}$ to $v_1$,
excluding $v_1$.\\

Having the integration variables shifted like this, we can finally
recast (\ref{Gl:ElementsOnTreeCanBeGaugedTo1}) into the following
form:

\begin{eqnarray}\label{Gl:InnerProductCanBeExpressedAsInnerProductOnFlowerGraph}
\Big\langle\Psi^t_{[g_1,\ldots,g_E]}\Big|\Psi^t_{[h_1,\ldots,h_E]}\Big\rangle\;
&=&\;\frac{e^{E/t}}{\pi^E}\;\sqrt\frac{\pi}{t}^{3E}\int_{SU(2)^V}d\m_H(k_1,\ldots
k_V)\\[5pt]\nonumber
&& \hspace{-50pt}\times\sum_{n_2,\ldots
n_{V}\in\Z}\prod_{l=2}^{V}\,\frac{d(k_l)\,-\,2\pi n_l}{\sin
d(k_l)}\;\exp\left[{-\frac{(d(k_l)\,-\,2\pi
n_l)^2}{t}}\right]\\[5pt]\nonumber
&&\hspace{-50pt}\times\sum_{n_{V},\ldots,n_E\in\Z}\prod_{m=V}^E\,\frac{z_m\,-\,2\pi
i n_m}{\sinh(z_m\,-\,2\pi i n_m)}\;\exp\left[{\frac{(z_m\,-\,2\pi i
n_m)^2}{t}}\right]
\end{eqnarray}

\noindent with

\begin{eqnarray}
\cosh z_m\;=\;\frac{1}{2}\Big((\bar k_m^{-1}\,\tilde
g_m\,\bar{\bar{k}}_m)^{\dag}\, k_1\,&&\tilde h_{m}\,k_1^{-1}\Big).
\end{eqnarray}

\noindent and $d(k):=d(1,k)$ is the geodesic distance between $k$
and $1\in SU(2)$.\\

 But with this we immediately see that
\begin{eqnarray}\label{Gl:RelateInnerProductsToThoseOnFlowerGraphs}
\Big\langle\Psi^t_{[g_1,\ldots,g_E]}\Big|\Psi^t_{[h_1,\ldots,h_E]}\Big\rangle\;
&=&\;\frac{e^{E/t}}{\pi^E}\;\sqrt\frac{\pi}{t}^{3E}\int_{SU(2)^{V-1}}d\m_H(k_2,\ldots
k_V)\\[5pt]\nonumber
&& \hspace{-50pt}\times\sum_{n_2,\ldots
n_{V}\in\Z}\prod_{l=2}^{V}\,\frac{d(k_l)\,-\,2\pi n_l}{\sin
d(k_l)}\;\exp\left[{-\frac{(d(k_l)\,-\,2\pi n_l)^2}{t}}\right]\\[5pt]\nonumber
&&\hspace{-50pt}\times\;\Big\langle\Psi^t_{[\bar k_V^{-1}\,\tilde
g_V\,\bar{\bar{k}}_V,\ldots,\bar k_E^{-1}\,\tilde
g_E\,\bar{\bar{k}}_E]}\Big|\Psi^t_{[\tilde h_V,\ldots, \tilde
h_E]}\Big\rangle
\end{eqnarray}

\noindent where the last term is the inner product between
gauge-invariant coherent states  on a $E-V+1$-flower graph, labeled
with the (equivalence class of the) $E-V+1$ elements $\bar
k_V^{-1}\,\tilde g_V\,\bar{\bar{k}}_V,\ldots,\bar k_E^{-1}\,\tilde
g_E\,\bar{\bar{k}}_E$ and $\tilde h_V,\ldots, \tilde h_E$
respectively. Thus, the inner product between the gauge-invariant
coherent states on arbitrary graphs can be related to the inner
product of states on flower graphs.\\

The appearance of the Gaussian factors containing the geodesic
distance of the elements $k_2,\ldots k_V$ to $1\in SU(2)$ leads to
the following idea concerning the behavior of the integral
(\ref{Gl:RelateInnerProductsToThoseOnFlowerGraphs}) in the limit of
small $t$, which is the ultimate interest in LQG, where this $t$ is
usually understood as some kind of semiclassicality scale, e.g. the
ratio between Planck- and classical scales. It is tempting to think
of the following:

Consider the limit $t\to 0$. Then, the Gaussians in
(\ref{Gl:RelateInnerProductsToThoseOnFlowerGraphs}) will damp away
all contributions to the integral not coming from the vicinity of
$k_2=\cdots=k_V$. So, in the limit of small $t$ one might, at least
asymptotically and up to a factor, say that
\begin{eqnarray}\label{Gl:Temptation}
\Big\langle\Psi^t_{[g_1,\ldots,g_E]}\Big|\Psi^t_{[h_1,\ldots,h_E]}\Big\rangle\;\longrightarrow\;
\Big\langle\Psi^t_{[\tilde g_V,\ldots,\tilde
g_E]}\Big|\Psi^t_{[\tilde h_V,\ldots,\tilde h_E]}\Big\rangle,
\end{eqnarray}

\noindent where the first inner product is between gauge-invariant
coherent states in (\ref{Gl:Temptation}) living on an arbitrary
graph, while the second one is between states on a (corresponding)
flower graph.\\

However, this deduction is, unfortunately, wrong: This is because
the inner product between gauge-invariant coherent states on the
$E-V+1$ flower graph contains the elements $k_2,\ldots k_V$ in the
exponential with power $1/t$. So, the reasoning that in the limit of
$t\to 0$ everything in the vicinity of $k_2=\ldots k_V=1$ is damped
away is not correct, since the integrand itself could be an
exponentially increasing function in the limit of $t\to 0$ there.
This can best be seen as an example. The above deduction states
that, in the limit of $t\to 0$, the main contribution to the
integral
\begin{eqnarray*}
\frac{1}{\sqrt{\pi t}}\;\int_{\R}dx\;e^{-\frac{x^2}{t}}\;F_t(x)
\end{eqnarray*}

\noindent comes from $x\approx 0$, since everywhere else the
integrand is exponentially small by the Gaussian. But, if for
instance $F_t(x)$ is given by $\exp((x+z)^2/2t)$ for some complex
number $z\in\C$, we the main contribution to the integral

\begin{eqnarray*}
\frac{1}{\sqrt{\pi
t}}\;\int_{\R}dx\;e^{-\frac{x^2}{t}}\;e^{\frac{(x+z)^2}{2t}}
\end{eqnarray*}

\noindent does not come from $x\approx 0$, so one cannot assume

\begin{eqnarray*}
\frac{1}{\sqrt{\pi
t}}\;\int_{\R}dx\;e^{-\frac{x^2}{t}}\;e^{\frac{(x+z)^2}{2t}}\;\longrightarrow\;e^{\frac{z^2}{2t}},
\end{eqnarray*}

\noindent rather, the integral is equal to $\exp(z^2/t)$ instead.
This is exactly the same situation as in
(\ref{Gl:RelateInnerProductsToThoseOnFlowerGraphs}), and is also the
reason why the idea of (\ref{Gl:Temptation}) is a priori not right.\\

Still, in the following chapter we will, for arbitrary graphs and in
the limit of $t\to 0$, compute the overlap of a gauge-invariant
coherent state labeled at $[1,\ldots,1]$ and another one, labeled at
a point close to $[1,\ldots, 1]$. By carefully performing the
limits, we will be able to obtain a similar result as in the case of
the flower graphs.


\subsection{Peakedness of $\Psi^t_{[1,\ldots,1]}$ on arbitrary graphs}

\noindent In this section, we will prove that for small
semiclassicality parameter $t$, the peakedness properties of the
state $\Psi^t_{[1,\ldots,1]}$ are qualitatively different than the
peakedness properties of generic gauge-invariant coherent states. We
have already shown this for states on $E$-flower graphs, where this
holds for arbitrary $t$. The strategy there was as follows: Our aim
was to show that the overlap
\begin{eqnarray}\label{Gl:OverlapInChapter7}
\frac{\Big|\Big\langle\Psi^t_{[1,\ldots,\,1]}\Big|\Psi^t_{[g_1,\ldots,g_E]}\Big\rangle\Big|^2}
{\Big\|\Psi^t_{[1,\ldots,\,1]}\Big\|^2\,\Big\|\Psi_{[g_1,\ldots,g_E]}\Big\|^2}
\end{eqnarray}

\noindent was, when expanding the $g_m$ around the identity
\begin{eqnarray*}
g_m\;=\;\exp\big(i\vec
z_m\cdot\vec\s\big)\;\approx\;1-\frac{z_m^2}{2}\,+\,i\vec z_m\cdot
\s,
\end{eqnarray*}

\noindent not a Gaussian. In particular, we showed that
\begin{eqnarray*}
(\ref{Gl:OverlapInChapter7})\;=\;1\;-\;O(\|\vec z_m\|^4),
\end{eqnarray*}

 \noindent rather than of the form $1-O(|\vec z_m|^2)$, as
would be expected from a Gaussian peak. We did this by noticing that
numerator and denominator in  (\ref{Gl:OverlapInChapter7}) are even
functions in the $\vec z_m$, so in the expansion only the even
powers occur. Furthermore, we were able to show that the quadratic
orders in the numerator and the denominator cancel, which shows that
at most the quartic orders contribute to the overlap.

In all our calculations it was crucial that we were dealing with a
state on a flower graph, since this allowed us to perform the gauge
integral analytically for the second order expansion. For arbitrary
graphs, as we have seen in the last section, things are more
difficult. Although we are able to relate the inner product of
states on arbitrary graphs to those on flower graphs, this will not
be enough to prove an equally strong result about the peakedness
properties of $\Psi^t_{[1,\ldots,1]}$ on arbitrary graphs. Still,
the second order of numerator and denominator in
(\ref{Gl:OverlapInChapter7}) can be evaluated in the limit $t\to 0$.
This shows that in the case of ultimate interest for LQG, where the
semiclassicality parameter $t$ will be, depending on the
application, about $t\approx 10^{-70}$, the states
$\Psi^t_{[1,\,\ldots ,\,1]}$ will have different peakedness
properties than states labeled by generic elements. As the plots for
the 3-bridge suggest, this statement holds true also for arbitrary
$t$, as in the case of the flower graphs, it is just that on
arbitrary graphs the integrals become too complicated to evaluate,
so the
limit $t\to 0$ is what we have to live with.\\

We proceed along similar lines as for the flower graphs: we expand
\begin{eqnarray}
\Big\langle\Psi^t_{[g_1,\ldots,g_E]}\Big|\Psi^t_{[h_1,\ldots,h_E]}\Big\rangle\;=\;\Big\langle\Psi^t_{[1,\ldots,1,\tilde
g_V,\ldots \tilde g_E]}\Big|\Psi^t_{[1,\ldots, 1,\tilde
h_V,\ldots,\tilde h_E]}\Big\rangle.
\end{eqnarray}

\noindent around $\tilde g_m\approx\tilde h_m\approx 1$, and
consider the second order contribution. We start with formula
(\ref{Gl:ElementsOnTreeCanBeGaugedTo1}):

\begin{eqnarray}\nonumber
\Big\langle\Psi^t_{[g_1,\ldots,g_E]}\Big|\Psi^t_{[h_1,\ldots,h_E]}\Big\rangle\;
&=&\;\frac{e^{E/t}}{\pi^E}\;\sqrt\frac{\pi}{t}^{3E}\int_{SU(2)^V}d\m_H(k_1,\ldots
k_V)\\[5pt]\nonumber
&& \hspace{-50pt}\times\sum_{n_1,\ldots
n_{V-1}\in\Z}\prod_{m=1}^{V-1}\,\frac{d(k_{b(m)},\,k_{f(m)})\,-\,2\pi
n_m}{\sin
d(k_{b(e_m)},\,k_{f(e_m)})}\;\exp\left[{-\frac{(d(k_{b(e_m)},\,k_{f(e_m)})\,-\,2\pi
n_m)^2}{t}}\right]\\[5pt]\label{Gl:NanuNochNeFormel}
&&\hspace{-50pt}\times\sum_{n_{V},\ldots,n_E\in\Z}\prod_{m=V}^E\,\frac{z_m\,-\,2\pi
i n_m}{\sinh(z_m\,-\,2\pi i n_m)}\;\exp\left[{\frac{(z_m\,-\,2\pi i
n_m)^2}{t}}\right].
\end{eqnarray}

\noindent We now choose the elements $g_1,\ldots,g_E$ and
$h_1,\ldots h_E$ all close to $1\in SL(2,\C)$. This, of course,
implies that also the gauge-fixed quantities $\tilde g_V,\ldots
\tilde g_E$ and $\tilde h_V,\ldots \tilde h_E$ are close to $1$. In
particular, we write
\begin{eqnarray}\label{Gl:ExpansionOfElementsInSL2,C}
\tilde g^{\dag}_m\;&=&\; \exp(i \vec
w_m\cdot\vec\sigma)\;\approx\;1\,-\,\frac{w_m^2}{2}\,+\,i\vec
w_m\cdot\vec \s\\[5pt]\nonumber
\tilde h_m\;&=&\; \exp(i \vec
z_m\cdot\vec\sigma)\;\approx\;1\,-\,\frac{z_m^2}{2}\,+\,i\vec
z_m\cdot\vec \s
\end{eqnarray}

\noindent for some vectors $\vec w_m,\,\vec z_m\in\C^3$ being close
to $0$, i.e. $\|\vec w_m\|,\,\|\vec z_m\|\,\ll 1$. This, of course,
also implies that $w_m,\,z_m$ are complex numbers with small
modulus:
\begin{eqnarray*}
|w_m|\,\ll\,1,\;\qquad|z_m|\,\ll\,1\qquad\text{for all
}m\,=\;V,\ldots,\, E.
\end{eqnarray*}

\noindent We now use the expansion
(\ref{Gl:ExpansionOfElementsInSL2,C}) to expand first

\begin{eqnarray*}
\cosh z_m\;=\;\frac{1}{2}\tr\;\big(\tilde
g_m^{\dag}\,k_{b(m)}\,\tilde h_m\,k_{f(m)}^{-1}\big),
\end{eqnarray*}

\noindent and then ultimately the inner product between
$\Psi^t_{[1,\ldots,\,1,\,\tilde g_V,\ldots,\, \tilde g_E]}$ and
$\Psi^t_{[1,\ldots,\,1,\,\tilde h_V,\ldots,\, \tilde h_E]}$ into
second order of $\vec w_m$ and $\vec z_m$. We will then see that in
the limit of small $t$, the second order term of the overlap between
these two states vanishes, leading to
\begin{eqnarray*}
\frac{\Big|\Big\langle\Psi^t_{[1,\ldots
1]}\,\Big|\,\Psi^t_{[1,\ldots,\,1,\,\tilde g_V,\ldots,\, \tilde
g_E]}\Big\rangle\Big|^2}{\Big\|\Psi^t_{[1,\ldots
1]}\Big\|^2\;\Big\|\Psi^t_{[1,\ldots,\,1,\,\tilde g_V,\ldots,\,
\tilde g_E]}\Big\|^2}\;=\;1\;-\;O(\|\vec w_m\|^4).
\end{eqnarray*}

\noindent We start with expanding the $\cosh$-term into second
order. With
\begin{eqnarray*}
k_{b(m)}\;=\;\exp\big(i\vec\phi_m\cdot\vec\s\big),\qquad
k_{f(m)}\;=\;\exp\big(i\vec\psi_m\cdot\vec\s\big)
\end{eqnarray*}

\noindent we get

\begin{eqnarray}\nonumber
\frac{1}{2}\tr\;\big(\tilde g_m^{\dag}\,k_{b(m)}\,\tilde
h_m\,k_{f(m)}^{-1}\big)\;&\approx&\;\frac{1}{2}\tr\;\big(k_{b(m)}\,k_{f(m)}^{-1}\big)\left(1-\frac{w_l^2+z_l^2}{2}\right)\;\\[5pt]\nonumber
&&+\;\left(\cos\phi_m\frac{\sin\psi_m}{\psi_m}\vec\psi_m\;-\;\cos\psi_m\frac{\sin\phi_m}{\phi_m}\vec\phi_m\right)\cdot(\vec
z_m+\vec w_m)\\[5pt]\label{Gl:BekotzteEntwicklung}
&&+\;\frac{\sin\phi_m}{\phi_m}\frac{\sin\psi_m}{\psi_m}(\vec\phi_m\times\vec\psi_m)(\vec
w_m-\vec z_m)\\[5pt]\nonumber
&&+\;\left(\cos\phi_m\frac{\sin\psi_m}{\psi_m}\vec\psi_m\;+\;\cos\psi_m\frac{\sin\phi_m}{\phi_m}\vec\phi_m\right)\cdot(\vec
z_m\times \vec w_m)\\[5pt]\nonumber
&&-\;\left(\cos\phi_m\cos\psi_m\,-\,\frac{\sin\phi_m}{\phi_m}\frac{\sin\psi_m}{\psi_m}\vec\phi_m\cdot\vec\psi_m\right)\,\vec
w_m\cdot\vec z_m\\[5pt]\nonumber
&&-\;\frac{\sin\phi_m}{\phi_m}\frac{\sin\psi_m}{\psi_m}(\vec
w_m\cdot\vec \psi_m)(\vec z_m\cdot\vec \phi_m).
\end{eqnarray}

\noindent Furthermore, if we expand the factors in
(\ref{Gl:ElementsOnTreeCanBeGaugedTo1}), we get
\begin{eqnarray}\nonumber
&&\frac{\arccosh (a+x)-2\pi i n}{\sinh\arccosh
a+x}\;e^{\frac{(\arccosh(a+x)-2\pi i
n)^2}{t}}\\[5pt]\label{Gl:RichtigBekotzteEntwicklung}
&&\qquad\;\approx\;\frac{\arccosh (a)-2\pi i n}{\sinh\arccosh
a}\;e^{\frac{(\arccosh(a)-2\pi i
n)^2}{t}}\\[5pt]\nonumber
&&\qquad\;\Bigg[1\;+\;\Big(a\,+\,2\frac{\sinh\arccosh a(\arccosh
a-2\pi
n i)}{t}\Big)x\\[5pt]\nonumber
&&\qquad\;+\;\frac{1}{2t^2\,\sinh^4\arccosh
a}\;\Big(-6a\sqrt{a^2-1}(\arccosh a-2\pi in) t +
4(a^2-1)(\arccosh(a)-2\pi i n)^2\\[5pt]\nonumber
&&\qquad\qquad\;+t(t-6+2 a^2(3+t))\Big)x^2\Bigg].
\end{eqnarray}

\noindent The next step would be to insert
(\ref{Gl:BekotzteEntwicklung}) into
(\ref{Gl:RichtigBekotzteEntwicklung}), then inserting this into
(\ref{Gl:ElementsOnTreeCanBeGaugedTo1}) and performing the gauge
integrals over $SU(2)^V$. While in the case for the $E$-flower graph
the gauge integral turned out to be trivial, here this is no longer
the case. On arbitrary graphs, this integral turns out to be too
complicated to solve directly.

Still, we are able to perform this calculation in the limit $t\to
0$, which is of ultimate interest for LQG. In this limit the terms
will simplify tremendously, so we will be able to produce the
desired result.\\

\noindent First, we look at the zero order term. This term amounts
to setting $\tilde g_m=\tilde h_m=0$ in (\ref{Gl:NanuNochNeFormel}),
hence the term of zero order is simply the norm of the coherent
state $\Psi^t_{[1,\ldots,1]}$. Next, let us consider the terms of
linear order in $\vec z_m,\,\vec w_m$. But from
(\ref{Gl:ElementsOnTreeCanBeGaugedTo1}) we immediately see that the
inner product (\ref{Gl:NanuNochNeFormel}) is an even function in
these variables. Thus, the linear terms in the numerator and the
denominator in (\ref{Gl:OverlapInChapter7}) will vanish, as will all
the odd order terms. So, the first terms that yield a nontrivial
contribution are the ones that are of second order in the $\vec w_m,
\vec z_m$. We will consider these now.

Before we insert all the terms of the expansions made above into
each other, it pays to look at how this integral behaves as $t\to
0$. The second order of (\ref{Gl:NanuNochNeFormel}) is of the form

\begin{eqnarray}\nonumber
(\text{$2^{\text{nd}}$
order})\;&=&\;\frac{e^{E/t}}{\pi^E}\;\sqrt\frac{\pi}{t}^{3E}\int_{SU(2)^V}d\m_H(k_1,\ldots
k_V)\;\\[5pt]\nonumber
&&\;\sum_{n_1,\ldots
n_E\in\Z}\prod_{m=1}^{E}\,\frac{d(k_{b(m)},\,k_{f(m)})\,-\,2\pi
n_m}{\sin
d(k_{b(m)},\,k_{f(m)})}\;\exp\left[{-\frac{(d(k_{b(m)},\,k_{f(m)})\,-\,2\pi
n_m)^2}{t}}\right]\\[5pt]\label{Gl:SecondOrder}
&&\times F_t(\vec z_V,\,\ldots,\,\vec z_E,\,\vec w_V,\,\ldots,\,\vec
w_E,\,k_1,\ldots,\, k_V).
\end{eqnarray}

\noindent Here, the function $F_t$ captures the complicated
expansion made above. Note however that $F_t$, although depending on
$t$, does not diverge exponentially, as $t\to 0$, in contrast to the
whole overlap in (\ref{Gl:ElementsOnTreeCanBeGaugedTo1}). There, the
fact that the integrand is actually containing an exponential with
$1/t$-dependence, spoiled the possibility of using the method of
stationary phase. If one does not want to compute the whole inner
product, but only the second order of (\ref{Gl:NanuNochNeFormel}),
we can now perform the limit of $t\to 0$. Of course, since $F_t$
contains inverse powers of $t$, the limit of $F_t$ does not exist.
But this could not have been expected, since the inner product
between two coherent states is no function of which the limit $t\to
0$ exist. But, as we will show now, the second derivative of
(\ref{Gl:NanuNochNeFormel}) is of the form that it cancels in the
numerator and the denominator in
(\ref{Gl:OverlapInChapter7}) asymptotically, as $t$ tends to $0$.\\

From the form (\ref{Gl:SecondOrder}) we immediately see that for
small $t$, the integrand is exponentially damped in regions where
not all of the $k_1,\,\ldots,\, k_V$ coincide. This is due to the
fact that the integrand contains Gaussians in the geodesic distance
between pairs of $k_1,\,\ldots,\,k_V$. Note that since the graph
$\g$ is connected, the integrand is really concentrated around
$k_1=\ldots=k_V$ for small $t$. From this we immediately conclude
that, up to orders of $O(t^{\infty})$, we can restrict our
calculations to the summand with $n_1=\ldots n_E=0$. Then we see
that the part of the integrand containing the exponentials
\begin{eqnarray*}
\sqrt{\frac{\pi}{t}}^{3(V-1)}\prod_{m=1}^{E}\exp\left[{-\frac{d(k_{b(m)},\,k_{f(m)})^2}{t}}\right]
\end{eqnarray*}

\noindent effectively behaves as a delta function in the limit $t\to
0$, times a constant $C$, which is given by
\begin{eqnarray}\label{Gl:PreFactorInFrontOfDelta}
C\;=\;\lim_{t\to
0}\;\sqrt{\frac{\pi}{t}}^{3(V-1)}\int_{SU(2)^V}d\m_H(k_1,\,\ldots,\,k_V)\prod_{m=1}^{E}\exp\left[{-\frac{d(k_{b(m)},\,k_{f(m)})^2}{t}}\right].
\end{eqnarray}

\noindent To compute $C$, we use the right-invariance of the Haar
measure and shift all integration variables $k_l$ other than $k_1$
by
\begin{eqnarray*}
k_l\;\longrightarrow\;k_l\,k_1.
\end{eqnarray*}

\noindent We note that this leaves all terms unchanged in which
$k_1$ does not appear, since the geodesic distance is invariant
under right translation. In those terms, in which $k_1$ does appear,
though, the transformation from the other element will effectively
cancel it, which leaves us with
\begin{eqnarray*}
C\;=\;\lim_{t\to
0}\;\sqrt{\frac{\pi}{t}}^{3(V-1)}\int_{SU(2)^V}d\m_H(k_1,\,\ldots,\,k_V)\prod_{m=1}^{E}\exp\left[{-\frac{d(k_{b(m)},\,k_{f(m)})^2}{t}}\right]_{\Bigg|_{k_1=1}}.
\end{eqnarray*}

\noindent Note that now the main contribution to this integral comes
from the vicinity of $k_1=k_2=\ldots=k_V=1$. Writing

\begin{eqnarray*}
k_l\;=\;\exp\big(i\vec\phi_l\cdot\vec\s\big)
\end{eqnarray*}

\noindent we recast the Haar measure of $SU(2)$ into
\begin{eqnarray*}
\int_{SU(2)}d\m_H(k)\;f(k)\;=\;\int_{B_{\pi}(0)}d^3\phi\;\frac{\sin^2\phi}{\phi^2}\;f\left(\exp\big(i\vec\phi\cdot\vec\sigma\big)\right)
\end{eqnarray*}

\noindent  with $B_{\pi}(0)=\{\vec\phi\;|\;\|\vec\phi\|<\pi\}$. Now
note that for $k_l\approx k_{l'}$ we have
\begin{eqnarray*}
d(k_l,\,k_{l'})^2\;=\;\arccos\left(\cos\phi_l\cos\phi_{l'}-\frac{\sin\phi_l}{\phi_l}
\frac{\sin\phi_{l'}}{\phi_{l'}}\vec\phi_l\cdot\vec\phi_{l'}\right)\;=\;\|\vec\phi_l-\vec\phi_{l'}\|^2\;+\;O\Big(\|\vec\phi_l-\vec\phi_{l'}\|^4\Big),
\end{eqnarray*}

\noindent so, in the limit $t\to 0$ we can write
\begin{eqnarray*}
C\;&=&\;\lim_{t\to
0}\;\sqrt{\frac{\pi}{t}}^{3(V-1)}\int_{B_{\pi}(0)^V}\prod_{l=1}^V\left[\frac{\sin^2\phi_l}{\phi^2_l}d^3\phi_l\right]
\prod_{m=1}^{E}\exp\left[{-\frac{\|\vec\phi_{b(m)}-\vec\phi_{f(m)}\|^2}{t}}\right]_{\Bigg|_{\vec\phi_1=0}}\\[5pt]
&=&\;\lim_{t\to
0}\;\sqrt{\frac{\pi}{t}}^{3(V-1)}\int_{B_{\pi}(0)^{V-1}}\prod_{l=2}^{V}\left[\frac{\sin^2\phi_l}{\phi^2_l}d^3\phi_l\right]
\prod_{m=1}^{E}\exp\left[{-\frac{\|\vec\phi_{b(m)}-\vec\phi_{f(m)}\|^2}{t}}\right]_{\Bigg|_{\vec\phi_1=0}}
\end{eqnarray*}

\noindent since the integration over $\vec\phi_1$ is trivial and the
Haar measure is normalized. Since we are integrating a Gaussian that
becomes more and more concentrated around $\vec\phi_l=0$ as $t\to
0$, we make an error of order $O(t^{\infty})$ if we extend the
integration range over all of $\R^{3(V-1)}$. In this limit, since
\begin{eqnarray*}
\frac{\sin^2{\phi_l}}{\phi^2_{l}}_{\Big|_{\vec\phi=0}}\;=\;1
\end{eqnarray*}

\noindent we have
\begin{eqnarray*}
C\;=\;\lim_{t\to
0}\;\sqrt{\frac{\pi}{t}}^{3(V-1)}\;\int_{\R^{3V}}d^{3V}\phi_l\;\exp\left[\sum_{m=1}^E\frac{\|\vec\phi_{b(m)}-\vec\phi_{f(m)}\|^2}{t}\right]\d^{(3)}(\vec\phi_1)
\end{eqnarray*}

\noindent with $\d^{(3)}$ being the three-dimensional delta
distribution. We have already encountered a similar integral in the
case of $G=U(1)$. There, this integral could be solved by
introducing the \emph{incidence matrix} $\l\in Mat(E\times V,\Z)$,
indicating which edges and vertices of the graph $\g$ are linked to
each other.

We note that the exponential function in the integral is invariant
under simultaneous shift of all integration variables by a constant
vector:
\begin{eqnarray*}
\vec\phi_l\;\longrightarrow\;\vec\phi_l+\vec a
\end{eqnarray*}

In this case, we can use a three-dimensional version of the lemma
that has been used in the $U(1)$-case \cite{GICS-I}:

\begin{Lemma}\label{Lem:WieKommtDieDeltaFunktionInDieFlascheIn3D}
Let $f:\R^{3n}\to \C$ be a function with the symmetry
\begin{eqnarray*}
f(\vec \phi_1+\vec a,\ldots,\vec \phi_n+\vec a)\;=\;f(\vec
\phi_1,\ldots,\vec\phi_n)\qquad\mbox{for all }\vec a\in\R^3
\end{eqnarray*}

\noindent such that $\vec \phi_1,\ldots,\vec\phi_{n-1}\to
f(\vec\phi_1,\ldots,\vec\phi_{n-1},0)$ is integrable. Then
\begin{eqnarray}
\int_{\R^{3(n-1)}}d^3\phi_2\cdots&&
d^3\phi_{n}\;f(0,\,\vec\phi_2,\vec\phi_2,\ldots,\vec\phi_{n})\\[5pt]\nonumber
&&\;=\;n^3\int_{\R^{3n}}d^3\phi_1\cdots
d^3\phi_n\;\d^{(3)}\left(\vec\phi_1+\cdots
+\vec\phi_n\right)\,f(\vec\phi_1,\ldots,\vec\phi_n).
\end{eqnarray}
\end{Lemma}

\noindent The proof of this lemma works completely analogous to the
one we have delivered in the case of the gauge-invariant coherent
states with gauge group $U(1)$.

With these ingredients, we can write the constant $C$ as

\begin{eqnarray}\nonumber
C\;&=&\;\lim_{t\to
0}\;\sqrt{\frac{\pi}{t}}^{3(V-1)}\,\int_{\R^{3V}}d^3\phi_1\ldots
d^3\phi_V\;\d^{(3)}\left(\vec\phi_1+\cdots
+\vec\phi_n\right)\;\exp\left[-\sum_{l,l'=1}^V\frac{\l_{ml}\l_{ml'}\vec\phi_l\cdot\vec\phi_{l'}}{t}\right]\\[5pt]\label{Gl:RewritingSU2CaseWithLambdaMatrix}
&=&\;\sqrt{\pi}^{{3}(V-1)}\,\int_{\R^{3V}}d^3\phi_1\ldots
d^3\phi_V\;\d^{(3)}\left(\vec\phi_1+\cdots
+\vec\phi_n\right)\;\exp\left[-\sum_{l,l'=1}^V\phi^T\Lambda\phi\right],
\end{eqnarray}

\noindent by scaling the integration range. Here, $\phi$ stands for
the collection $\phi^I_l$, where $l=1,\ldots V$ go though the list
of vertices, while $I=1,2,3$ denotes the component of $\vec\phi_l$.

Since the graph $\g$ we are considering is connected, Kirchhoff's
theorem \cite{GRAPH} tells us that the matrix
\begin{eqnarray*}
\L_{ll'}^{IJ}\;=\;(\l\l^T)_{ll'}\,\d^{IJ}
\end{eqnarray*}

\noindent has three zero-eigenvectors, spanned by
\begin{eqnarray*}
(\vec a,\ldots\vec a)\;\in\;\R^3\otimes\R^V
\end{eqnarray*}

\noindent for all $\vec a\in\R^3$. The three-dimensional
delta-function in (\ref{Gl:RewritingSU2CaseWithLambdaMatrix})
ensures that the integration ranges over the orthogonal complement
of the zero space of $\L$. But Kirchhoff's theorem tells us that all
other eigenvalues $\m_2,\ldots\m_V$ of $\l\l^T$ are positive, and
their product divided by $V$ gives

\begin{eqnarray*}
G\;=\;\frac{1}{V}\prod_{l=2}^V\m_l,
\end{eqnarray*}

\noindent where $G$ is the number of different possible maximal
trees in $\g$. If we know the eigenvalues of $\l\l^T$, we also know
the ones of $\L$, and can perform the integration
(\ref{Gl:RewritingSU2CaseWithLambdaMatrix}), which only consists of
multiple Gaussians. Finally, we arrive at
\begin{eqnarray}\label{Gl:FactorInFrontOfDelta}
C\;=\;\pi^{3(V-1)}\sqrt\frac{V}{G}^{3}.
\end{eqnarray}

\noindent So, in the limit of $t\to 0$, we see that part of the
integrand (\ref{Gl:SecondOrder}) behaves as a delta distribution. As
already mentioned, $F_t$ does not converge for $t\to 0$, but from
(\ref{Gl:BekotzteEntwicklung}) and
(\ref{Gl:RichtigBekotzteEntwicklung}) we see that $F_t$ is a sum of
terms that are proportional to negative powers of $t$. So, although
the integral (\ref{Gl:SecondOrder}) does not converge for $t\to 0$,
we will compute it asymptotically, which will be enough to see that
the second order of numerator and denominator in
(\ref{Gl:OverlapInChapter7}) cancel in the limit $t\to 0$.

With (\ref{Gl:FactorInFrontOfDelta}) and
(\ref{Gl:PreFactorInFrontOfDelta}), we are able to evaluate
(\ref{Gl:SecondOrder}) asymptotically:

\begin{eqnarray*}
(\text{$2^{\text{nd}}$
order})\;&\sim&\;\frac{e^{\frac{E}{t}}}{\pi^{3E}}\frac{\sqrt{\pi}^{3(E+V-1)}}{\sqrt{t}^{3(E-V+1)}}\sqrt\frac{V}{G}^3\;\int_{SU(2)}d\m_H(k)F_t(\vec
z_V,\,\ldots,\,\vec z_E,\,\vec w_V,\,\ldots,\,\vec
w_E,\,k,\,\ldots,\,k).
\end{eqnarray*}

\noindent So, in the limit $t\to 0$ the integration over all
variables $k_1,\ldots,k_V$ is restricted to the integration over the
submanifold of all $k_1=\ldots=k_V=k$ being equal. On this set,
however, the expansions (\ref{Gl:BekotzteEntwicklung}) and
(\ref{Gl:RichtigBekotzteEntwicklung}) simplify tremendously. These
simplification amount to setting $\vec\phi_l=\vec\psi_l$ in
(\ref{Gl:BekotzteEntwicklung}) and setting $a=1$ (and $n=0$) in
(\ref{Gl:RichtigBekotzteEntwicklung}). With this, one can readily
check that

\begin{eqnarray}\label{Gl:FinaleSuperSimpleFormel}
(\text{$2^{\text{nd}}$
order})\;&\sim&\;-\;\left(1+\frac{t}{3}\right)\frac{e^{\frac{E}{t}}}{\pi^{3E}}\frac{\sqrt{\pi}^{3(E+V-1)}}{\sqrt{t}^{3(E-V+1)}}\sqrt\frac{V}{G}^3\;\frac{
z_m^2+ w_m^2}{t},
\end{eqnarray}

\noindent exactly as in the case for the flower graph. So, as has
happened there, the quadratic orders in numerator and denominator of
(\ref{Gl:OverlapInChapter7}) cancel, and we arrive at
\begin{eqnarray*}
\frac{\Big|\Big\langle\Psi^t_{[1,\ldots,\,1]}\Big|\Psi^t_{[g_1,\ldots,g_E]}\Big\rangle\Big|}
{\Big\|\Psi^t_{[1,\ldots,\,1]}\Big\|^2\,\Big\|\Psi_{[g_1,\ldots,g_E]}\Big\|^2}\;\longrightarrow\;1\;-\;O(\|\vec
w_m\|^4).
\end{eqnarray*}

\noindent Here, the arrow denotes that the quadratic order vanishes
in the limit $t\to 0$, so that at most the quartic order remains.
So, for small $t$, one can expect the gauge-invariant coherent state
$\Psi_{[1,\ldots,1]}^t$ to be no Gaussian.\\

\noindent A couple of remarks are in order:

First, note that all of the above considerations could also have
been carried out with any distribution of $\pm 1$ instead of just
$1$. In particular, all that we have derived above for
$\Psi^t_{[1,\ldots,1]}$ is equally true for all
$\Psi^t_{[\pm1,\ldots,\pm1]}$. That is, states that are labeled by
points on which the gauge group $SU(2)^V$ acts trivial on
$SL(2,\C)^E$ in the case of the $E$-flower graph. For arbitrary
graphs, the orbit of any distribution $\pm1,\ldots,\pm1$ along the
leaves of a graph and $1$ on the edges of the tree $\t$ is left
invariant by the gauge transformations $k_1=k_2=\ldots k_V\equiv k$,
unlike generic points in $SL(2,\C)^E$. So, the gauge orbit of
$[\pm1,\ldots,\pm1]$ has three dimensions less than orbits of
generic points. In particular, the gauge-invariant states have
different peakedness properties when labeled at these points, where
the gauge orbits do not have the full dimension. These points
correspond to the singular points on the orbifold which consists of
the gauge orbits.


Second, while the theorem about the qualitatively different
peakedness behavior of $\Psi^t_{[\pm1,\ldots,\pm1]}$ could be shown
for finite $t$ in the case of the $E$-flower graph, for arbitrary
graphs we could only establish this theorem in the limit $t\to 0$.

However, we believe this theorem even to be true for arbitrary,
finite $t$. There are two hints that support this conjecture: First,
it is true for $E$-flower graphs. For arbitrary graphs it could not
be shown due to the complicated form of the remaining integral
(\ref{Gl:RelateInnerProductsToThoseOnFlowerGraphs}), but this does
not mean it is not true. Rather, the degenerate points of the gauge
orbits are present in the gauge orbit space for every graph, and are
generic for non-abelian gauge theories. So flower graphs do not seem
to be special in this respect.

The second reason why we believe this to be true for arbitrary
graphs is that the flatness of the overlap function could also be
seen numerically for the 3-bridge graph, even for values of $t$ that
are not incredibly tiny. So, we think that states labeled on gauge
orbits degenerate to a point exhibit a qualitatively different
peakedness behavior than states labeled on generic gauge orbits.

\section{Summary and Conclusion}
\subsection{Summary of the work}
\noindent This work constitutes the second part of a pair of
articles that investigate the gauge-invariant coherent states for
LQG. The first article investigated the simpler model of abelian
gauge group $G=U(1)$, from which $G=U(1)^3$, which has been employed
in LQG, can be immediately obtained. This article considered the
much more complicated, but also more realistic case of $G=SU(2)$.

One of the results of this work is to show the peakedness properties
of the gauge-invariant coherent states. For the simplest example of
a $1$-flower graph this could be done analytically and showed
interesting features. The gauge-invariant states are labeled by
gauge orbits, and the overlap (\ref{Gl:FinalerUeberlapp}) between
two gauge-invariant coherent states exhibits a peak structure at the
point where both gauge orbits coincide. The width of this peak is
proportional to the semiclassicality parameter $t$. Due to the fact
that the space of the gauge-orbits is no manifold (rather, it is an
'orbifold'), the peakedness of the states is no clean Gaussian, but
a more complicated function, that still tends to a Gaussian if $t$
goes to zero. This shows that states that are labeled by gauge
orbits have useful semiclassical properties. In particular, the
limit $t\to 0$ corresponds to the classical limit $\hbar\to 0$, in
which the state approaches the classical gauge-invariant state
determined by the gauge orbit. This stays true for states labeled at
degenerate points of the orbifold, but in this case the state never
approaches a Gaussian, not even for small $t$. Rather, the overlap
exhibits a plateau that is much flatter than a Gaussian. In
particular, the second derivatives at the maximum of the peak vanish
along with all the odd derivatives, leading to a function that has a
$e^{-x^4/t}$-profile, rather than a Gaussian one.

Similar features could be established numerically for more
complicated graphs, in particular the 2-flower-, the 3-bridge- and
the tetrahedron graph. To investigate the overlaps on these graphs,
we used a gauge-fixing procedure to separate the gauge-invariant
degrees of freedom from the ones that are pure gauge. Not only did
this show a lot about the general procedure how this can be done, it
also enabled us to work entirely with gauge-invariant variables,
which made the peakedness properties for these states transparent.
In fact, also on the 2-flower-,  the 3-bridge- and the tetrahedron
graph the peakedness properties for generic points of the space of
gauge orbits could be seen. Additionally, the states on the
2-flower- and the 3-bridge graph showed the same change of
peakedness structure at points that correspond to degenerate gauge
orbits, i.e. orbits in $SL(2,\C)^E$ under the gauge action
$SL(2,\C)^V$, whose dimension is less than $6V$.

Apart from the special graphs, we have also derived some results for
arbitrary graphs. First, we were able to prove that the 'flattening'
of the overlap of states labeled at degenerate gauge orbits, which
consist only of a point, is generic for $E$-flower graphs, i.e. for
graphs with one vertex and $E$ edges all emerging and ending at that
vertex. So this is not only a coincidence because of the simple
graphs we have chosen, but rather this is true for any $E$-flower
graph.

Second, we have generalized the gauge-fixing procedure that helped
us in our numerical examples to extract the gauge-invariant
information from the overlap expression for states on arbitrary
graphs. This enabled us to establish a relation between the inner
product of states on arbitrary graphs with $E$ edges and $V$
vertices with states on $E-V+1$-flower graphs. In particular,
(\ref{Gl:RelateInnerProductsToThoseOnFlowerGraphs}) shows this
relation.

After that, we used this relation to extend the theorem about the
peakedness properties of states labeled by degenerate gauge orbits
from flower- to arbitrary graphs. Unfortunately, in this case the
theorem could only be established in the limit $t\to 0$, since only
in this limit the expression became tractable. However, there are
hints that the theorem is in fact true for finite $t$ as well,
although the proof seems to be much harder than for the limit $t\to
0$.\\

%

Although the gauge-invariant coherent states labeled by degenerate
orbits are not Gaussian-peaked but have a peak that is much flatter,
this does not spoil the semiclassical properties of gauge-invariant
coherent states. First, the states are still peaked, the peak
profile is just not the nice, clean Gaussian that one is used to from
the harmonic oscillator coherent states or (approximately) from the
complexifier coherent states on graphs. But still, the width of the
peak is proportional to (a fractional power of) $t$, which indicates
that the limit $t\to 0$ corresponds to the semiclassical limit, in
which the state approaches a point in classical, gauge-invariant
phase space.

It is in fact not surprising at all that the peakedness of the
states labeled at degenerate gauge orbits is qualitatively
different. From a mathematical point of view the degenerate gauge
orbits correspond to singular points ('edges' or 'corners') in the
gauge-invariant phase space, which is, as already pointed out, no
manifold. But also from a physical point of view this is not
disturbing: The case for instance, where all edge labels are $1\in
SL(2,\C)$ corresponds to a state which is labeled by the physical
distribution of the Ashtekar connection $A_a^I=0$ and $E_I^a=0$
along these edges \cite{GCS1, CCS}. So, this case corresponds to a
highly degenerate metric. It could have guessed that these states
exhibit a qualitatively different behavior than states that are
labeled by elements which approximate, say, flat Minkowski space.

This feature will become important in the case of
diffeomorphism-invariant coherent states, where there are many
different types of degenerate gauge orbits, corresponding to
symmetrical metric configurations. Since one is particularly
interested in these situations (i.e. Minkowski space), one should
expect peculiar peakedness properties for diff-invariant coherent
states labeled by configurations corresponding to these symmetric
situations.\\

Another point investigated in this article concerns the complexifier
coherent states, and is of a more mathematical nature. For the inner
product between two states a formula could be found that depends
entirely on the geometry of the complexified gauge group. In
particular, for $G=U(1)$ as well as for $G=SU(2)$, an expression
could be derived that involves the complex lengths of geodesics on
$G^{\C}$ (\ref{Gl:InnerProductAsSumOverGeodesics}). In particular,
if both states are labeled on $G\subset G^{\C}$, then the inner
product is a sum over all geodesics, involving terms proportional to
Gaussians in the length of these geodesic measured by the Killing
metric. For the case of $G=U(1)$ this was rather trivial and seemed
to be a coincidence, while the corresponding formula for $G=SU(2)$
came more as a surprise. It raises the hope that a similar formula
can be shown for CCS on arbitrary compact Lie groups $G$. There are
in fact hints that support this conjecture:

Remember that the complexifier coherent states (choosing $\hat
C=-\Delta$ as the complexifier) labeled by points on $G$ are nothing
but solutions of the heat equation \cite{HALL1, GCS2}). In
particular, the norm of the CCS $\psi^t_g$ is equal to 1 as $g\in
G$. Moreover, the inner product is given by
\begin{eqnarray}
\langle\,\psi^t_g|\psi_{g'}^t\,\rangle\;=\;\psi^{2t}_{1}(g^{-1}g'),
\end{eqnarray}

\noindent so it also solves the heat equation (in $t$). But the heat
equation can be thought of releasing a random walker at $1\in G$,
letting it walk along $G$ for some finite time $t$, and then measure
the probability distribution $\rho(t,g)=\psi^t_1(g)$ of where he is
on $G$. Obviously, the smaller the allowed time $t$ or farther the
geodesic distance (defined by the Killing metric on $G$, which is
positive definite) between $1$ and $g$, the smaller the probability
$\rho(t,g)$. In particular, on $\R^n$ this probability is given by a
Gaussian in the geodesic distance, which might also be true on
arbitrary $G$. Furthermore, the sum over all geodesics from $1$ to
$g$ arises naturally, since it encodes the different ways the random
walker could have taken on $G$ to walk from $1$ to $g$, since the
topology of $G$ will in general be nontrivial.

These considerations lead to the possibility  that one could be able
to define a generalization of
(\ref{Gl:InnerProductAsSumOverGeodesics}) to arbitrary compact Lie
groups. We hope to be able to address  this point in some future
work.

\subsection{Conclusion and outlook}

\noindent In \cite{GICS-I} and the present article we have shown
that the kinematical complexifier coherent states, which are
convenient tools for investigating the semiclassical limit of the
kinematical sector of LQG, can be projected to the gauge-invariant
subspace, and the resulting gauge-invariant coherent states are
suitable for addressing semiclassical issues on the gauge-invariant
sector.

It may seem quite discouraging that the Gauss invariant coherent states for LQG considered here are difficult to handle
analytically. However, one should keep in mind that in this paper we only investigated the integral formula 
(\ref{Gl:EichinvariantesSkalarprodukt}) rather than the sum over intertwiners such as 
(\ref{Gl:EichinvarianteKohaerenteZustaendeNachIntertwinernZerlegt}). It may well be that using asymptotic formulae 
for large spin for Clebsch Gordon coefficients, $6j$-symbols, etc. (see e.g. \cite{ASYM1, ASYM2}) one can gain more analytical control.
Also it may be that we overlooked some clever technique that allows to simplify the gauge integrals. We hope to come back to this point in some future publication.  In any case, numerically the Gauss invariant states are well under control, although one needs to write an adapted code to handle arbitrary complicated graphs.\\

The next obvious step will be to consider the action of the
diffeomorphism group on the set of (gauge-invariant) coherent
states. In particular, the projection of the gauge-invariant
coherent states to the diffeomorphism-invariant Hilbert space
$\H_{\text{diff}}$ via some sort of rigging map, in order to arrive
at \emph{diffeomorphism-invariant coherent states}. This space is -
other than the gauge-invariant Hilbert space - not completely under
control, in particular, there is no unique definition of
$\H_{\text{diff}}$ \cite{INTRO, ALLMT}. So, this task will be
significantly more challenging than the - conceptually quite clear -
definition of the gauge-invariant coherent states. But  this could
grant a way to investigate all the different possibilities to define
$\H_{\text{diff}}$, and maybe even distinguish some of them as more
suitable than others. Furthermore, as soon as the
diffeomorphism-invariant coherent states are defined, approximations
and semiclassical techniques for the graph-changing version of the
master constraint \cite{PHOENIX, QSD8} become available, which
encodes the Hamiltonian constraints and can be defined on
$\H_{\text{diff}}$.

\section*{Acknowledgements}\nonumber

\noindent BB would like to thank Christian B\"ar for knowledge about
cohomology classes, Bianca Dittrich for the discussion about
degenerate gauge orbits, Hendryk Pfeiffer for the support and
discussions, and Klaus Wirthm\"uller for his time and patience.
Research at the Perimeter Institute for Theoretical Physics is
supported by the Government of Canada through NSERC and by the
Province of Ontario.

\appendix

\section*{Appendix}

\section{Cohomology groups with values in non-abelian
groups}\label{Ch:CohomologyNonabeilan}

\noindent In \cite{GICS-I}, we have shown the definition for
simplicial cohomology with values in abelian groups. The definition
is in fact standard and works exactly the same way as for, say $\Z$
or $\R$. However, defining cohomology classes with values in
non-abelian groups is not quite that simple. In particular, due to
the non-abelianness of $G$, the coboundary operator $\d$ can only
sensibly defined for one-dimensional cell-complexes. Furthermore, it
is no group homomorphism any longer, which results in the
corresponding cohomology "groups" to being no groups at all.
Finally, the resulting spaces are no manifolds any longer, but
rather spaces with "edges" and "corners". These singular points will
turn out to have some implications of the peakedness behavior of the
coherent states labeled with these points.

In the following, we will give the definition for cohomology with
values in non-abelian groups. Although this definition seems to be
common knowledge, we have not found anything about it in the
literature.

In the following, we assume the reader to be familiar with the
concepts of homology and cohomology. Throughout the text, we will
use the notation

\begin{eqnarray}
A^B\;:=\;\{f:B\to A\text{ any map}\}
\end{eqnarray}

\noindent for the set of maps from any set $B$ to any set $A$. Let
$\g$ be a directed graph consisting of a set of vertices $V(\g)$ and
a set of edges $E(\g)$. With $V:=|V(\g)|$ and $E:=|E(\g)|$, the sets
$G^{V(\g)}$ and $G^{E(\g)}$ are the set of all maps from the set of
vertices or, respectively, edges to $G$. These spaces are naturally
groups and the staring point for the definition of cohomology. What
is needed in order to define the cohomology groups now are the
coboundary operators $\d$ in the sequence

\begin{eqnarray*}
\{1\}\;\stackrel{\d}{\longmapsto}\;G^{V(\g)}\;\stackrel{\d}{\longmapsto}\;G^{E(\g)}\;\stackrel{\d}{\longmapsto}\;\{1\}.
\end{eqnarray*}

\noindent With these operators, the natural way to define cohomology
classes would be by dividing out images from kernels:

\begin{eqnarray}\label{Gl:DefinitionZerothCohomology}
H^0(\g,G)\;&=&\;\frac{\ker\;\d:G^{V(\g)}\to
G^{E(\g)}}{\text{img}\,\d:\{1\}\to
G^{V(\g)}}\;=\;\ker\;\d:G^{V(\g)}\to
G^{E(\g)},\\[5pt]\label{Gl:DefinitionFirstCohomology}
H^1(\g,G)\;&=&\;\frac{\ker\;\d:G^{E(\g)}\to
\{1\}}{\text{img}\,\d:G^{V(\g)}\to
G^{E(\g)}}\;=\;\frac{G^{E(\g)}}{\text{img}\,\d:G^{V(\g)}\to
G^{E(\g)}}.
\end{eqnarray}

\noindent The map $\d:\{1\}\to G^{V(\g)}$ makes sense, trivially, as
well as $\d:G^{E(\g)}\to \{1\}$.  However, the naive definition for
$\d:G^{V(\g)}\to G^{E(\g)}$, which works in the abelian case,

\begin{eqnarray}\label{Gl:WannabeCoboundaryOperator}
\d(k_{v_1},\ldots,k_{v_V})\;\longmapsto\;\left(k_{b(e_1)}k_{f(e_1)}^{-1},\ldots,k_{b(e_E)}k_{f(e_E)}^{-1}\right),
\end{eqnarray}

\noindent  does not make $\d$ into a group homomorphism, since $G$
is not abelian, as one can readily see. But if $\d:G^{V(\g)}\to
G^{E(\g)}$ is no group homomorphism, then its image is not a
subgroup of $G^{E(\g)}$, hence it is not clear what the quotient
(\ref{Gl:DefinitionFirstCohomology}) should be.

The only way to generalize the definition of $\d:G^{V(\g)}\to
G^{E(\g)}$ to the case of non-abelian $G$ goes as follows. Every
subgroup of $G^{E(\g)}$ defines a (say, right) action on
$G^{E(\g)}$, simply by (right) multiplication. In fact, this is the
way in which one usually uses the image of $\d:G^{V(\g)}\to
G^{E(\g)}$ in the definition of $H^1(\g, G)$, since $H^1(\g,G)$ is
simply the space of orbits under the action of the group
$\d(G^{V(\g)})\subset G^{E(\g)}$. So, it is sufficient to define
$\d$ to be a map from $G^{V(\g)}$ to the set of group homomorphisms
of $G^{E(\g)}$ into itself. In other words, $\d$ defines an action
of $G^{V(\g)}$ on $G^{E(\g)}$. This definition can easily be
generalized to the non-abelian case, via:
\begin{eqnarray}\label{Gl:DefinitionOfDeltaInNonabelianCase}
\d:\;G^{V(\g)}\;&\longrightarrow&\;\text{Hom
}(G^{E(\g)},\,G^{E(\g)})\\[5pt]\nonumber
\d(k_{v_1},\ldots,k_{v_V})&\;:=\;&\Big((h_{e_1},\ldots,h_{e_E})\,\to\,(k_{b(e_1)}h_{e_1}k^{-1}_{f(e_1)},\ldots,k_{b(e_E)}h_{e_E}k^{-1}_{f(e_E)})\Big).
\end{eqnarray}

\noindent Of course, for the case of abelian $G$, this definition
reduces to (\ref{Gl:WannabeCoboundaryOperator}). In particular, for
abelian $G$ the homomorphism
(\ref{Gl:DefinitionOfDeltaInNonabelianCase}) is induced by a right
multiplication of an element in $G^{E(\g)}$, which is not the case
for non-abelian $G$. With this definition, it is possible to
generalize the definition of the first two cohomology groups
$H^0(\g,G)$ and $H^1(\g,G)$ to the case of non-abelian $G$:

Define, in analogy to (\ref{Gl:DefinitionZerothCohomology}) and
(\ref{Gl:DefinitionFirstCohomology})
\begin{eqnarray}\label{Gl:DefinitionZerothCohomologyNonabelian}
H^0(\g,G)\;&:=&\;\ker\;\d:\;G^{V(\g)}\;\longrightarrow\;\text{Hom
}(G^{E(\g)},\,G^{E(\g)})\\[5pt]\label{Gl:DefinitionFirstCohomologyNonabelian}
H^1(\g,G)\;&:=&\;\frac{G^{E(\g)}}{\text{img}\,\d:G^{V(\g)}\to
\text{Hom }(G^{E(\g)}, G^{E(\g)})}.
\end{eqnarray}

\noindent In the case of abelian groups, $H^0(\g,G)$ consist of all
elements in $G^{V(\g)}$ that are mapped to the unit element in
$G^{E(\g)}$. Here, $H^0(\g,G)$ consist of all elements in
$G^{V(\g)}$ that are mapped to the trivial group homomorphism of
$G^{E(\g)}$ into itself. It can readily be seen from
(\ref{Gl:DefinitionOfDeltaInNonabelianCase}) that $H^0(\g,G)$
consists only of one element, in particular the unit element in
$G^{V(\g)}$:
\begin{eqnarray*}
H^0(\g,G)\;\simeq\;\{1\}.
\end{eqnarray*}

\noindent This shows the first important difference between
cohomology with values in abelian groups and cohomology in
nonabelian groups: In the abelian case, the coboundary operator $\d$
maps an element in $G^{V(\g)}$ of the form $(k,\ldots,k)$ with $k\in
G$ to the unit element, whereas in the nonabelian case, this element
is mapped to a nontrivial element of $\text{Hom
}(G^{E(\g)},G^{E(\g)})$.

The quotient in (\ref{Gl:DefinitionFirstCohomologyNonabelian}) is,
in the nonabelian case, no quotient between a group and one of its
subgroups, but rather the set of equivalence classes of the manifold
$G^{E(\g)}$ under the action of $G^{V(\g)}$ on $G^{E(\g)}$, defined
by (\ref{Gl:DefinitionOfDeltaInNonabelianCase}).

Here we can see the second important difference between cohomologies
with values in abelian and nonabelian groups: The action of
$G^{V(\g)}$ is not effective on each of the orbits in $G^{E(\g)}$,
i.e. the orbits are not all diffeomorphic to each other. This can be
seen easily: Let $Z\subset G$ be the center of $G$, i.e. the set of
group elements commuting with every other group element. Since $z\in
Z$ commutes with all elements in $G$, one can see by
(\ref{Gl:DefinitionOfDeltaInNonabelianCase}) that the orbit of
$(z,\ldots,z)$ under the action of $G^{V(\g)}$ consists just of one
point $\{(z,\ldots,z)\}$, in contrast to any element not being in
the center.

This has two serious consequences: First, for nonabelian $G$ the set
$H^1(\g,G)$ does not inherit any group structure from $G^{E(\g)}$,
while for abelian $G$ this is always the case. Second, giving
$H^1(\g,G)$ the quotient atlas of $G^{E(\g)}$ under the action of
$G^{V(\g)}$ does not turn it into a manifold at all. Rather, it will
be a manifold with certain singular points and higher dimensional
singular manifolds, which correspond to the elements of the center
of $Z^{|E(\g)|}$.\\

The easiest example for such an occurrence is the case of $G=SU(2)$
and a graph $\g$ being just one vertex and one edge starting and
ending at that vertex. The action of $G^{V(\g)}\simeq SU(2)$ on
$G^{E(\g)}\simeq SU(2)$ is then simply given by conjugation:
\begin{eqnarray*}
\d k\,\cdot\,h\;=\;khk^{-1}.
\end{eqnarray*}

\noindent It can easily be seen that the only invariant information
of $SU(2)$ elements $g$ under conjugation is the trace $\tr\, g$,
which is a number $\tr\, g\in[-2,2]$ in a closed interval. So
$H^1(\g,G)\simeq\,[-2,2]$, which is clearly neither a group nor a
manifold, but has a boundary consisting of $\{-2,2\}$, which are the
orbits of $\{-1,1\}\in SU(2)$ under conjugation.\\

Still, although $H^1(\g,G)$ is no Lie group or even a manifold, one
can still characterize it with the help of a gauge-fixing procedure
similar to the one employed in \cite{GICS-I} to describe abelian
cohomology.\\

For this, we choose a maximal tree $\t\subset\g$. Remember that this
means that $\t$ contains no loops and meets every vertex of $\g$ :
$V(\t)=V(\g)$. In particular, between any two vertices in $\g$,
there is a unique path which takes only edges from $\t$. Note that
$\t$ has $V$ vertices and $V-1$ edges.

Every element $h\in G^{E(\g)}$ can be seen as a distribution of
elements of $G$ among the edges of $\g$. We now define, for each
element $h=(h_{e_1},\ldots,h_{e_E})$ of $G^{E(\g)}$ an element $k\in
G^{V(\g)}$ such that $\d(k)\cdot h$ is an element of $G^{E(\g)}$
with the unit element $1\in G$ among all edges in $\t$.

Choose a vertex $v\in V(\g)$. Then, for every $w\in V(\g)$ define

\begin{eqnarray}\label{Gl:NochEineEichfixierungNichtabelsch}
k_w\;=\;h_{e_1}^{\pm 1}h_{e_{2}}^{\pm 1}\cdots h_{e_n}^{\pm 1},
\end{eqnarray}

\noindent where $e_{1},\ldots e_n$ are the edges in $\t$
constituting the unique path from $v\in V(\g)$ to $w\in V(\g)$ in
$\t$, and the either the element $h_{e_j}$ or its inverse
$h_{e_j}^{-1}$ has to be taken, depending on whether the path from
$v$ to $w$ goes with or against the orientation of $e_j$,
respectively.

This defines an element $k=(k_{v_1},\ldots,k_{v_V})\in G^{V(\g)}$.
We claim that the distribution of elements among edges, which is
given by $\d k\cdot h$, has the neutral element $1\in G$ at every
edge which belongs to $\t$. To see this, consider
(\ref{Gl:DefinitionOfDeltaInNonabelianCase}). Let furthermore $e\in
E(\t)$ be an edge of $\t$.  If the path from $v$ to $b(e)$ does not
pass through $e$, then the path from $v$ to $f(e)$ does, and it does
so with the orientation of $e$. In this case, by
(\ref{Gl:NochEineEichfixierungNichtabelsch})
\begin{eqnarray}
k_{f(e)}\;=\;k_{b(e)}\,h_e.
\end{eqnarray}

\noindent In the case that the path from $v$ to $b(e)$ passes
through $e$, then it first passes the vertex $f(e)$, i.e. goes
opposite to the orientation of $e$. In this case, by
(\ref{Gl:NochEineEichfixierungNichtabelsch}), we have
\begin{eqnarray}
k_{b(e)}\;=\;k_{f(e)}\,h_e^{-1}.
\end{eqnarray}

\noindent In both cases, the element of $\d k\cdot h$ at $e$ is
given by
\begin{eqnarray}
k_{b(e)}\,h_e\,k_{f(e)}^{-1}\;=\;1.
\end{eqnarray}

\noindent This shows that, as in the abelian case, where a similar
procedure could be used, each representant of an element
$[h_{e_1},\ldots, h_{e_E}]$ in $H^1(\g,G)$ is equivalent to one
$[\tilde h_{e_1},\ldots, \tilde h_{e_E}]$ with $\tilde h_e=1$ for
$e$ being an edge in $\t$. Note that, since $G$ is not abelian, the
order of elements in (\ref{Gl:NochEineEichfixierungNichtabelsch})
now plays a crucial role.

In the abelian case, the resulting distributions of elements in $G$
along the leaves of $\g$, i.e. the edges that do not belong to $\t$,
was unique. This allowed us to identify $H^1(\g,G)$ with $\text{Hom
}(\pi_1(\g),G)$, where $\pi_1(\g)$ denotes the first fundamental
group of $\g$. In the non-abelian case however, the element
$\d(k,\ldots,k)$, which leaves the distribution of $1\in G$ along
the edges of $\t$ invariant, does no longer leave the values of
$\d(k)\cdot h$ along the leaves invariant. Rather, it acts as a
conjugation on these elements. Hence, fixing the elements on the
edges of $\t$ to $1$ determines the set of elements along the edges
not in $\t$ only up to global conjugation:
\begin{eqnarray*}
(h_{e_1},\ldots,h_{e_L})\;\sim\;(kh_{e_1}k^{-1},\ldots,kh_{e_L}k^{-1})\qquad
k\in G.
\end{eqnarray*}

Thus, $H^1(\g,G)$ is not diffeomorphic to $\text{Hom
}(\pi_1(\g),G)\simeq G^L$, but can rather be identified by this
space modulo global conjugation:
\begin{eqnarray*}
H^1(\g,G)\;\simeq\;G^L/G\;\simeq\;\text{Hom }(\pi_1(\g),G)\,/\,G.
\end{eqnarray*}

\noindent Although $H^1(\g,G)$ can be constructed in this way, there
is no way to define the higher cohomology groups $H^k(\g,G)$, $k>1$,
in a similar manner. The case of $k=1$ is still tractable, since
every $1$-cell $e\in E(\g)$ has a boundary consisting (at most) of
two points. together with the orientation of $e$, this allowed for a
canonical way of letting the elements of $G^{V(\g)}$ act on
$G^{E(\g)}$, in particular via
(\ref{Gl:NochEineEichfixierungNichtabelsch}). But, as soon as the
boundary of a $k$-cell consists of more than two $k-1$-cells, there
is no canonical way of ordering the elements in
(\ref{Gl:NochEineEichfixierungNichtabelsch}). So, a general
definition of $H^k(K,G)$ for non-abelian groups and $k>1$ is not
possible.


\subsection{Gauge-invariant functions}

\noindent The notion of gauge-invariant cylindrical functions fits
nicely into the framework of cohomology. Remember that a
gauge-variant function on a graph $\g$ is given by a function of a
number of copies of the gauge group $G$:
\begin{eqnarray*}
f:\underbrace{G\times\cdots\times G}_{\mbox{one for each edge in
}E(\g)}\;\to\;\C
\end{eqnarray*}

\noindent that is square-integrable with respect to the product
Haar-measure $d\m_H^{\otimes|E(\g)|}$. These function constitute the
Hilbert space $\H_{\g}$, and with the notions of the previous
sections, we identify this space to be
\begin{eqnarray*}
\H_{\g}\;=\;L^2\big(G^{E(\g)},\;d\m_H^{\otimes|E(\g)|}\big).
\end{eqnarray*}

\noindent The gauge transformed $f$ is determined by letting the
gauge group $G$ act on every vertex $v\in V(\g)$ by the following
rule:
\begin{eqnarray*}
\a_{k_{v_1},\ldots,k_{v_V}}f\;\big(h_{e_1},\ldots,h_{e_E}\big)\;=\;f\big(k_{b(e_1)}^{-1}h_{e_1}
k_{f(e_1)},\;\ldots\;,k_{b(e_E)}^{-1}h_{e_E} k_{f(e_E)}\big),
\end{eqnarray*}

\noindent where $b(e)$ and $f(e)$ are the vertices sitting at the
beginning and the end of the edge $e$ respectively. \\

Not only do we recognize the gauge transformation group as the space
$G^{V(\g)}$ from the previous section, one can see readily the
connection between the gauge transformation $\a$ and the coboundary
operator $\d$:
\begin{eqnarray*}
(\a_{g_1,\ldots,g_V}f)(h_1,\ldots,h_E)\;=\;f\big(\d(g_1,\ldots,g_V)\cdot
(h_1,\ldots,h_E)\big),
\end{eqnarray*}

\noindent where $\cdot$ means group multiplication in $G^{E(\g)}$
for the abelian case, or action of $\text{Hom
}(G^{E(\g)},G^{E(\g)})$ on $G^{E(\g)}$ in the non-abelian case. So,
the gauge-invariant functions on the graph $\g$ are just the
functions on the group $G^{E(\g)}$ that are invariant under the
action of $\d(G^{V(\g)})$.
We conclude that the set of gauge-invariant functions is equal to
the set $L^2\big(H^1(\g,G),\,d\m\big)$, where the measure $d\m$ is
the quotient measure of $d\m_H^{\otimes|E(\g)|}$ under the action of
the gauge transformation group ${G^{V(\g)}}$. Although $H^1(\g,G)$
might not be a manifold, it still is a compact Hausdorff space,
which makes the corresponding $L^2$-space tractable.

Note that, since the number of leaves, i.e. the number of elements
which freely generate the first fundamental group $\pi_1(\g)$ of a
graph $\g$, is $E-V+1$. Since, for abelian gauge groups, the first
cohomology group is $\text{Hom }(\pi_1(\g),G)$, it is a manifold of
dimension $\dim G(E-V+1)$. However, for non-abelian gauge groups,
the first cohomology group is $\text{Hom }(\pi_1(\g),G)\,/\,G$.
Apart from the points where the orbits of the conjugation are
degenerate, this describes a manifold with only $\dim G(E-V)$
dimensions. This is another important difference between abelian and
non-abelian gauge groups.


\end{document}